\documentclass[longauth]{aa} 
\pdfoutput=1
\usepackage{graphicx}
\usepackage{txfonts}
\usepackage{pdflscape}

\usepackage{natbib,twoopt}
\usepackage{amsmath} 
\usepackage[breaklinks=true]{hyperref} 
\bibpunct{(}{)}{;}{a}{}{,} 
\makeatletter
\newcommandtwoopt{\citeads}[3][][]{\href{http://adsabs.harvard.edu/abs/#3}%
{\def\hyper@linkstart##1##2{}%
\let\hyper@linkend\@empty\citealp[#1][#2]{#3}}}
\newcommandtwoopt{\citepads}[3][][]{\href{http://adsabs.harvard.edu/abs/#3}%
{\def\hyper@linkstart##1##2{}%
\let\hyper@linkend\@empty\citep[#1][#2]{#3}}}
\newcommandtwoopt{\citetads}[3][][]{\href{http://adsabs.harvard.edu/abs/#3}%
{\def\hyper@linkstart##1##2{}%
\let\hyper@linkend\@empty\citet[#1][#2]{#3}}}
\newcommandtwoopt{\citeyearads}[3][][]%
{\href{http://adsabs.harvard.edu/abs/#3}
{\def\hyper@linkstart##1##2{}%
\let\hyper@linkend\@empty\citeyear[#1][#2]{#3}}}
\makeatother

\usepackage{color}
\definecolor{mygreen}{RGB}{0,128,0}

\hypersetup{colorlinks=true,linkcolor=blue,citecolor=blue,urlcolor=blue}

\begin{document}

   \title{Gaia Data Release 1}

   \subtitle{Astrometry -- one billion positions, two million proper motions and parallaxes}


\author{
L.~Lindegren\inst{\ref{inst:lund}} 
\and 
U.~Lammers\inst{\ref{inst:esac}} 
\and 
U.~Bastian\inst{\ref{inst:ari}} 
\and 
J.~Hern{\'a}ndez\inst{\ref{inst:esac}} 
\and 
S.~Klioner\inst{\ref{inst:tud}} 
\and
D.~Hobbs\inst{\ref{inst:lund}} 
\and 
A.~Bombrun\inst{\ref{inst:hespaceesac}} 
\and 
D.~Michalik\inst{\ref{inst:lund}} 
\and 
M.~Ramos-Lerate\inst{\ref{inst:vitrocisetesac}} 
\and A.~Butkevich\inst{\ref{inst:tud}}
\and G.~Comoretto\inst{\ref{inst:vegaesac}}
\and E.~Joliet\inst{\ref{inst:nasa},\ref{inst:hespaceesac}} 
\and B.~Holl\inst{\ref{inst:geneva}} 
\and A.~Hutton\inst{\ref{inst:auroraesac}} 
\and P.~Parsons\inst{\ref{inst:theserverlabsesac}}
\and H.~Steidelm\"{u}ller\inst{\ref{inst:tud}}
%
%
%
%
%
%
\and U.~Abbas\inst{\ref{inst:oato}} 
\and M.~Altmann\inst{\ref{inst:ari},\ref{inst:paris}}
\and A.~Andrei\inst{\ref{inst:rio}} 
\and S.~Anton\inst{\ref{inst:porto},\ref{inst:lisbon}}
\and N.~Bach\inst{\ref{inst:auroraesac}} 
\and C.~Barache\inst{\ref{inst:paris}}
\and U.~Becciani\inst{\ref{inst:catania}} 
\and J.~Berthier\inst{\ref{inst:imcce}}
\and L.~Bianchi\inst{\ref{inst:eurix}}
\and M.~Biermann\inst{\ref{inst:ari}}
\and S.~Bouquillon\inst{\ref{inst:paris}}
\and G.~Bourda\inst{\ref{inst:bordeaux1},\ref{inst:bordeaux2}}
\and T.~Br{\"u}semeister\inst{\ref{inst:ari}}
\and B.~Bucciarelli\inst{\ref{inst:oato}}
\and D.~Busonero\inst{\ref{inst:oato}} 
\and T.~Carlucci\inst{\ref{inst:paris}}
\and J.~Casta\~neda\inst{\ref{inst:ieec}}
\and P.~Charlot\inst{\ref{inst:bordeaux1},\ref{inst:bordeaux2}}
\and M.~Clotet\inst{\ref{inst:ieec}}
\and M.~Crosta\inst{\ref{inst:oato}}
\and M.~Davidson\inst{\ref{inst:uoe}} 
\and F.~de~Felice\inst{\ref{inst:padova}}
\and R.~Drimmel\inst{\ref{inst:oato}} 
\and C.~Fabricius\inst{\ref{inst:ieec}}  
\and A.~Fienga\inst{\ref{inst:oca}}
\and F.~Figueras\inst{\ref{inst:ieec}}
\and E.~Fraile\inst{\ref{inst:rheaesac}}
\and M.~Gai\inst{\ref{inst:oato}} 
\and N.~Garralda\inst{\ref{inst:ieec}} 
\and R.~Geyer\inst{\ref{inst:tud}}
\and J.J.~Gonz\'alez-Vidal\inst{\ref{inst:ieec}}  
\and R.~Guerra\inst{\ref{inst:esac}} 
\and N.C.~Hambly\inst{\ref{inst:uoe}} 
\and M.~Hauser\inst{\ref{inst:ari}}
\and S.~Jordan\inst{\ref{inst:ari}}
\and M.G.~Lattanzi\inst{\ref{inst:oato}}
\and H.~Lenhardt\inst{\ref{inst:ari}}
\and S.~Liao\inst{\ref{inst:oato},\ref{inst:shanghai}}
\and W.~L{\"o}ffler\inst{\ref{inst:ari}}
\and P.J.~McMillan\inst{\ref{inst:lund}}
\and F.~Mignard\inst{\ref{inst:oca}} 
\and A.~Mora\inst{\ref{inst:auroraesac}} 
\and R.~Morbidelli\inst{\ref{inst:oato}}
\and J.~Portell\inst{\ref{inst:ieec}} 
\and A.~Riva\inst{\ref{inst:oato}}
\and M.~Sarasso\inst{\ref{inst:oato}}
\and I.~Serraller\inst{\ref{inst:gmvesac}, \ref{inst:ieec}} 
\and H.~Siddiqui\inst{\ref{inst:vegaesac}}
\and R.~Smart\inst{\ref{inst:oato}}
\and A.~Spagna\inst{\ref{inst:oato}}
\and U.~Stampa\inst{\ref{inst:ari}} 
\and I.~Steele\inst{\ref{inst:liverpool}}
\and F.~Taris\inst{\ref{inst:paris}}
\and J.~Torra\inst{\ref{inst:ieec}}
\and W.~van Reeven\inst{\ref{inst:auroraesac}}
\and A.~Vecchiato\inst{\ref{inst:oato}}
\and S.~Zschocke\inst{\ref{inst:tud}}
%
\and J.~de~Bruijne\inst{\ref{inst:estec}} 
\and G.~Gracia\inst{\ref{inst:poesac}} 
\and F.~Raison\inst{\ref{inst:praesepeesac},\ref{inst:eso}} 
\and T.~Lister\inst{\ref{inst:las-cumbres}} 
\and J.~Marchant\inst{\ref{inst:liverpool}} 
\and R.~Messineo\inst{\ref{inst:altec}} 
\and M.~Soffel\inst{\ref{inst:tud}} 
\and J.~Osorio\inst{\ref{inst:porto}} 
\and A.~de~Torres\inst{\ref{inst:hespaceesac}} 
\and W.~O'Mullane\inst{\ref{inst:esac}} 
}



\institute{
Lund Observatory, Department of Astronomy and Theoretical Physics, Lund University, Box 43, SE-22100, Lund, Sweden\ \email{lennart@astro.lu.se}
\label{inst:lund}
\and 
ESA, European Space Astronomy Centre, Camino Bajo del Castillo s/n, 28691 Villanueva de la Ca{\~n}ada, Spain
\label{inst:esac}
\and
Astronomisches Rechen-Institut, Zentrum f\"ur Astronomie der Universit\"at Heidelberg, M\"onchhofstra{\ss}e 14, D-69120 Heidelberg,
Germany
\label{inst:ari}
\and
Lohrmann-Observatorium, Technische Universit\"{a}t Dresden, Mommsenstrasse 13, 01062 Dresden, Germany
\label{inst:tud}
\and
HE Space Operations BV for ESA/ESAC, Camino Bajo del Castillo s/n, 28691 Villanueva de la Ca{\~n}ada, Spain
\label{inst:hespaceesac}
\and
Vitrociset Belgium for ESA/ESAC, Camino Bajo del Castillo s/n, 28691 Villanueva de la Ca{\~n}ada, Spain
\label{inst:vitrocisetesac}
\and
Telespazio Vega UK Ltd for ESA/ESAC, Camino Bajo del Castillo s/n, 28691 Villanueva de la Ca{\~n}ada, Spain
\label{inst:vegaesac}
\and
NASA/IPAC Infrared Science Archive, California Institute of Technology, Mail Code 100-22, 770 South Wilson Avenue, Pasadena, CA, 91125, USA
\label{inst:nasa}
\and
Observatoire Astronomique de l'Universit{\'e} de Gen{\`e}ve, Sauverny, 
Chemin des Maillettes 51, CH-1290 Versoix, Switzerland
\label{inst:geneva}
\and
Aurora Technology for ESA/ESAC, Camino Bajo del Castillo s/n, 28691 Villanueva de la Ca{\~n}ada, Spain
\label{inst:auroraesac}
\and
The Server Labs S.L. for ESA/ESAC, Camino Bajo del Castillo s/n, 28691 Villanueva de la Ca{\~n}ada, Spain
\label{inst:theserverlabsesac}
\and
Istituto Nazionale di Astrofisica, Osservatorio Astrofisico di Torino, Via Osservatorio 20, Pino Torinese, Torino, 10025, 
Italy
\label{inst:oato}
\and
SYRTE, Observatoire de Paris, PSL Research University, CNRS, Sorbonne Universités, UPMC Univ. Paris 06, LNE, 61 avenue de l’Observatoire, 75014 Paris, France
\label{inst:paris}
\and
GEA-Observatorio National/MCT,Rua Gal. Jose Cristino 77, CEP 20921-400,
Rio de Janeiro, Brazil
\label{inst:rio}
\and
Universidade do Porto, Rua do Campo Alegre, 687, Porto, 4169-007, Portugal
\label{inst:porto}
\and
Institute of Astrophysics and Space Sciences
Faculdade de Ciencias, Campo Grande, PT1749-016 Lisboa, Portugal
\label{inst:lisbon}
\and 
INAF, Osservatorio Astrofisico
di Catania, Catania, Italy
\label{inst:catania}
\and
IMCCE, Institut de Mecanique Celeste et de Calcul des Ephemerides, 77 avenue Denfert-Rochereau, 75014 Paris, France
\label{inst:imcce}
\and
EURIX S.r.l., via Carcano 26, Torino, 10153, Italy
\label{inst:eurix}
\and
Univ. Bordeaux, LAB, UMR 5804, F-33270, Floirac, France
\label{inst:bordeaux1}
\and
CNRS, LAB, UMR 5804, F-33270, Floirac, France
\label{inst:bordeaux2}
\and
Institut de Ci\`encies del Cosmos, Universitat de Barcelona (IEEC-UB), Mart\'i Franqu\`es 1, 
E-08028 Barcelona, Spain\ \label{inst:ieec}
\and
Institute for Astronomy, School of Physics and Astronomy, University of Edinburgh, 
Royal Observatory, Blackford Hill, Edinburgh, EH9~3HJ, 
United Kingdom
\label{inst:uoe}
\and
University of Padova, Via Marzolo 8, Padova, I-35131, Italy
\label{inst:padova}
\and
Observatoire de la C\^{o}te d'Azur,	BP 4229, Nice Cedex 4, 06304, France
\label{inst:oca}
\and
RHEA for ESA/ESAC, Camino Bajo del Castillo s/n, 28691 Villanueva de la Ca{\~n}ada, Spain
\label{inst:rheaesac}
\and
Shanghai Astronomical Observatory, Chinese Academy of Sciences, 80 Nandan Rd, 200030 Shanghai, China
\label{inst:shanghai}
\and
GMV for ESA/ESAC, Camino Bajo del Castillo s/n, 28691 Villanueva de la Ca{\~n}ada, Spain
\label{inst:gmvesac}
\and
Astrophysics Research Institute, ic2 - Liverpool Science Park, 146 Brownlow Hill, Liverpool L3 5RF, United Kingdom
\label{inst:liverpool}
\and
ESA, European Space Research and Technology Centre, Keplerlaan 1, 2200 AG, Noordwijk, The Netherlands
\label{inst:estec}
\and
Gaia Project Office for DPAC/ESA, Camino Bajo del Castillo s/n, 28691 Villanueva de la Ca{\~n}ada, Spain
\label{inst:poesac}
\and
Praesepe for ESA/ESAC, Camino Bajo del Castillo s/n, 28691 Villanueva de la Ca{\~n}ada, Spain
\label{inst:praesepeesac}
\and
European Southern Observatory, Karl-Schwarzschild-Strasse 2, 85748 Garching, Germany
\label{inst:eso}
\and
Las Cumbres Global Optical Telescope Network, 6740 Cortona Dr. 102, Goleta, CA 93117, United States of America 
\label{inst:las-cumbres}
\and
Altec, Corso Marche 79, Torino, 10146 Italy
\label{inst:altec}
}



   \date{ }

 
\abstract
  {Gaia Data Release 1 (Gaia DR1) contains astrometric results for more than 1~billion stars 
  brighter than magnitude 20.7 based on observations collected by the Gaia satellite 
  during the first 14~months of its operational phase.}  
  {We give a brief overview of the astrometric content of the data release and of the 
  model assumptions, data processing, and validation of the results.}  
  {For stars in common with the Hipparcos and Tycho-2 catalogues, complete astrometric
  single-star solutions are obtained by incorporating positional information from the earlier
  catalogues. For other stars only their positions are obtained, essentially by neglecting their 
  proper motions and parallaxes. The results are validated by an analysis of the residuals, 
  through special validation runs, and by comparison with external data.}  
  {For about two million of the brighter stars (down to magnitude $\sim$11.5) we obtain 
  positions, parallaxes, and proper motions to Hipparcos-type precision or better.
  For these stars, systematic errors depending for example on position and colour are at a 
  level of $\pm 0.3$~milliarcsecond (mas).
  For the remaining stars we obtain positions at epoch J2015.0 accurate to 
  $\sim$10~mas. Positions and proper motions are given in a reference 
  frame that is aligned with the International Celestial Reference Frame (ICRF) 
  to better than 0.1~mas at epoch
  J2015.0, and non-rotating with respect to ICRF to within 0.03~mas~yr$^{-1}$.
  The Hipparcos reference frame is found to rotate with respect to the Gaia DR1 frame at 
  a rate of 0.24~mas~yr$^{-1}$.}  
  {Based on less than a quarter of the nominal mission length and on very provisional 
  and incomplete calibrations, the quality and completeness of the astrometric data in 
  Gaia DR1 are far from what is expected for the final mission products. The present results 
  nevertheless represent a huge improvement in the available fundamental stellar data and 
  practical definition of the optical reference frame.}

   
   \titlerunning{Gaia Data Release 1 -- Astrometry} 
   \authorrunning{L.~Lindegren et al.}

   \maketitle

%

\section{Introduction}

This paper describes the first release of astrometric data from the European Space Agency 
mission Gaia \citep{2016GaiaP}. 
The first data release \citep{2016GaiaB}
contains provisional results 
based on observations collected during the first 14 months since the start of nominal 
operations in July 2014. The initial treatment of the raw Gaia data \citep{2016GaiaF}
provides the main input to the astrometric data processing outlined below.

The astrometric core solution, also known as the astrometric global iterative solution (AGIS),
was specifically developed to cope with the high accuracy requirements, large data volumes, 
and huge systems of equations that result from Gaia's global measurement principle. 
A detailed pre-launch description was given in \citetads{2012A&A...538A..78L}, hereafter 
referred to as the AGIS paper. The present solution is largely based on the models and
algorithms described in that paper, with further details on the software implementation 
in \citetads{2011ExA....31..215O}. Nevertheless, comparison with real data and a 
continuing evolution of concepts have resulted in many changes. One purpose of 
this paper is to provide an updated overview of the astrometric processing as applied to 
Gaia Data Release 1 (Gaia DR1). A specific feature of Gaia DR1 is the incorporation of earlier
positional information through the Tycho-Gaia astrometric solution
(TGAS; \citeads{2015A&A...574A.115M}).
  
It is important to emphasise the provisional nature of the astrometric results in this first
release. Severe limitations are set by the short time period on which the solution is based, 
and the circumstance that the processing of the raw data -- including the image 
centroiding and cross-matching of observations to sources -- had not yet benefited 
from improved astrometry. Some of the known problems are discussed in 
Sect.~\ref{sec:problems}.  These shortcomings will successively be eliminated in 
future releases, as more observations are incorporated in the solution, and as the raw data are 
re-processed using improved astrometric parameters, attitude, and modelling of the instrument
geometry.

\section{Astrometric content of the data release}
\label{sec:content}

The content of Gaia DR1 as a whole is described in \citet{2016GaiaB}. 
The astrometric content consists of two parts:
\begin{enumerate}
\item
The \emph{primary data set} contains positions, parallaxes, and mean proper motions 
for 2\,057\,050 of the brightest stars. This data set was derived by combining the Gaia observations 
with earlier positions from the Hipparcos 
(\citeads{1997ESASP1200.....E}, \citeads{2007ASSL..350.....V}) 
and \mbox{Tycho-2} \citepads{2000A&A...355L..27H} catalogues, and mainly includes 
stars brighter than visual magnitude 11.5. The typical uncertainty is 
about 0.3~milliarcsec (mas) for the positions and parallaxes, and about 
1~mas~yr$^{-1}$ for the proper motions. For the subset of 93\,635 stars where 
Hipparcos positions at epoch J1991.25 were incorporated in the solution, the proper 
motions are considerably more precise, about 0.06~mas~yr$^{-1}$ 
(see Table~\ref{tab:statSummary} for more statistics). 
The positions and proper motions are given in the International Celestial Reference System 
(ICRS; \citeads{1995A&A...303..604A}), 
which is non-rotating with respect to distant quasars. The parallaxes are absolute in the sense
that the measurement principle does not rely on the assumed parallaxes of background
sources. Moreover, they are independent of previous determinations such as the Hipparcos parallaxes. 
The primary data set was derived using the primary solution
outlined in Sect.~\ref{ssec:primary}, which is closely related to both TGAS 
and the Hundred Thousand Proper Motions (HTPM) project
(Mignard 2009, unpublished; \citeads{2014A&A...571A..85M}).
\item
The \emph{secondary data set} contains approximate positions in the ICRS (epoch J2015.0) 
for an additional 1\,140\,622\,719 stars and extragalactic sources, mainly
brighter than magnitude 20.7 in Gaia's unfiltered ($G$) band. This data set was derived using 
the secondary solution outlined in Sect.~\ref{ssec:secondary}, which essentially
neglects the effects of the parallax and proper motion during the 14 months of Gaia observations.
The positional accuracy is therefore limited by these effects, which typically amount 
to a few mas but could be much larger for some stars
(see Table~\ref{tab:statSummarySecondary} for statistics).
\end{enumerate}
Gaia DR1 therefore contains a total of 1\,142\,679\,769 sources.
Neither data set is complete to any particular magnitude limit. The primary data set
lacks the bright stars ($G\lesssim 6$) not nominally observed by Gaia, plus a number of
stars with high proper motion (Sect.~\ref{sec:results_primary}). The magnitude limit
for the secondary data set is very fuzzy and varies with celestial position. A substantial 
fraction of insufficiently observed sources is missing in both data sets.

\section{Observations and their modelling }

\begin{figure}
\begin{center}
\resizebox{\hsize}{!}{\includegraphics{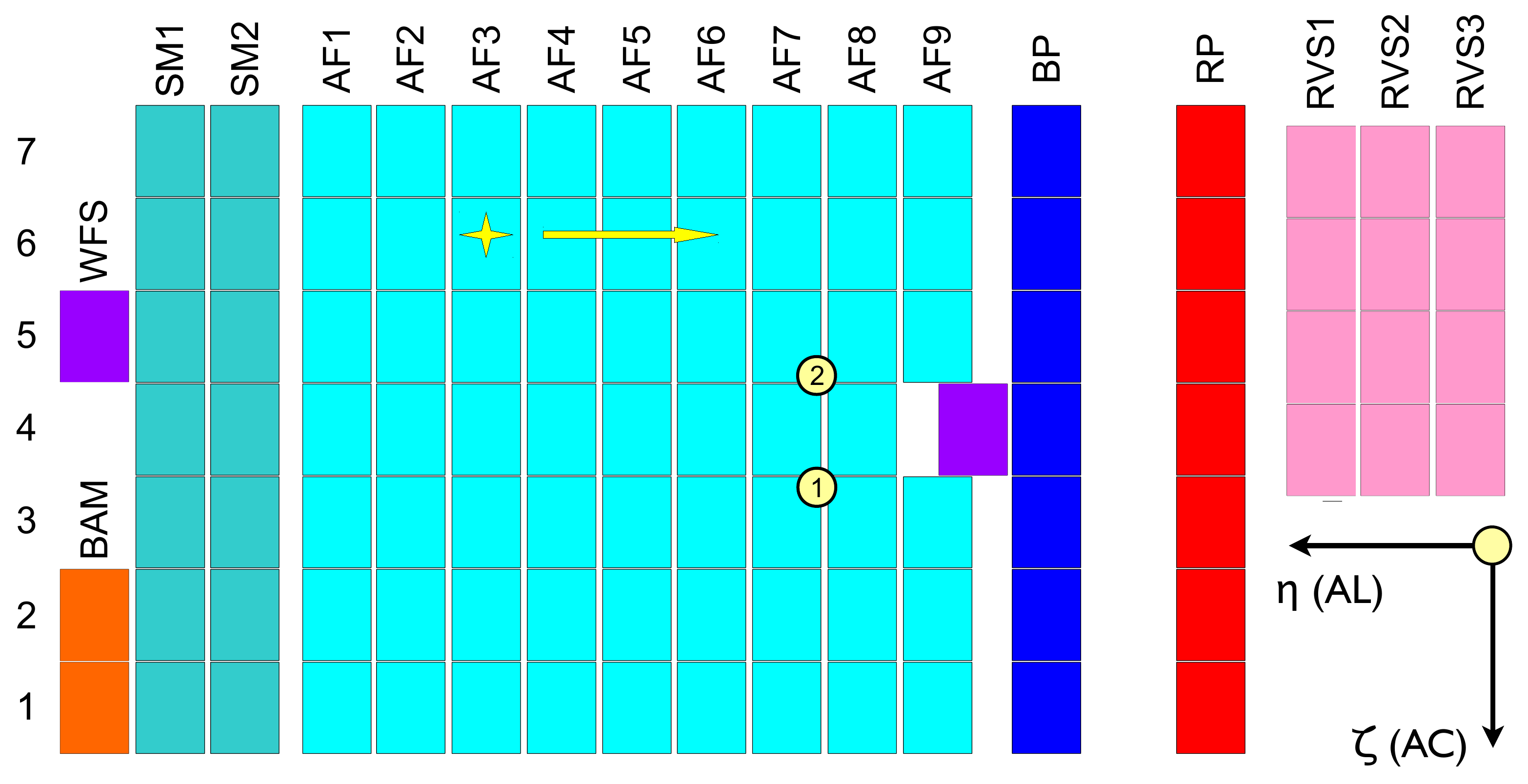}}
\caption{Layout of the CCDs in Gaia's focal plane. 
Star images move from left to right in the diagram.
As the images enter the field of view, they are detected
by the sky mapper (SM) CCDs and astrometrically observed by the
62 CCDs in the astrometric field (AF). Basic-angle variations are 
interferometrically measured using the basic angle monitor (BAM) 
CCD in row~1 (bottom row in figure). The BAM CCD in row~2 is available 
for redundancy. Other CCDs are used for the red and blue photometers 
(BP, RP), radial velocity spectrometer (RVS), and wavefront sensors (WFS).
The orientation of the field angles $\eta$ (along-scan, AL) and $\zeta$
(across-scan, AC) is shown at bottom right. The actual origin $(\eta,\zeta)=(0,0)$ 
is indicated by the numbered yellow circles 1 (for the preceding field of
view) and 2 (for the following field of view).
}\label{fig:fpa} 
\end{center}
\end{figure}

\subsection{Input data for the astrometric solutions}
\label{sec:input}

The main input data for the astrometric solutions are the astrometric elementary records,
generated by the initial data treatment \citep{2016GaiaF}. 
Each record holds the along-scan (AL) and 
across-scan (AC) coordinates for the transit of a source over the sky mapper and astrometric 
CCDs (Fig.~\ref{fig:fpa}), along with the measured fluxes and ancillary information such as the source identifier
obtained by cross-matching the observation with the current source list. 
The record normally 
contains ten AL coordinate estimates, i.e.\ one from the sky mapper and nine 
from the astrometric CCDs; the number of AC measurements ranges from one to ten 
depending on the window class assigned to the source by the onboard detection algorithm.%
\footnote{Only a small window of CCD pixels around each detected source is transmitted to the 
ground. The choice of window size, and the binning of pixels in the AC direction, uses one of three 
distinct schemes known as window classes. This results in different one- or two-dimensional 
samplings of the image, depending on the detected flux level. Window class 0, selected for bright 
sources ($G\lesssim 13$), gives two-dimensional images from which both the AL and AC coordinates 
can be determined. Window classes 1 ($13\lesssim G\lesssim 16$) and 2 ($G\gtrsim 16$) give 
one-dimensional images of 18 and 12 samples, respectively, from which only the AL coordinate can 
be determined.\label{footnote1}} 
Most observations in the primary data set
are of window class 0 and contain ten AC measurements per record, while the mostly faint sources 
in the secondary data set have much fewer AC observations. 
The sky mapper observations were not used in the astrometric solutions for Gaia DR1.

The fundamental AL astrometric observation is the precise time at which the
centroid of an image passes the fiducial observation line of a CCD  (see Sect.~\ref{sec:cal}). 
This observation time initially refers to the timescale defined by the onboard clock, i.e.\ the 
onboard mission timeline (OBMT), but later transformed to the barycentric coordinate time (TCB) 
of the event by means of the time ephemeris (Sect.~\ref{sec:ephem}). The OBMT provides a 
convenient and unambiguous way of labelling onboard events, and will be used below
to display, for example, the temporal evolution of calibration parameters. It is then expressed
as the number of nominal revolutions of exactly 21600~s OBMT from an arbitrary origin. 
For the practical interpretation of the plots the following approximate relation between
the OBMT (in revolutions) and TCB at Gaia (in Julian years) can be used:
\begin{equation}
\text{TCB} \simeq \text{J}2015.0 + (\text{OBMT} - 1717.6256~\text{rev})/(1461~\text{rev}) \, .
\end{equation}     
The time interval covered by the observations used for Gaia DR1 starts at  
OBMT 1078.3795~rev = J2014.5624599~TCB (approximately 2014~July~25, 10:30:00~UTC), 
and ends at 
OBMT 2751.3518~rev = J2015.7075471~TCB (approximately 2015~September~16, 16:20:00~UTC),
thus spanning 418~days albeit with a number of gaps (see Sect.~\ref{sec:attitude}).

\begin{figure}
\resizebox{\hsize}{!}{\includegraphics{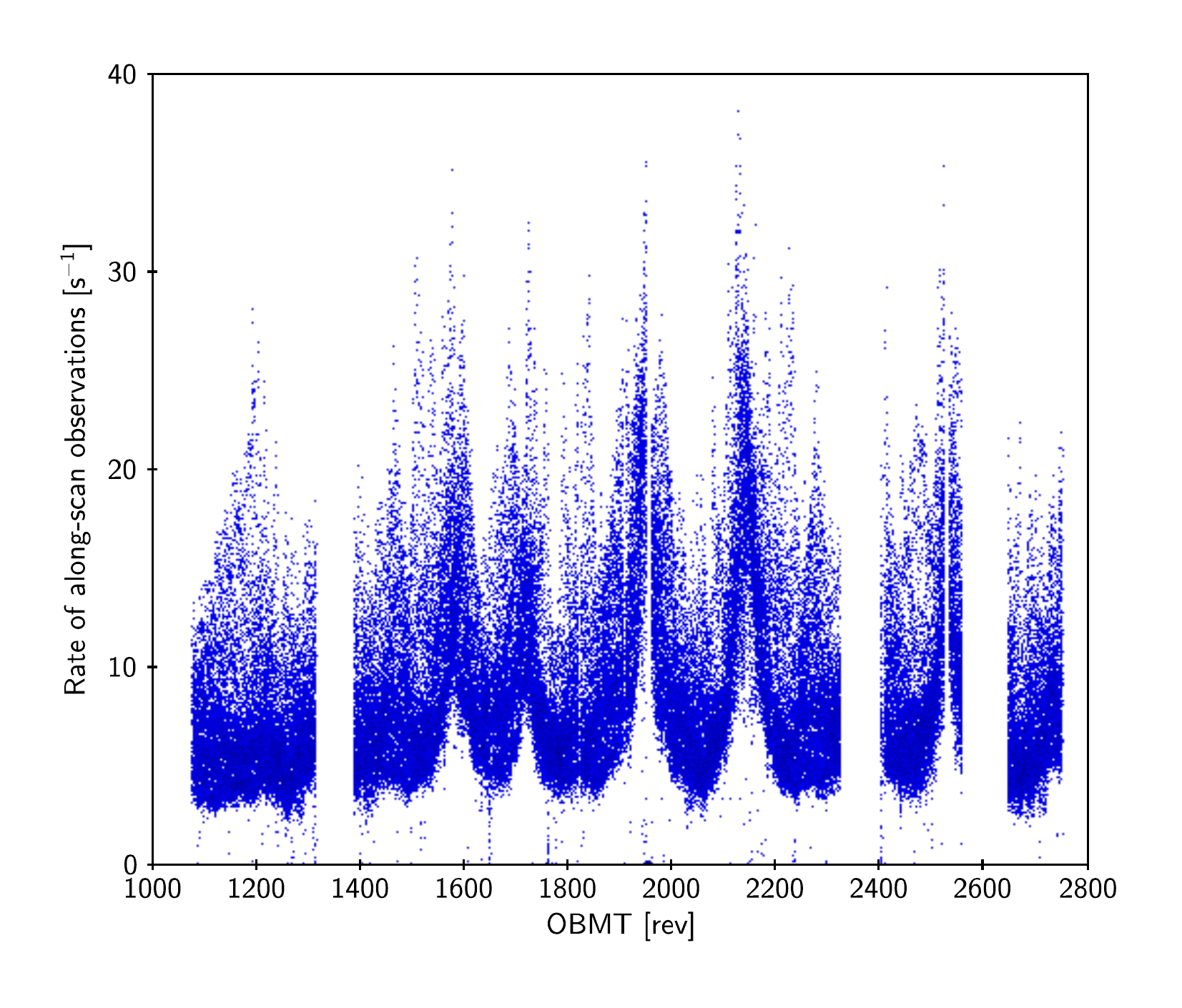}} 
\caption{Rate of AL CCD observations input to the primary solution
(mean rate per 30~s interval).
Time is expressed in revolutions of the onboard mission timeline (OBMT; see text).
The three major gaps were caused by decontamination and refocusing activities
(Sect.~\ref{sec:attitude}).}
\label{fig:obsRate}
\end{figure}

In the primary solution we processed nearly 35~million elementary
records, containing some 265~million AL astrometric observations, and a similar 
number of AC observations, for 2.48~million sources. 
Figure~\ref{fig:obsRate} shows how the rate of these observations varied with time. 
Peak rates occurred when the scans were roughly along the Galactic plane. 
On average about 107 AL observations (or 12 field-of-view transits) were processed 
per source. The actual number of observations per source varies owing to the scanning 
law and data gaps (see Fig.~\ref{fig:stats1}b).
For the secondary solution, a total of $1.7\times 10^{10}$ astrometric elementary records 
were processed.

Auxiliary input data used in the solutions include the initial Gaia source list 
(IGSL; \citeads{2014A&A...570A..87S}),
ephemerides (Sect.~\ref{sec:ephem}), and positions at epoch J1991.25 taken from the 
Hipparcos catalogue (\citeads{2007ASSL..350.....V}, as retrieved from CDS) and the
\mbox{Tycho-2} catalogue \citepads{2000A&A...355L..27H}.

\subsection{Celestial reference frame}
\label{sec:frame}

The Gaia data processing is based on a consistent theory of relativistic
astronomical reference systems and involves rigorous modelling of
observable quantities. Various components of the model are gathered in the
Gaia relativity model
(GREM; \citeads{2003AJ....125.1580K}, \citeyearads{2004PhRvD..69l4001K}).
The primary coordinate system used for the astrometric processing of Gaia data is the
Barycentric Celestial Reference System  
(BCRS; \citeads{2003AJ....126.2687S}). The BCRS has its origin at the solar-system barycentre
and its axes are aligned with the ICRS. The time-like coordinate of the BCRS
is TCB. The motions of Gaia and other solar-system
objects are thus described by the space-like coordinates of the BCRS, 
$x(t)$, $y(t)$, $z(t)$, using TCB as the independent time variable $t$. The motions 
of all objects beyond the solar system are also parametrised in terms of BCRS 
coordinates (Sect.~\ref{sec:sources}), but here the independent time variable $t$ should 
be understood as the time at which the event would be observed at the solar-system barycentre, 
i.e.\ the time of observation corrected for the R{\o}mer delay. This convention is necessitated
by the in general poor knowledge of distances beyond the solar system.

The reference frame for the positions and proper motions in Gaia DR1 is in practice
defined by the global orientation of positions at the two epochs J1991.25 and J2015.0.
From the construction of the Hipparcos and \mbox{Tycho-2} catalogues, the positions of
stars around epoch J1991.25, as given in these catalogues, represent the best available
realisation of the optical reference frame at that epoch, with an estimated uncertainty 
of 0.6 mas in each axis (Vol.~3, Ch.~18.7 in \citeads{1997ESASP1200.....E}). On the other
hand, by using the Gaia observations of quasars with positions in 
the ICRS accurately known from Very Long Baseline Interferometry (VLBI), 
it was possible to align the global system of positions in Gaia DR1 to the ICRS
with an estimated uncertainty of $<\,$0.1~mas at epoch J2015.0 (Sect.~\ref{sec:alignment}). 
From the 23.75~yr time difference between these epochs it follows that the resulting 
proper motion system should have no global rotation with respect to ICRS at an 
uncertainty level of about 0.03~mas~yr$^{-1}$.

The Gaia observations of quasars over several years will eventually permit a non-rotating 
optical reference frame to be determined entirely from Gaia data, independent of the 
Hipparcos reference frame, and to a much higher accuracy than in the current release.

\subsection{Astrometric modelling of the sources}
\label{sec:sources}

The basic astrometric model is described in Sect.~3.2 of the AGIS paper and assumes
uniform space motion relative to the solar-system barycentre. In a regular AGIS solution 
this is applicable only to the subset of well-behaved ``primary sources'', used 
to determine the attitude, calibration, and global parameters, while ``secondary sources'' 
may require more complex modelling. In Gaia DR1
the basic model is applied to all stellar and extragalactic objects, which are thus treated 
effectively as single stars. The distinction between primary and secondary sources is
instead based on the type of prior information incorporated in the solutions 
(Sects.~\ref{ssec:primary} and \ref{ssec:secondary}).

In the basic model the apparent motion of a source, as seen by Gaia, is completely 
specified by six kinematic parameters, i.e.\ the standard five astrometric parameters 
$(\alpha,\,\delta,\,\varpi,\,\mu_{\alpha*},\,\mu_\delta)$, defined below,
and the radial velocity $v_r$. For practical reasons $v_r=0$ is assumed in Gaia DR1 for
all objects, meaning that perspective acceleration is not taken into account
(see below). All the parameters refer to the reference epoch 
$t_\text{ep}=\text{J2015.0~TCB}$.

The time-dependent coordinate direction from Gaia towards an object beyond 
the solar system is therefore modelled, in the BCRS, as the unit vector%
\footnote{$\langle\rangle$ denotes vector normalisation: 
$\langle\vec{a}\rangle=\vec{a}\left|\vec{a}\right|^{-1}$.} 
\begin{equation}\label{eq:source1}
\vec{\bar{u}}(t) = \Bigl\langle \vec{r} + (t_\text{B}-t_\text{ep})\left(\vec{p}\mu_{\alpha*}+
\vec{q}\mu_\delta+\vec{r}\mu_r\right) - \vec{b}_\text{G}(t)\varpi/A_\text{u}\Bigr\rangle \, ,
\end{equation} 
where $t$ is the time of observation (TCB); $\vec{p}$, $\vec{q}$, and $\vec{r}$
are orthogonal unit vectors defined in terms of the astrometric parameters $\alpha$ and $\delta$,
\begin{equation}\label{eq:source2}
\vec{p}=\begin{bmatrix}-\sin\alpha\\ \cos\alpha\\ 0 \end{bmatrix} , \quad
\vec{q}=\begin{bmatrix}-\sin\delta\cos\alpha\\ -\sin\delta\sin\alpha\\ \cos\delta \end{bmatrix} , \quad
\vec{r}=\begin{bmatrix} \cos\delta\cos\alpha\\ \cos\delta\sin\alpha\\ \sin\delta \end{bmatrix} ; \quad
\end{equation}
$t_\text{B}=t+\vec{r}'\vec{b}_\text{G}(t)/c$ is the time of observation corrected for the 
R{\o}mer delay ($c=$~speed of light); $\vec{b}_\text{G}(t)$ is the barycentric position of Gaia
at the time of observation; and $A_\text{u}$ is the astronomical unit.%
\footnote{The prime in $\vec{r}'$ stands for the transpose of the vector or matrix.
$\vec{r}'\vec{b}$ is therefore the scalar product of vectors $\vec{r}$ and $\vec{b}$.}
$\mu_{\alpha*}=\mu_\alpha\cos\delta$ and $\mu_\delta$ are the components of proper motion along 
$\vec{p}$ (towards increasing $\alpha$) and $\vec{q}$ (towards increasing $\delta$), 
respectively, and $\varpi$ is the parallax. $\mu_r=v_r\varpi/A_\text{u}$ is the 
radial proper motion related to the perspective acceleration discussed below.

The modelling of stellar proper motions neglects all effects that
could make the apparent motions of stars non-linear in the $\sim$24~year interval
between the Hipparcos/Tycho observations and the Gaia observations. 
Thus, orbital motion in binaries and perturbations 
from invisible companions are neglected, as well as the perspective secular changes 
caused by non-zero radial velocities.
The published proper motions should therefore be 
interpreted as the mean proper motions over this time span, rather than as the 
instantaneous proper motions at the reference epoch J2015.0. The published positions,
on the other hand, give the barycentric directions to the stars at J2015.0. 

Perspective acceleration is a purely geometrical effect caused by the changing distance 
to the source and changing angle between the velocity vector and the line of sight
(e.g.\ \citeads{1981ASSL...85.....V}). It is fully accounted for in Eq.~(\ref{eq:source1})
by means of the term containing $\mu_r$. The perspective
acceleration (in mas~yr$^{-2}$) is proportional to the product of the star's 
parallax, proper motion, and parallax, and is therefore very small except for some
nearby, high-velocity stars (cf.\ \citeads{2012A&A...546A..61D}). In the current
astrometric solutions it is effectively ignored by assuming $v_r=0$, and hence
$\mu_r=0$, for all objects. 
This is acceptable for Gaia DR1 provided that the resulting proper motions are 
interpreted as explained above. In future releases perspective acceleration will be 
taken into account, whenever possible, using radial-velocity data from Gaia's 
onboard spectrometer \citep[RVS;][]{2016GaiaP}.

\subsection{Relativistic model and auxiliary data}
\label{sec:ephem}

The coordinate direction $\vec{\bar{u}}$ introduced in
Sect.~\ref{sec:sources} should be transformed into the observed
direction $\vec{u}$ (also known as proper direction) as seen
by Gaia. This is done using the previously mentioned GREM
(\citeads{2003AJ....125.1580K}, \citeyearads{2004PhRvD..69l4001K}).

The transformation essentially consists of two steps. First, the
light propagation from the source to the location of Gaia is modelled
in the BCRS. In this process, the influence of the gravitational field
of the solar system is taken into account in full detail. It includes the
gravitational light-bending caused by the Sun, the major planets and the
Moon. Both post-Newtonian and major
post-post-Newtonian effects are included. For observations close to
the giant planets the effects of their quadrupole gravitational fields
are taken into account in the post-Newtonian approximation. The
non-stationarity of the gravitational field, caused by the
translational motion of the solar-system bodies, is also properly
taken into account. On the other hand, no attempt is made to account 
for effects of the gravitational field outside the solar system. This plays 
a role only in cases when its influence is variable on timescales comparable 
with the duration of observations, e.g.\ in various gravitational lensing
phenomena.

The second step is to compute the observed direction $\vec{u}$ from the 
computed BCRS direction of light propagation at the location of Gaia. To this
end, a special physically adequate (local) proper reference system for
the Gaia spacecraft, known as the centre-of-mass reference system (CoMRS),
is used as explained in \citetads{2004PhRvD..69l4001K}.  
At this step we take into account the stellar aberration caused by the 
BCRS velocity of Gaia's centre of mass, as well as certain smaller 
general-relativistic effects. 

The model requires several kinds of auxiliary data. These include the
Gaia ephemeris (the BCRS position and velocity of Gaia),
the solar-system ephemeris (the positions and velocities of all
gravitating bodies of the solar system used in the model), and the time
ephemeris used to convert the reading of the Gaia onboard clock into 
TCB. 

The Gaia ephemeris is provided by the European Space Operation Centre
(ESOC) based on radiometric observations of the spacecraft and using
standard orbit reconstruction procedures \citep{2016GaiaP}.
The Gaia orbit determination satisfies the accuracy requirements imposed 
by Gaia DR1: the uncertainty of the BCRS velocity of Gaia is believed to 
be considerably below 10~mm~s$^{-1}$. For future releases, the Gaia orbit 
will be verified in a number of ways at the level of 1~mm~s$^{-1}$, which is 
needed to reach the accuracy goal of the project. 

The solar system
ephemeris used in the Gaia data processing is the INPOP10e ephemeris
\citep{FiengaEtAl2016} parametrised by TCB. 
The time ephemeris for the Gaia clock is constructed from special
time-synchronisation observations of the spacecraft \citep{2016GaiaP},
using a consistent relativistic model for the proper time 
of the Gaia spacecraft. 

The CoMRS also provides a consistent definition of the spacecraft
attitude in the relativistic context. The reference system that is aligned
with the instrument axes is known as the scanning reference system
(SRS; \citeads{2012A&A...538A..78L}). The attitude discussed in
Sect.~\ref{sec:attitude} represents a pure spatial rotation between
CoMRS and SRS.

\subsection{Attitude model}
\label{sec:attitude}

The attitude model is fully described in Sect.~3.3 of the AGIS paper. It uses cubic 
splines to represent the four components of the attitude quaternion as functions of time.
The basic knot sequence for the present solutions is regular with a knot interval of 30~s. 
Knots of multiplicity four
are placed at the beginning and end of the knot sequence, allowing the spline to be discontinuous
at these points, and similarly around imposed data gaps. Such gaps were introduced 
around the fourth and fifth mirror decontaminations \citep{2016GaiaP}, 
spanning OBMT 1316.490--1389.113 and 
2324.900--2401.559~rev, respectively, and in connection with the refocusing of the 
following field of view at OBMT 1443.963--1443.971~rev, and of the preceding field of view 
at OBMT 2559.0--2650.0~rev (see Fig.~\ref{fig:obsRate}).
Additional gaps were placed around 45 micrometeoroid hits identified in provisional solutions. 
These gaps are typically less than 10~s, but reach 1--1.5 min in some cases. 
The total number of knots is 980\,666, yielding 3\,922\,648 attitude parameters. 

A longer knot interval of 180~s was used in the first phases of the solution (phase A and B
in Fig.~\ref{fig:primaryConv}). At the end of phase B, a spline with 30~s knot interval was 
fitted to the attitude estimate at that point, and the iterative solution continued with the 
shorter interval. This procedure speeds up the convergence considerably
without degrading the final, converged solution. 

As described in Sect.~5.2.4 of the AGIS paper, the attitude updating uses a regularisation
parameter $\lambda$ to constrain the updated quaternion to unit length. The adopted value
is $\lambda = \sqrt{10^{-7}}$.

\subsection{Geometric instrument calibration model}
\label{sec:cal}

The astrometric instrument consists of the optical telescope with
two viewing directions (preceding and following field of view), together with
the sky mapper (SM) and astrometric field (AF) CCDs, see Fig.~\ref{fig:fpa}. 
The geometric calibration of the instrument provides an accurate 
transformation from pixel coordinates on the CCDs to the
field angles $(\eta,\,\zeta)$. Depending on the field of view in which an object
was observed, the field angles define its observed direction in the SRS at the 
time of observation, $t$. The observation time is the precise instant when 
the stellar image crosses a fiducial observation line on the CCD.
The AL calibration describes the geometry of the observation line as
a function $\eta(\mu)$ of the AC pixel coordinate $\mu$.
The latter is a continuous variable covering the 1966 pixel columns, running from 
$\mu=13.5$ at one edge of the CCD (minimum $\zeta$) to 
$\mu=1979.5$ at the opposite edge (maximum $\zeta$).
$\eta(\mu)$ additionally depends on a number of parameters including the CCD index 
($n$), field-of-view index ($f$), CCD gate ($g$), and time. The temporal dependence 
is described by means of discrete calibration intervals, $t_j \le t < t_{j+1}$.
In the current configuration these intervals are not longer than 3~days, but have additional 
breakpoints inserted at appropriate times, e.g.\ when a significant jump 
is seen in the onboard metrology signal (see Sect.~\ref{sec:bam}).

The current model also includes a dependence on the window class ($w$)
of the observation (see footnote~\ref{footnote1}). 
Ideally the location of the image centroid should 
not depend on the size of the window used to calculate the centroid, 
i.e.\ on the window class. (Nor should it depend on, for example, the colour and magnitude
of the star.) However, this can only be achieved after the CCD image line-spread
function (LSF) and point-spread function (PSF) have been
calibrated using astrometric, attitude, and geometric calibration information 
from a previous AGIS.
Since this outer processing loop has not yet been closed, a dependence on the window class is
introduced in the geometric calibration model as a temporary measure.  

The detailed specification of the calibration model and all its dependencies
is made in the framework of the generic calibration model briefly described
in Sect.~3.4 of the AGIS paper. The model used for the current astrometric 
solution is further explained in Appendix~\ref{sec:calibration_parameters}.

\subsection{Use of onboard metrology (BAM)}
\label{sec:bam}

Integrated with the Gaia instrument is a laser-interferometric device, the basic angle monitor
(BAM; \citealt{2016GaiaP}), which measures variations 
of the basic angle on timescales from minutes to days. Line-of-sight variations are
monitored by means of two interference patterns, one per field of view, projected on
a dedicated CCD next to the sky mappers (Fig.~\ref{fig:fpa}). An example of
the line-of-sight variations is given in the top part of Fig.~\ref{fig:bamRaw}, which
shows fringe positions derived from the interference pattern in the preceding field of 
view. The basic angle variations are calculated as the 
differential line-of-sight variation between the two fields of view.

Because the BAM was not designed for long-term stability, it measures reliably only
the relative variations on timescales shorter than a few days.
The absolute value of the basic angle ($\Gamma$) and its evolution 
on longer timescales are routinely determined in the astrometric solution as part 
of the geometric instrument calibration (Sect.~\ref{sec:cal}). On the relevant (short) 
timescales, the variations of the basic angle, reconstructed from the BAM data, 
exhibit very significant periodic patterns 
\citep[amplitude $\sim$1~mas; see][]{2016GaiaP} as well as discontinuities, trends,
and other features, all of which may be used to correct the astrometric observations. For
Gaia DR1 a somewhat conservative approach has been adopted, in which only the most
prominent features of the BAM signal are taken into account in the astrometric solution.
These include the major discontinuities and the regular part of the periodic variations. The
discontinuities are taken into account by appropriate choice of calibration boundaries 
as described in Sect.~\ref{sec:cal}. The corrections derived from the periodic variations 
of the BAM signal are discussed in Appendix~\ref{sec:calibration_bam}.

\section{Astrometric solutions}
\label{sec:solutions}

\begin{figure} 
\resizebox{\hsize}{!}{\includegraphics{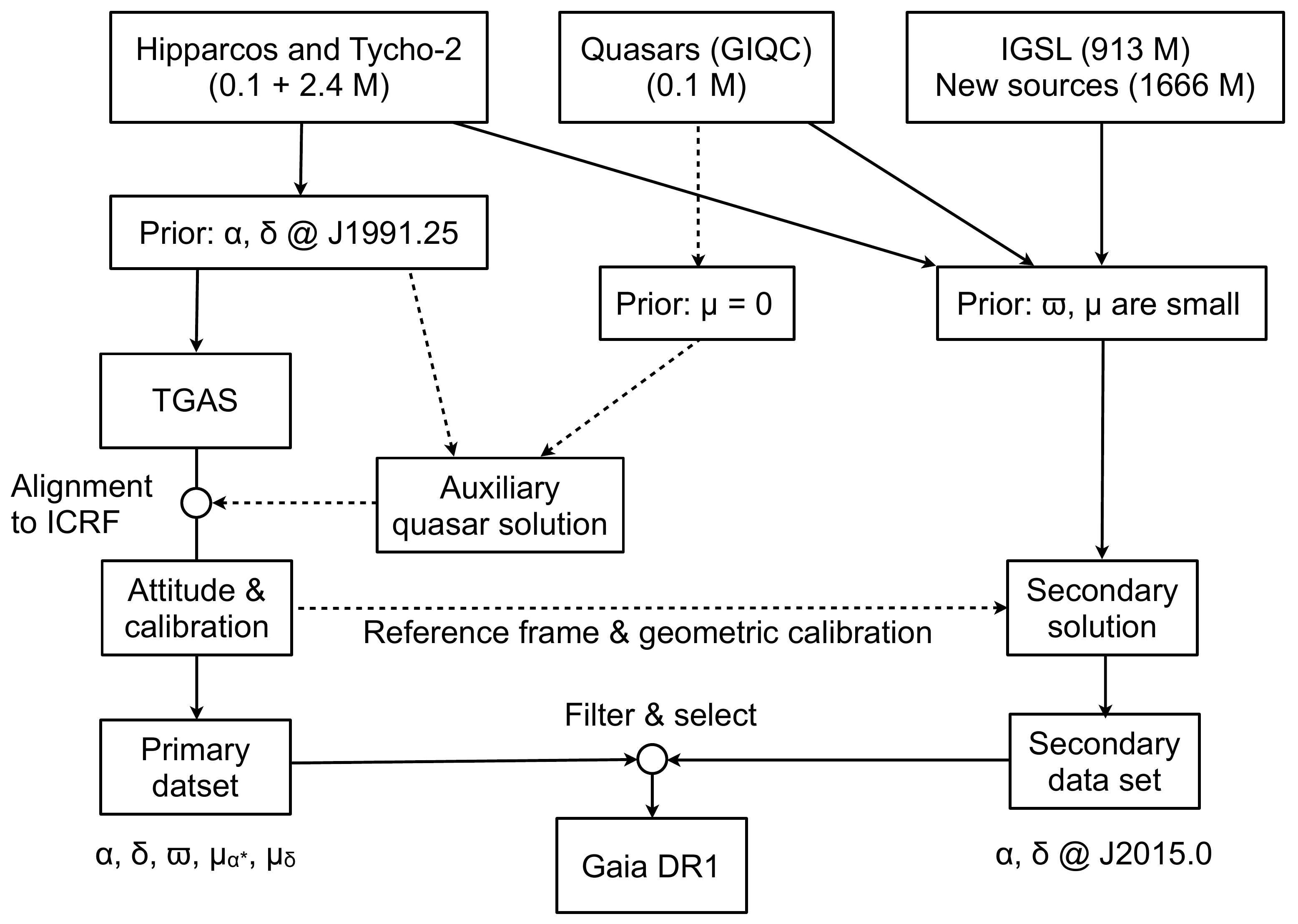}} 
\caption{Logic of the astrometric solutions contributing to Gaia DR1.
The top boxes show the number of sources input to the solutions, and the different
priors used to constrain the solutions. The number of sources finally kept in Gaia DR1
is substantially smaller. The TGAS and the auxiliary quasar solutions work in the Hipparcos
reference frame; the final primary and secondary solutions are aligned with
ICRF. 
All sources are also treated by the secondary solution; in the end, a decision is made
for each source whether to select the primary or secondary solution for Gaia DR1, or none. 
The results of the auxiliary quasar solution 
are only used for alignment, calibration, and validation purposes. The quasar 
results in Gaia DR1 come, with few exceptions, from the secondary solution.}
\label{fig:flowChart}
\end{figure}

The astrometric results in Gaia DR1 come from several interdependent solutions,
as illustrated in Fig.~\ref{fig:flowChart} and detailed below.

\subsection{Primary solution (TGAS)}
\label{ssec:primary}

The primary solution for Gaia DR1 uses the positions of $\sim$114\,000 sources from the 
re-reduced  Hipparcos catalogue \citepads{2007ASSL..350.....V}, and an additional 2.36~million positions 
from the \mbox{Tycho-2} catalogue \citepads{2000A&A...355L..27H} as prior information for a joint 
Tycho-Gaia astrometric solution (TGAS; \citeads{2015A&A...574A.115M}). 
Only the positions at J1991.25 (for the Hipparcos stars) or 
at the effective \mbox{Tycho-2} observation epoch (taken to be the mean of the 
$\alpha$ and $\delta$ epochs) were used, together with the uncertainties and correlations 
given in the catalogues. It is important that the parallaxes from the Hipparcos catalogue and
the proper motions from the Hipparcos and \mbox{Tycho-2} catalogues were not used.%
\footnote{The formalism of TGAS requires that the prior astrometric 
parameters have finite variances. The prior uncertainties are therefore set to 1~arcsec for the 
parallaxes and to 10~arcsec~yr$^{-1}$ for the proper motions. This gives negligible ($<10^{-6}$)
weight to the prior information on these parameters compared with their posterior estimates.}
This ensures that the calculated parallaxes and proper motions in the primary solution are 
independent of the corresponding
values in the earlier catalogues, which can therefore usefully be compared with the new results
(see Appendices~\ref{app:precision} and \ref{app:external}).

\begin{figure} 
\resizebox{0.94\hsize}{!}{\includegraphics{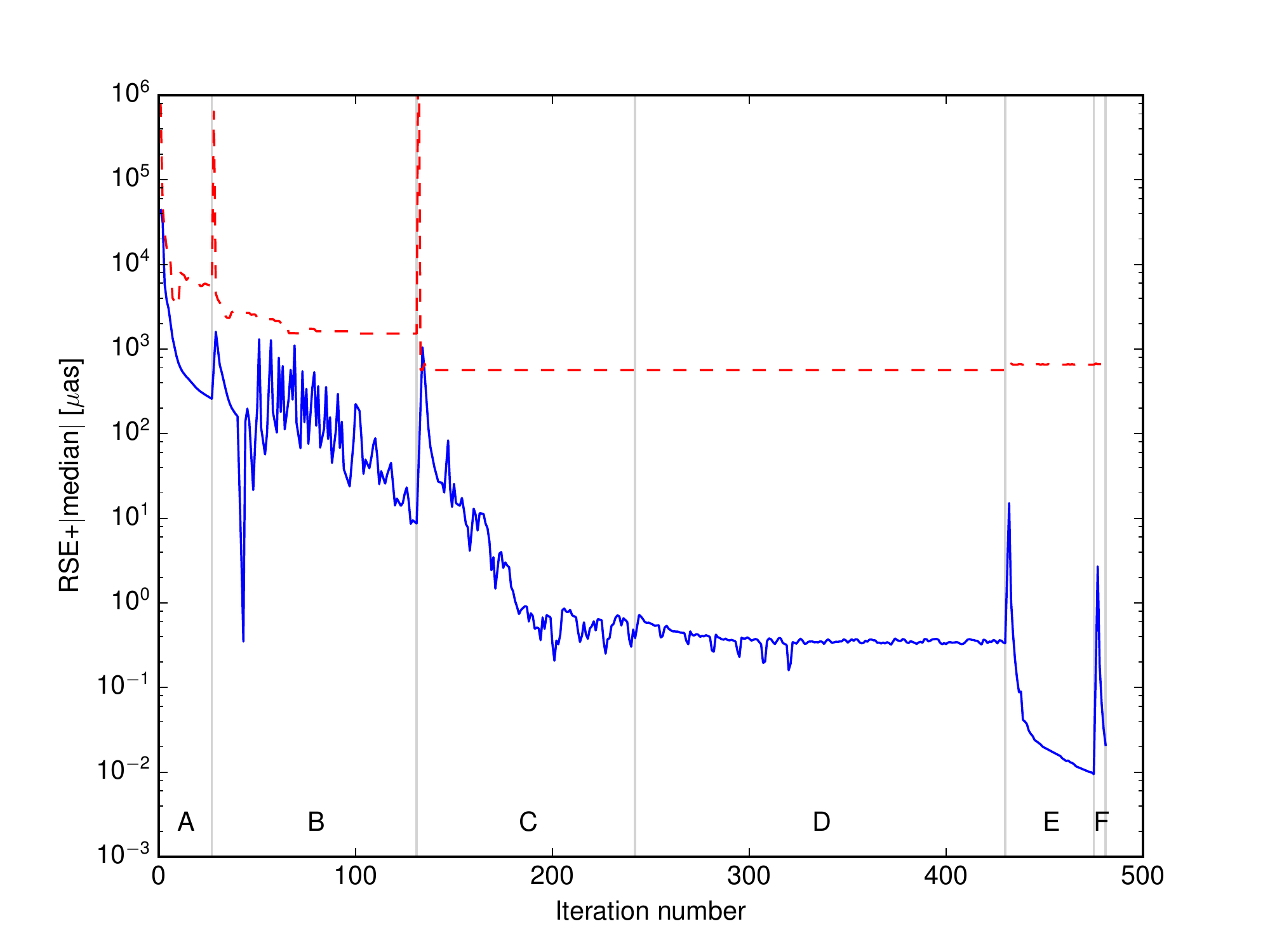}} 
\caption{Convergence of the iterative astrometric solution for the primary data set. 
The solid curve is the typical size of parallax updates in each iteration; the dashed 
curve is the typical size of AL residuals (in $\mu$as). The letters refer to
the main phases of the iterative scheme:
A -- iterations with 180~s attitude knot interval (Sect.~\ref{sec:attitude})
and simplified calibration model; 
B -- iterations with full calibration model;
C -- using 30~s attitude knot interval;
D -- Hipparcos alignment active (see text);
E -- auxiliary quasar solution (Sect.~\ref{sec:quasars}) using a fixed attitude;
F -- source and calibration updates after final alignment of the attitude to the ICRS
(Sect.~\ref{sec:alignment}).}
\label{fig:primaryConv}
\end{figure}

The primary solution cyclically updates the source, attitude, and calibration parameters, using 
a hybrid scheme alternating between so-called simple iterations and the conjugate gradient 
algorithm (\citeads{2012A&A...538A..77B}, \citeads{2012A&A...538A..78L}. 
While the conjugate gradient method 
in general converges much faster than simple iterations, the latter method allows the 
minimisation problem to be modified between
iterations, which is necessary in the adaptive weighting scheme used to identify outliers
and to estimate the excess source noise and excess attitude noise. The iterative solution
for the primary data set of Gaia DR1 was done in several phases, using successively more
detailed modelling, as summarised in Fig.~\ref{fig:primaryConv}. For example, a simplified 
calibration model was used during the first phase (A), and a longer attitude knot interval
was used in the first two phases (A and B), compared with all subsequent phases. 

In phase D the source and attitude parameters were aligned with the 
Hipparcos reference frame after each iteration. This was done by applying a global
rotation to the TGAS positions at epoch J2015.0, such that for the Hipparcos stars they
were globally consistent, in a robust least-squares sense, with the positions obtained by 
propagating the Hipparcos catalogue to that epoch. By construction, the TGAS positions 
extrapolated back to J1991.25 coincide with the Hipparcos positions used as priors at 
that epoch. Therefore, at this stage of the processing, both the TGAS positions and the 
TGAS proper motions were strictly in the Hipparcos reference frame.
The subsequent auxiliary quasar solution in phase E (Sect.~\ref{sec:quasars}) computed the
positions and parallaxes of the quasars, as well as the calibration parameters for window
class~1 and 2 ($G\gtrsim 13$; see footnote~\ref{footnote1})  
needed in the secondary solution (Sect.~\ref{ssec:secondary}). The attitude, however, was
not updated in the auxiliary quasar solution, which was therefore kept in the Hipparcos reference 
frame during this phase. As explained in Sect.~\ref{sec:alignment}, the final reference 
frame of Gaia DR1 was obtained by a further small rotation applied in phase F.

The iteration scheme described above uses both AL and AC observations
with their formal uncertainties provided by the initial data processing. However, we found
that the resulting parallax values depend in a systematic way on the uncertainties assigned
to the AC observations. The origin of this effect is not completely understood, although it is
known that the AC measurements are biased, owing to the rudimentary PSF calibration 
used in the pre-processing of the current data sets. 
To eliminate the effect in the present solution we artificially increased the AC formal 
standard uncertainties 1000 times in the last source update in phase E (for the quasars) and
in phase F (for the primary data set).%
\footnote{In the secondary solution (Sect.~\ref{ssec:secondary}) a smaller factor of 3 was used, 
which roughly brings the formal AC uncertainties into agreement with the residual AC scatter.
For this solution it was harmless, and sometimes helpful, to use the AC observations, as no
parallaxes were determined and no attitude update was made.}

As shown by the dashed curve in Fig.~\ref{fig:primaryConv}, the width of the residual distribution 
does not decrease significantly after the first $\sim$150 iterations. (The slight increase from 
phase E is caused by the addition of the faint quasars, which on average have larger residuals 
than the TGAS sources.) However, the subsequent few hundred iterations in phase C and D, 
during which the updates (solid curve) continue to decrease, are extremely 
important for reducing spatially correlated errors. 
It is difficult to define reliable convergence criteria even for idealised simulations 
\citepads{2012A&A...538A..77B}, but the typical updates in parallax should 
be at least a few orders of magnitude smaller than the aimed-for precision.
In the present solution the final updates are typically well below 1~microarcsec ($\mu$as). 
During phase E there was a further rapid decrease of the 
updates, down to ${\sim}0.01~\mu$as. However, since the attitude parameters were 
not updated in phase E, it is probable that truncation errors remain at roughly 
the same level as at the end of phase D.

\paragraph{Uncertainty estimates}
It is known from simulations (e.g.\ Sect.~7.2 in the AGIS paper) that the formal 
uncertainties of the astrometric parameters calculated in AGIS underestimate
the actual errors. One reason for this is that the covariances are computed from the truncated 
$5\times 5$ normal matrices of the individual sources, thus ignoring the contributing 
uncertainties from other unknowns such as attitude and calibration parameters.
The relation between formal and actual uncertainties may under certain
conditions be derived from a statistical comparison with an independent data set. 
The Hipparcos parallaxes offer such a possibility, which is explored in 
Appendix~\ref{app:precision}. For the Gaia DR1 parallaxes of Hipparcos sources the
following inflation factor is derived:
\begin{equation}\label{eq:infl}
F\equiv\sigma_\varpi/\varsigma_\varpi \simeq \sqrt{a^2 + (b/\varsigma_\varpi)^2} \, .
\end{equation} 
Here $\varsigma_\varpi$ is the formal parallax uncertainty calculated in 
the source update of AGIS (i.e.\ from the inverse $5\times 5$ normal matrix of the
astrometric parameters), 
$\sigma_\varpi$ is the actual parallax uncertainty estimated from a comparison 
with the Hipparcos parallaxes, and $a=1.4$, $b=0.2$~mas are constants 
(see Fig.~\ref{fig:app2}).
Although this relation was derived only for a subset of the sources (i.e.\ Hipparcos entries)
and for one specific parameter (parallax), it has been applied, for lack of any better 
recipe, to all the sources and all astrometric parameters in the primary data set. 
This was done by applying the factor $F^2$, calculated individually for each source,
to its $5\times 5$ covariance matrix. This leaves the correlation coefficients among
the five astrometric parameters unchanged. All astrometric uncertainties for the primary 
solution quoted in this paper refer to the inflated values 
$\sigma_{\alpha *}=F\varsigma_{\alpha *}$, etc., except when explicitly stated otherwise.

\subsection{Auxiliary quasar solution} 
\label{sec:quasars}

Some 135\,000 quasars from the Gaia initial quasar catalogue
(GIQC; Andrei et al.\ \citeyearads{2009A&A...505..385A}, 
\citeyearads{2012sf2a.conf...61A}, \citeyearads{2014jsrs.conf...84A})
were included towards the end of the solution (phase~E in Fig.~\ref{fig:primaryConv}).
By assuming that these sources have negligible proper motions 
(the prior was set to $0\pm 0.01$~mas~yr$^{-1}$ in each component)
it was possible to
solve the positions and parallaxes for most of them as described by
\citetads{2016A&A...586A..26M}. At the end of phase E these objects had 
positions and parallaxes with median (inflated) standard uncertainties of about 1~mas.
Their proper motions, although formally solved as well, are not meaningful 
as they merely reflect the prior information: they are practically zero.
Because the attitude was not updated in the auxiliary quasar solution,
the quasar positions were obtained in the same reference frame as the
preceding TGAS (at the end of phase~D).  

The resulting positions and parallaxes are used for two purposes:
(i) the positions for a subset of sources with accurately known positions
from VLBI 
are used to align the Gaia DR1 reference frame with the extragalactic radio
frame as described in Sect.~\ref{sec:alignment}; and (ii) as described
in Appendix~\ref{sec:validation_quasars} the observed parallaxes for the 
whole set of ${\sim}10^5$ quasars provide a valuable check of the 
parallax zero point and external accuracy of the solutions.

The quasar solution also provides the geometric instrument calibration for 
the fainter sources observed using window class~1 and 2 (see footnote~\ref{footnote1}). 
This part of the calibration is needed for the secondary solution 
(Sect.~\ref{ssec:secondary}), but cannot be obtained in the primary solution 
of the brighter sources, which are normally observed using window class~0.

The positions and parallaxes from the auxiliary quasar solution are not
contained in Gaia DR1. The positions for these objects are instead
computed in the secondary solution (Sect.~\ref{ssec:secondary}) and
become part of the secondary data set along with data for other quasars 
and most of the Galactic stars. The secondary solution does not constrain 
the proper motion of the quasars to a very small value, as in the auxiliary
quasar solution, and the resulting positions are therefore slightly different.
After correction for the different reference frames of Gaia DR1 and the
auxiliary quasar solution (Sect.~\ref{sec:alignment}), the RSE difference%
\footnote{The robust scatter estimate (RSE) is consistently used in this
paper as a robust measure of the scale or dispersion of a distribution. 
RSE is defined as 
${\left(2\sqrt{2}\,{\rm erf}^{-1}\bigl(4/5\bigr)\right)}^{-1}\approx0.390152$ 
times the difference between the 90th and 10th percentiles, which for a normal
distribution equals the standard deviation. Similarly, the median is generally
used as a robust measure of the location or centre of a distribution.\label{footnote:RSE}}
between the quasar positions in the two solutions is 1.22~mas in right ascension 
and 0.94~mas in declination, with median differences below 0.05~mas.

\subsection{Alignment to the ICRF} 
\label{sec:alignment}

Ideally, the alignment procedure should define a celestial coordinate system for the 
positions and proper motions in Gaia DR1 that (i) is non-rotating with respect 
to distant quasars; and (ii) coincides with the ICRF
at J2015.0. (Because the ICRF is also non-rotating, the two frames should then coincide at all
epochs.) For Gaia DR1 the time interval covered by the observations is too short to
constrain the spin of the reference frame by means of the measured proper motions of quasars,
as will be done for future releases.
Instead, a special procedure was devised, which relies on the assumption that the 
Hipparcos catalogue, at the time of its construction, was carefully aligned with the ICRF 
\citepads{1997A&A...323..620K}. Since the primary solution takes the Hipparcos positions 
at J1991.25 as priors, it should by construction be properly aligned with the ICRF at that epoch. 
However, this is not sufficient to constrain the spin of the Gaia DR1 reference frame. 
For that we must also require that the quasar positions at J2015.0 are consistent with 
ICRF2. It may seem surprising that the combination of stellar positions at J1991.25 with 
quasar positions at J2015.0 can be used to constrain the spin, given that the two sets of 
objects do not overlap. However, this is achieved by the auxiliary quasar solution, 
in which the observations of both kinds of objects are linked by a single set of attitude and 
calibration parameters. The practical procedure is somewhat more complicated, as it uses 
the Hipparcos reference frame as a provisional intermediary for the proper motions.

The current physical realisation of the ICRS at radio wavelengths is ICRF2
(\citeads{2009ITN....35....1M}, \citeads{2015AJ....150...58F}), 
which contains precise VLBI positions 
of 3414 compact radio sources, of which 295 are defining sources.
Among the sources in the auxiliary quasar solution (Sect.~\ref{sec:quasars}) 
we find 2191 objects with acceptable astrometric quality ($\epsilon_i < 20$~mas 
and $\sigma_\text{pos,max} < 100$~mas; cf.\ Eq.~\ref{eq:formal2}) that,
based on positional coincidence (separation $<$\,150~mas), are likely to be 
the optical counterparts of ICRF2 sources. 
(The remaining $\sim$1200 ICRF2 sources may have optical counterparts that are
too faint for Gaia.) As described in Sect.~\ref{sec:quasars}, 
the positions computed in the auxiliary quasar solution are
expressed in a provisional reference frame aligned with the Hipparcos reference frame.
They are here denoted $(\alpha_\text{H},\,\delta_\text{H})$ to distinguish them from
the corresponding positions $(\alpha,\,\delta)$ in the final Gaia DR1 reference frame.
The VLBI positions of the matched ICRF2 sources are denoted 
$(\alpha_\text{ICRF},\,\delta_\text{ICRF})$.
The position differences for the matched sources are generally less than 10~mas, 
and exceed 50~mas for less than a percent of the sources.

If the orientation of the optical positions with respect to the ICRF2 is modelled by
an infinitesimal solid rotation, we have
\begin{equation}\label{eq:posAlign}
\left. \begin{aligned}
(\alpha_\text{H}-\alpha_\text{ICRF})\cos\delta &= 
(\vec{\varepsilon}\times\vec{r})'\vec{p} = \phantom{-}\vec{q}'\vec{\varepsilon} \\
\delta_\text{H}-\delta_\text{ICRF}  &=
(\vec{\varepsilon}\times\vec{r})'\vec{q} = -\vec{p}'\vec{\varepsilon}
\end{aligned}\quad\right\} \, , \quad
\end{equation}
where $\vec{p}$ and $\vec{q}$ are given by Eq.~(\ref{eq:source2}) and 
$\vec{\varepsilon}$ is a vector
whose components are the rotation angles around the ICRS axes.
Equation~(\ref{eq:posAlign}) involves approximations that break down for sources close 
to the celestial poles, or if $|\vec{\varepsilon}|$ is too large. None of these conditions 
apply in the present case. Rigorous formulae are given in Sect.~6.1 of the AGIS paper. 

A robust weighted least-squares estimation of the orientation parameters, based on
the 262 defining sources in ICRF2 with separation $<$\,150~mas, gives
\begin{equation}\label{eq:align1}
\vec{\varepsilon} \equiv \begin{bmatrix}
\varepsilon_X \\ \varepsilon_Y \\ \varepsilon_Z
\end{bmatrix} =  
\begin{bmatrix}
-2.990 \\ +4.387 \\ +1.810
\end{bmatrix}~\text{mas}\, . 
\end{equation}
The robust fitting retains 260 of the defining sources.
The uncertainty, estimated by bootstrap resampling \citep{efron1994introduction},
is about 0.04~mas in each component.
For comparison, a solution based instead on the 1929 non-defining sources in ICRF2 
gives $\vec{\varepsilon}=[-2.933,\,+4.453,\,+1.834]'$~mas. Using both defining and
non-defining sources, but taking only one hemisphere at a time
($\pm X$, $\pm Y$, $\pm Z$), gives solutions that never differ from Eq.~(\ref{eq:align1})
by more than 0.15~mas in any component. These tests suggest that the result (\ref{eq:align1}) 
is robust at the 0.1~mas level. Figure~\ref{fig:icrfRes} shows the distribution of positional 
residuals with respect to this solution.
The median total positional residual $(\Delta\alpha{*}^2+\Delta\delta^2)^{1/2}$ is 
0.61~mas for the 262 matched defining sources, and 1.27~mas for the 1929 non-defining 
sources. The 90th percentiles are, respectively, 2.7~mas and 7.2~mas.
Additional statistics are given in Appendix~\ref{sec:validation_quasars}.

The reference frame of Gaia DR1 is defined by its orientations at the two epochs J1991.25 (set
by the Hipparcos reference frame at that epoch) and J2015.0 (set by the Gaia observations
of ICRF2 sources). Assuming that the Hipparcos positions were accurately aligned at the
earlier epoch, the result in Eq.~(\ref{eq:align1}) implies that the Hipparcos reference 
frame has a rotation relative to ICRF2 of $\vec{\omega}=(23.75~\text{yr})^{-1}\vec{\varepsilon}$
or
\begin{equation}\label{eq:omega}
\begin{bmatrix}
\omega_X \\ \omega_Y \\ \omega_Z
\end{bmatrix} \simeq  
\begin{bmatrix}
-0.126 \\ +0.185 \\ +0.076
\end{bmatrix}~\text{mas~yr$^{-1}$}\, . 
\end{equation}
This has an uncertainty of about 0.03~mas~yr$^{-1}$ in each axis, mainly from the
uncertainty of the orientation of the Hipparcos reference frame at J1991.25, estimated 
to be 0.6~mas in each axis (Vol.~3, Ch.~18.7 in \citeads{1997ESASP1200.....E}), divided 
by the epoch difference. To put the Hipparcos proper motions on the 
Gaia DR1 reference frame therefore requires the correction
\begin{equation}\label{eq:align3}
\left. \begin{aligned}
\mu_{\alpha*} &= \mu_{\alpha*\text{H}} - \vec{q}'\vec{\omega}\\
\mu_{\delta} &= \mu_{\delta\text{H}} \phantom{j}+ \vec{p}'\vec{\omega}
\end{aligned}\quad\right\} \, .
\end{equation}
It can be noted that the inferred rotation in Eq.~(\ref{eq:omega}) is well within the claimed 
uncertainty of the spin of the Hipparcos reference frame, which is 0.25~mas~yr$^{-1}$ 
per axis (Vol.~3, Ch.~18.7 in \citeads{1997ESASP1200.....E}). 

Subsequent iterations of the primary data set (phase F in Fig.~\ref{fig:primaryConv})
and the secondary solution (Sect.~\ref{ssec:secondary}) used a fixed attitude
estimate, obtained by aligning the attitude from phase D with the Gaia DR1 reference
frame. This was done by applying the time-dependent rotation
$\vec{\varepsilon}+(t-t_\text{ep})\vec{\omega}$, where $t_\text{ep}=\text{J2015.0}$.  
The procedure for rotating the attitude is described in Sect.~6.1.3 of the AGIS paper.
With this transformation the axes of the positions in Gaia DR1 and those of the ICRF2 
are aligned with an estimated uncertainty of 0.1~mas at epoch J2015.0.

\subsection{Secondary solution}
\label{ssec:secondary}

At the end of the primary and quasar solutions (Sects.~\ref{ssec:primary}--\ref{sec:quasars})
the final attitude estimate is aligned with ICRF2 to within a fraction of a mas, and calibration 
parameters consistent with this attitude are available for all magnitudes (different gates and 
window classes). The secondary solution uses this fixed set of attitude and calibration parameters
to estimate the positions of sources in the secondary data set. Contrary to the primary solution, 
this can be done one source at a time, as it does not involve complex iterations between the 
source, attitude, and calibration parameters.

For Gaia DR1 the sources in the secondary data set are all treated as single stars. The astrometric model
is therefore the same as for the primary sources (Sect.~\ref{ssec:primary}) with five parameters per
source. Lacking a good prior position at some earlier epoch, as for the \mbox{Tycho-2} stars, it is
usually not possible to reliably disentangle the five astrometric parameters of a given star
based on the observations available for the current release. 
Therefore, only its position at epoch J2015.0 is estimated. The neglected parallax and proper motion 
add some uncertainty to the position, which is included in the formal positional uncertainties. 
The latter are calculated using the recipe in \citetads{2015A&A...583A..68M}, based on a realistic model 
of the distribution of stellar parallaxes and proper motions as functions of magnitude and Galactic
coordinates. The inflation factor in Eq.~(\ref{eq:infl}) is not
applicable to these uncertainties and was not used for the secondary data set.

\section{Results}
\label{sec:results}

\begin{table*}
\caption{Statistical summary of the 2~million sources in the primary data set of Gaia DR1.
\label{tab:statSummary}}
\centering
\small
\setlength{\tabcolsep}{7pt}
\vspace{-2mm}
\begin{tabular}{lrrrrrrrl}
\hline\hline
\noalign{\smallskip}
 & \multicolumn{3}{c}{All primary sources}
 && \multicolumn{3}{c}{Hipparcos subset} \\
Quantity 
& 10\% & 50\% & 90\% && 10\% & 50\% & 90\% 
& Unit\\
\noalign{\smallskip}
\hline
\noalign{\smallskip}
Standard uncertainty in $\alpha$ ($\sigma_{\alpha*}=\sigma_{\alpha}\cos\delta$) &  0.147 &  0.254 &  0.601 &&  0.158 &  0.224 &  0.391 & mas \\ 
Standard uncertainty in $\delta$ ($\sigma_{\delta}$) &  0.140 &  0.233 &  0.530 &&  0.150 &  0.218 &  0.378 & mas \\ 
Standard uncertainty in $\varpi$ ($\sigma_{\varpi}$) &  0.242 &  0.322 &  0.644 &&  0.229 &  0.283 &  0.499 & mas \\ 
Standard uncertainty in $\mu_{\alpha*}$ ($\sigma_{\mu\alpha*}$) &  0.500 &  1.132 &  2.671 &&  0.035 &  0.064 &  0.129 & mas~yr$^{-1}$ \\ 
Standard uncertainty in $\mu_\delta$  ($\sigma_{\mu\delta}$) &  0.441 &  0.867 &  1.957 &&  0.031 &  0.056 &  0.109 & mas~yr$^{-1}$ \\ 
Semi-major axis of error ellipse in position ($\sigma_\text{pos,\,max}$) &  0.203 &  0.319 &  0.753 &&  0.196 &  0.263 &  0.475 & mas \\ 
Semi-major axis of error ellipse in proper motion ($\sigma_\text{pm,\,max}$) &  0.715 &  1.322 &  3.189 &&  0.038 &  0.069 &  0.137 & mas~yr$^{-1}$ \\ 
Excess source noise ($\epsilon_i$) &  0.299 &  0.478 &  0.855 &&  0.347 &  0.572 &  1.185 & mas \\ 
Number of field-of-view transits input to the solution ($N$) &    8 &   15 &   25 &&    7 &   14 &   25 &    \\ 
Number of good CCD observations AL used in the solution ($n_\text{good}$) &   57 &   99 &  185 &&   51 &   93 &  180 &    \\ 
Fraction of bad CCD observations AL ($n_\text{bad}/(n_\text{good}+n_\text{bad})$) &  0.0 &  0.0 &  2.0 &&  0.0 &  0.0 &  1.8 & \% \\ 
Normalised difference to Hipparcos proper motion ($\Delta Q$) & -- & -- & -- &&  0.33 &  2.35 & 11.32 &    \\ 
Magnitude in Gaia's unfiltered band ($G$) &   9.27 &  11.04 &  12.05 &&   6.84 &   8.28 &   9.70 & mag \\
\noalign{\smallskip}
\hline
\end{tabular}
\tablefoot{
Columns headed 10\%, 50\%, and 90\% give the lower decile, median, and upper decile
of the quantities for all 2\,057\,050~primary sources, and for the subset of 93\,635 sources 
in common with the Hipparcos catalogue \citepads{2007ASSL..350.....V}.
See footnote~\ref{footnote:downweighting} for the definition of good and bad CCD observations.
}
\end{table*}

\begin{figure*}
\resizebox{\hsize}{!}{%
\includegraphics{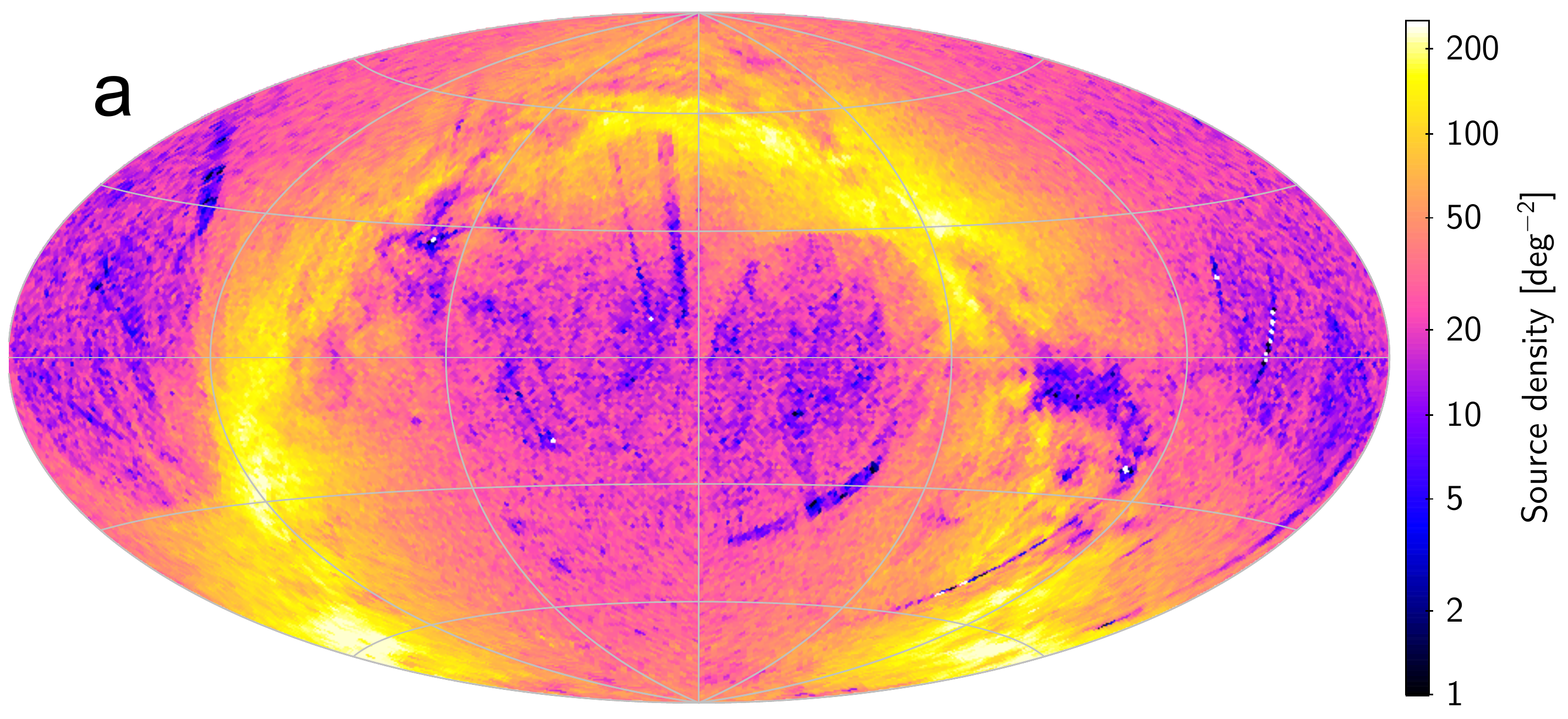}
\includegraphics{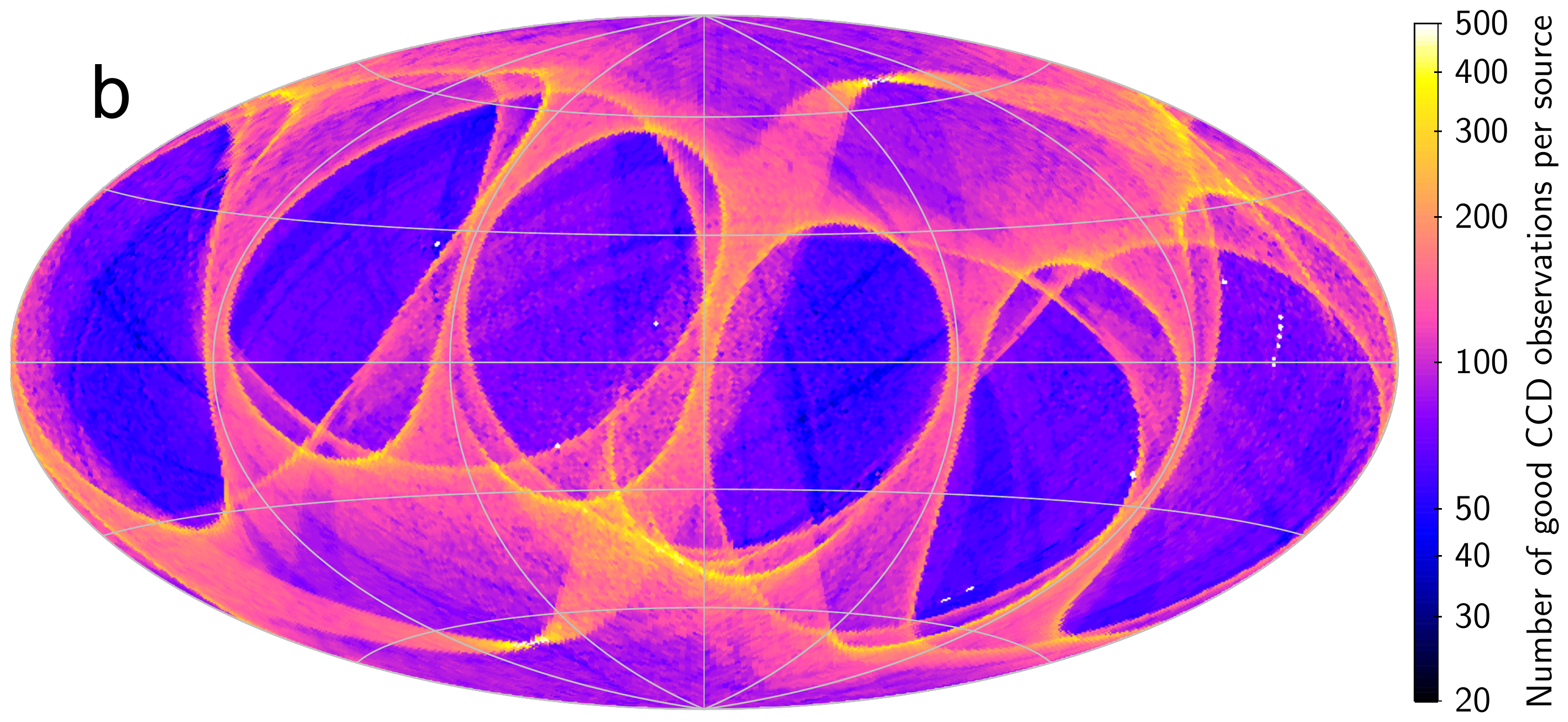}
\includegraphics{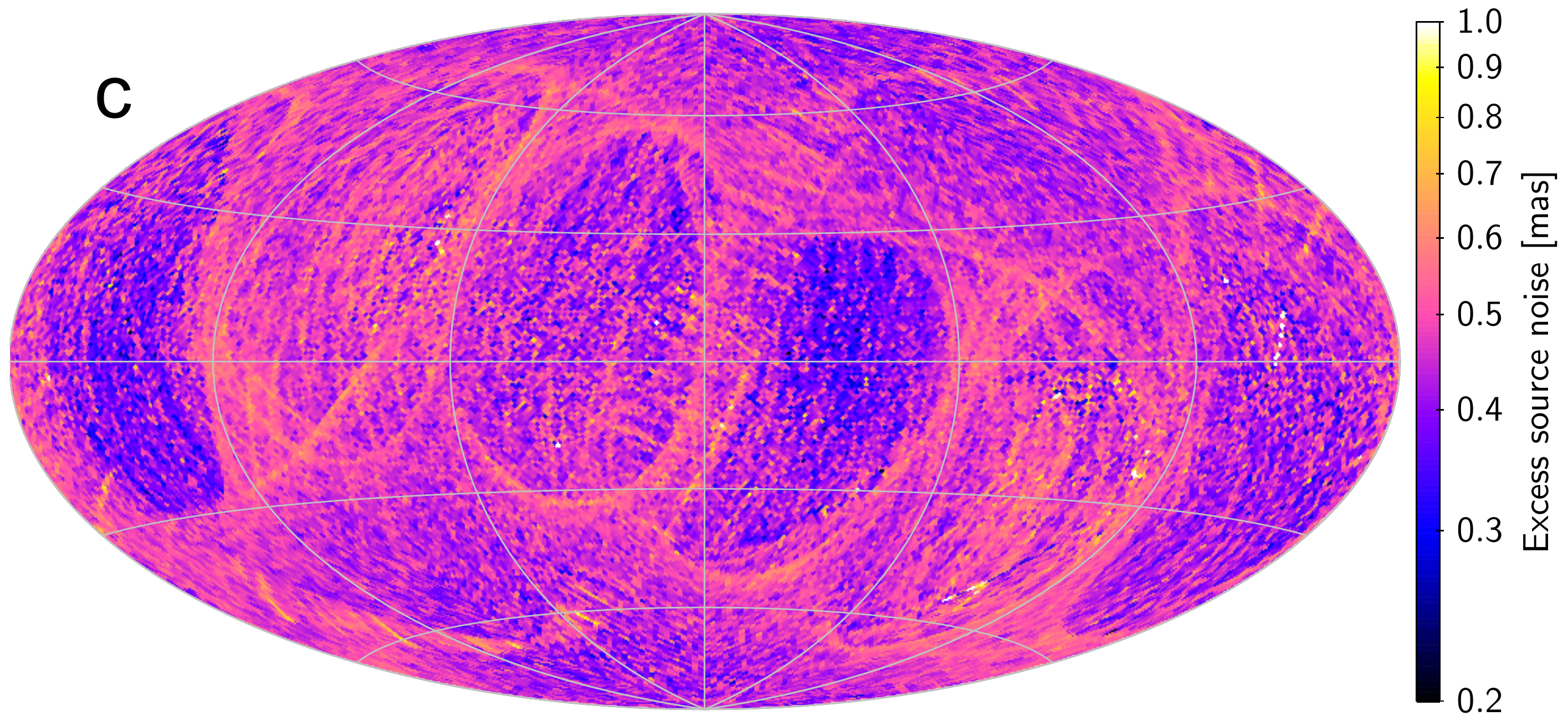}
} 
\caption{Summary statistics for the 2~million sources in the primary data set of Gaia DR1:
(\textbf{a}) density of sources;
(\textbf{b}) number of good CCD observations per source;
(\textbf{c})  excess source noise.
The maps use an Aitoff projection in equatorial 
(ICRS) coordinates, with origin $\alpha=\delta=0$ at the centre and $\alpha$ 
increasing from right to left. The mean density ({\bf a}) and median values ({\bf b} and {\bf c}) 
are shown for sources in cells of about 0.84~deg$^2$. A small number of empty cells are shown
in white.}
\label{fig:stats1}
\end{figure*}

\begin{figure*}
\resizebox{\hsize}{!}{%
\includegraphics{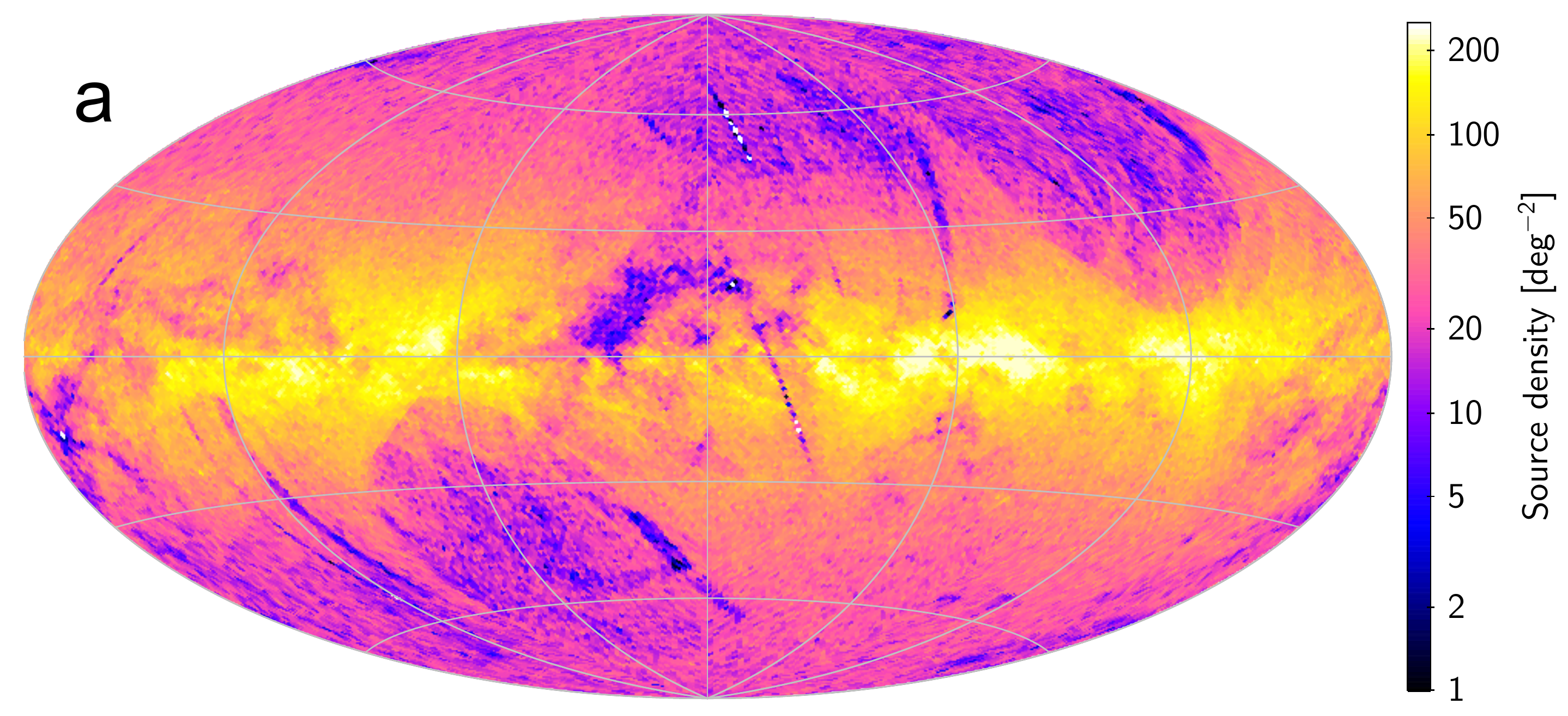}
\includegraphics{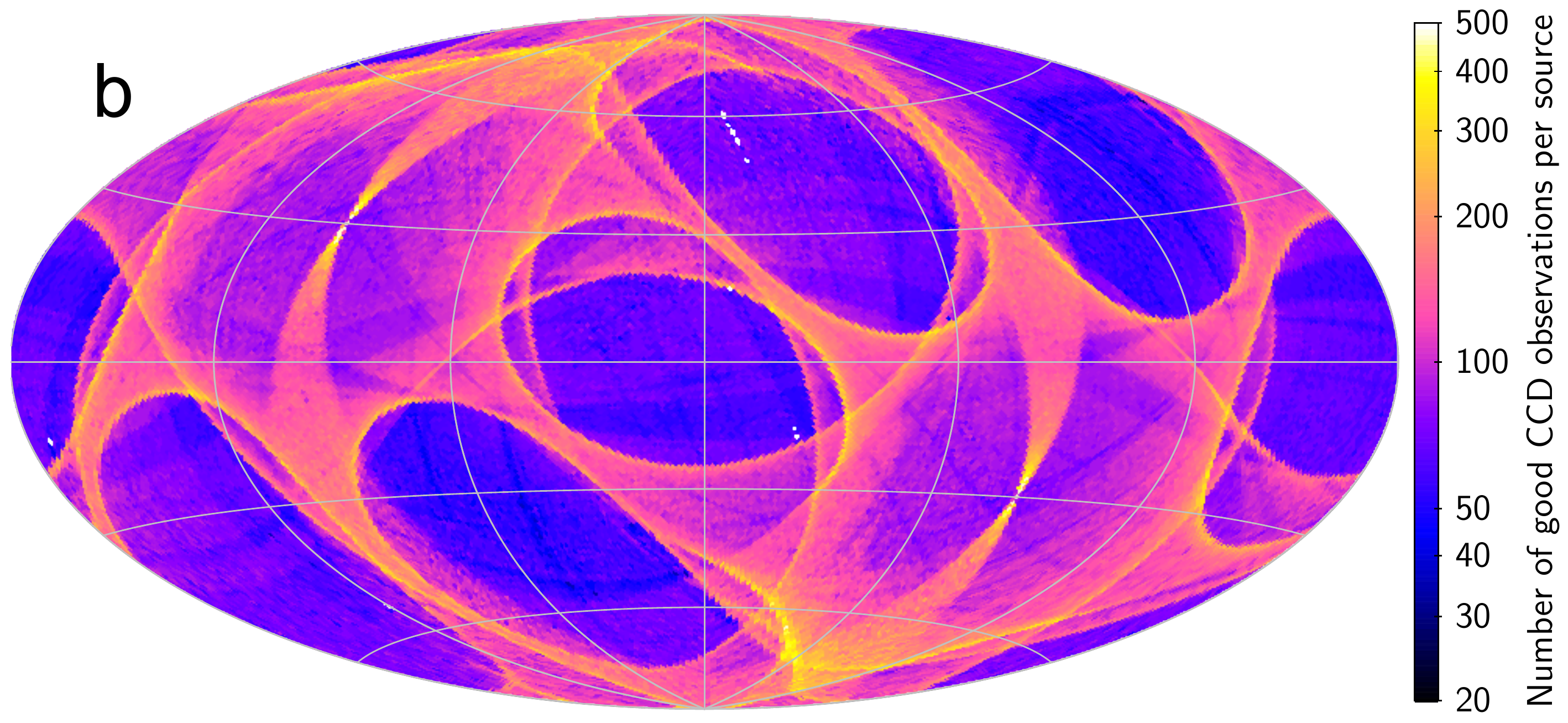}
\includegraphics{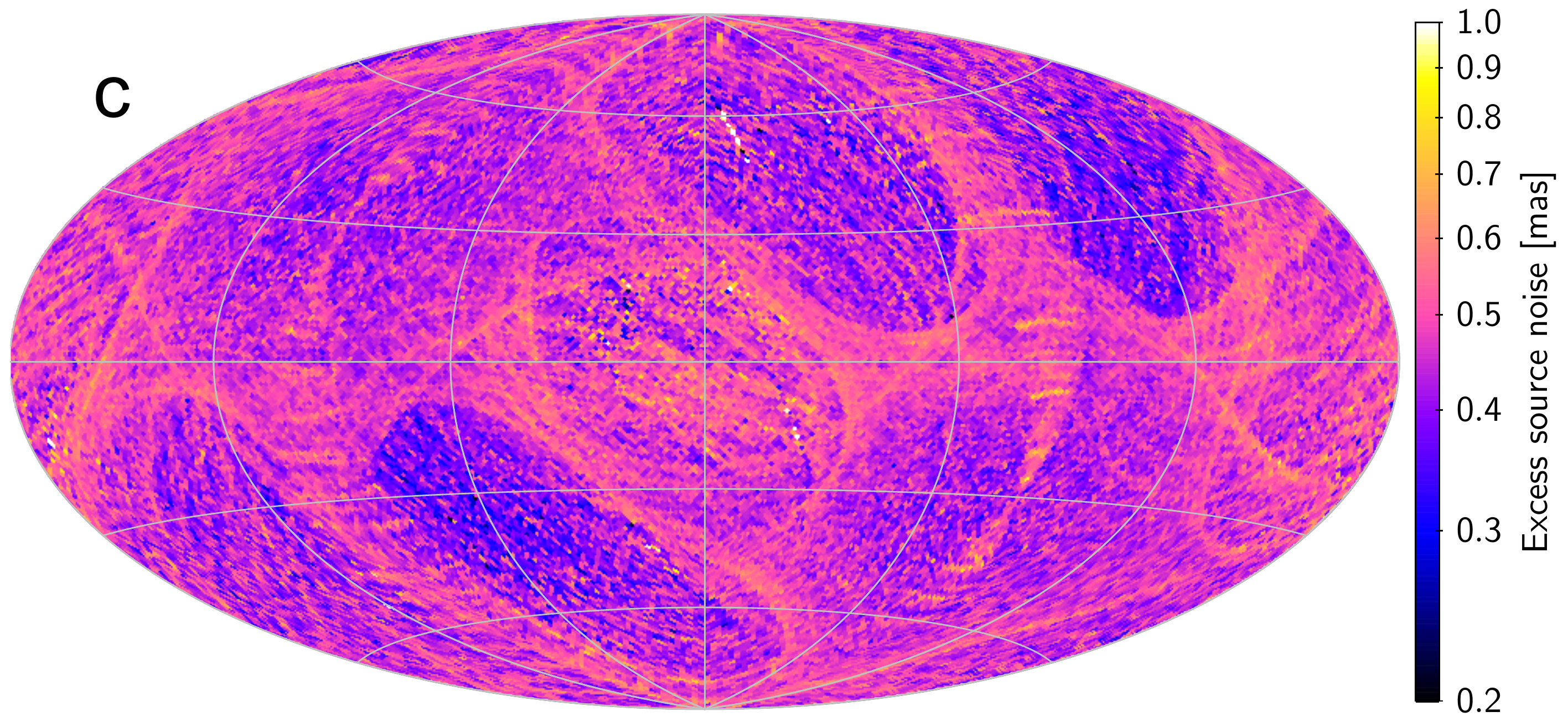}
} 
\caption{Summary statistics for the 2~million sources in the primary data set of Gaia DR1:
(\textbf{a}) density of sources;
(\textbf{b}) number of good CCD observations per source;
(\textbf{c}) excess source noise.
These maps use an Aitoff projection in Galactic coordinates, with origin 
$l=b=0$ at the centre and $l$ increasing from right to left.
The mean density ({\bf a}) and median values ({\bf b} and {\bf c}) 
are shown for sources in cells of about 0.84~deg$^2$. A small number of empty cells are shown
in white.}
\label{fig:stats2}
\end{figure*}

\subsection{Primary data set}
\label{sec:results_primary}
 
For each source the primary solution gives the five astrometric parameters  
$\alpha$, $\delta$, $\varpi$, $\mu_{\alpha *}$, and $\mu_\delta$ together 
with various statistics indicating the quality of the results. The most
important statistics are
\begin{itemize}
\item
the standard uncertainties of the astrometric parameters: 
$\sigma_{\alpha *}=\sigma_{\alpha}\cos\delta$, $\sigma_{\delta}$, 
$\sigma_\varpi$, $\sigma_{\mu\alpha *}$, and $\sigma_{\mu\delta}$;
\item
the ten correlation coefficients among the five parameters:
$\rho(\alpha,\delta)$, $\rho(\alpha,\varpi)$, etc.;
\item
the number of field-of-view transits of the source 
used in the solution: $N$;  
\item
the number of good and bad CCD observations%
\footnote{As described in Sect.~5.1.2 of the AGIS paper, an observation is never
rejected but is downweighted in the solution if it gives a large residual.
$n_\text{bad}$ is the number of CCD observations for which the downweighting
factor $w_l < 0.2$. According to Eq.~(66) in the AGIS paper this means that 
the absolute value of the residual exceeds $3\ln 5\simeq 4.83$ times the total 
uncertainty of the residual, computed as the quadratic sum of the formal standard 
uncertainty of the observation ($\sigma_l$), the excess attitude noise, and the 
excess source noise. $n_\text{good}$ is the number of CCD observations for which 
$w_l \ge 0.2$ (absolute residual less than 4.83 times the total uncertainty);
$n_\text{good}+n_\text{bad}$ is the total number of CCD observations of the
source.\label{footnote:downweighting}}
of the source: $n_\text{good}$, $n_\text{bad}$;
\item
the excess source noise: $\epsilon_i$. This is meant to
represent the modelling errors specific to a given source, i.e.\ deviations
from the astrometric model in Eq.~(\ref{eq:source1}) caused, for example,
by binarity (see Sect.~3.6 in the AGIS paper). Thus, it should ideally be zero 
for most sources. In the present primary solution nearly all 
sources obtain significant excess source noise ($\sim$0.5~mas) from the high level of  
attitude and calibration modelling errors. An unusually large value of
$\epsilon_i$ (say, above 1--2~mas) could nevertheless indicate that the source is
an astrometric binary or otherwise problematic.
\end{itemize}
Additional statistics can be calculated from the standard uncertainties
and correlation coefficients. These include the semi-major axes of the
error ellipses in position and proper motion. Let 
$C_{00}=\sigma_{\alpha *}^2$,
$C_{11}=\sigma_{\delta}^2$, and
$C_{01}=\sigma_{\alpha *}\sigma_{\delta}\rho(\alpha,\delta)$ 
be elements of the $5\times 5$ covariance matrix of the astrometric
parameters. The semi-major axis of the error ellipse in position is
\begin{equation}\label{eq:sigmaPos}
\sigma_\text{pos,\,max} = \sqrt{\frac{1}{2}(C_{00}+C_{11}) +
\frac{1}{2}\sqrt{(C_{11}-C_{00})^2+4C_{01}^2}} \, ,
\end{equation}
with a similar expression for the semi-major axis of the error ellipse 
in proper motion, $\sigma_\text{pm,\,max}$, using the covariance
elements $C_{33}$, $C_{44}$, and $C_{34}$.%
\footnote{The semi-minor axis is obtained by taking the negative sign
of the inner square root in Eq.~(\ref{eq:sigmaPos}). The position angle 
of the major axis (in the range $-90^\circ$ to $90^\circ$) is obtained
as $\theta=\text{atan2}(2C_{01}, C_{11}-C_{00})/2$.}

For the subset in common with the Hipparcos catalogue one additional
statistic is computed: $\Delta Q$, which measures the difference between
the proper motion derived in the primary (TGAS) solution and the proper motion
given in the Hipparcos catalogue.%
\footnote{The quantity $\Delta Q$ was introduced by \citetads{2014A&A...571A..85M}
in the context of the HTPM project, but the present definition differs from the one
in that paper in that only the proper motion differences are considered here.}
It is computed as
\begin{equation}\label{eq:DeltaQ}
\Delta Q = 
\begin{bmatrix} 
\Delta\mu_{\alpha *} & \Delta\mu_{\delta} 
\end{bmatrix}
\left(\vec{C}_\text{pm,\,T}+\vec{C}_\text{pm,\,H}\right)^{-1}
\begin{bmatrix} 
\Delta\mu_{\alpha *} \\ \Delta\mu_{\delta} 
\end{bmatrix} \, ,
\end{equation}
where $\Delta\mu_{\alpha *}=\mu_{\alpha *\text{T}}-\mu_{\alpha *\text{H}}$
and $\Delta\mu_{\delta}=\mu_{\delta\text{T}}-\mu_{\delta\text{H}}$ are the
proper motion differences, with T and H designating the values from respectively
TGAS and the Hipparcos catalogue. 
$\vec{C}_\text{pm,\,T}$ is the $2\times 2$
covariance submatrix of the TGAS proper motions and $\vec{C}_\text{pm,\,H}$
the corresponding matrix from the Hipparcos catalogue. The new reduction of the
raw Hipparcos data by \citetads{2007ASSL..350.....V} was used, as retrieved from CDS,
with covariances computed as described in Appendix~B of \citetads{2014A&A...571A..85M}.
For the calculation in Eq.~(\ref{eq:DeltaQ}) 
the Hipparcos proper motions were first transformed to the Gaia DR1 reference 
frame by means of Eq.~(\ref{eq:align3}) and then propagated to epoch J2015.0,
assuming zero radial velocity. $\Delta Q$ is therefore sensitive to all deviations
from a purely linear tangential proper motion, including perspective effects.
If the proper motion errors in TGAS and in the Hipparcos catalogue are
independent and Gaussian with the given covariances, then $\Delta Q$ is
expected to follow a chi-squared distribution with two degrees of freedom,
i.e.\ $\text{Pr}(\Delta Q > x)=\exp(-x/2)$.

\begin{landscape}
\begin{figure}
\resizebox{\hsize}{!}{\includegraphics[angle=0]{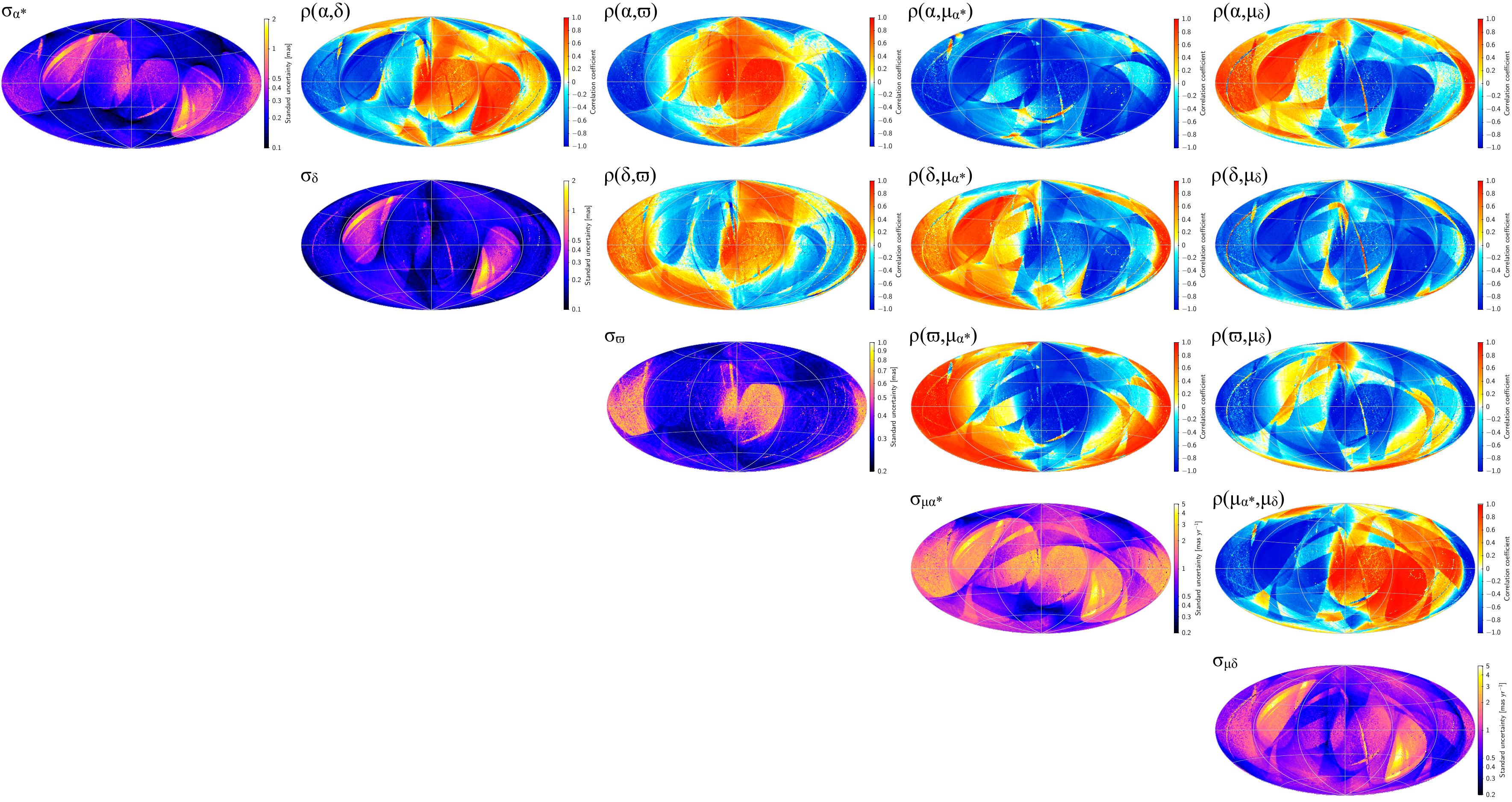}}
\caption{Summary statistics for the 2~million sources in the primary data set. 
The five maps along the main diagonal show, from top-left to bottom-right, 
the standard uncertainties in $\alpha$, $\delta$, $\varpi$, $\mu_{\alpha*}$, $\mu_\delta$. 
The ten maps above the diagonal show the correlation coefficients, in the range $-1$ to $+1$,
between the corresponding parameters on the main diagonal. 
All maps use an Aitoff projection in equatorial (ICRS) coordinates, with origin $\alpha=\delta=0$ 
at the centre and $\alpha$ increasing from right to left. Median values are shown in
cells of about 0.84~deg$^2$.}
\label{fig:statSigmaCorr}
\end{figure}
\end{landscape}

The primary solution gives astrometric results for about 2.48~million sources.
Unreliable solutions are removed by accepting only sources with
\begin{equation}\label{eq:formal1}
\sigma_\varpi < 1~\text{mas} \quad \text{and} \quad \sigma_\text{pos,\,max} < 20~\text{mas} \, .
\end{equation}
Here $\sigma_\varpi$ is the standard uncertainty in parallax from Eq.~(\ref{eq:infl}),
and $\sigma_\text{pos,\,max}$ is the semi-major axis of the error ellipse in position
at the reference epoch (J2015.0). 
The second condition removes a small fraction of stars with extremely elongated error ellipses. 

Applying the filter in Eq.~(\ref{eq:formal1}) results in a set of 2\,086\,766~sources
with accepted primary solutions. However, for a source to be included in Gaia DR1
it must also have valid photometric information. The primary data set therefore 
gives astrometric parameters for 2\,057\,050~sources
together with their estimated standard uncertainties, correlations 
among the five parameters, and other quality indicators.
A statistical summary is presented in Table~\ref{tab:statSummary}. Separate statistics 
are given for the subset of Hipparcos sources, which have rather different uncertainties 
in proper motion owing to the more accurate positions at the Hipparcos epoch.
Figures~\ref{fig:stats1}--\ref{fig:statSigmaCorr} show the variation of some statistics
with celestial position. The distribution of $\Delta Q$ for the Hipparcos subset
is discussed in Appendix~\ref{sec:hipparcos}.

In the primary data set, the standard uncertainties of the positions at epoch J2015.0 and of
the parallaxes are dominated by attitude and calibration errors in the Gaia observations. They therefore 
show little or no systematic dependence on magnitude. For the proper motions, on the other hand, 
the dominating error source is usually the positional errors at J1991.25 resulting from the Hipparcos and 
Tycho-2 catalogues. The uncertainties in proper motion therefore show a magnitude dependence 
mimicking that of the positional uncertainties in these catalogues.

To preserve the statistical integrity
of the data set, no filtering was applied based on the actual values of the astrometric
parameters. Thus, the primary data set contains 30\,840 (1.5\%) negative parallaxes.
The most negative parallax is $-24.82\pm 0.63$~mas, but even this provides
valuable information, e.g.\ that there are parallaxes that are wrong by at least
40~times the stated uncertainty.
However, owing to a technical issue in the construction of the initial source list,
several nearby stars with high proper motion are missing in the Hipparcos 
subset of Gaia DR1. In particular, the 19 Hipparcos stars with total proper motion
$\mu>3500$~mas~yr$^{-1}$
are missing, including the five nearest stars HIP~70891 (Proxima Cen), 
71681 ($\alpha^2$~Cen), 71683 ($\alpha^1$~Cen), 87937 (Barnard's star), and
HIP~54035. ($\alpha^{1}$ and $\alpha^{2}$~Cen would in any case have been 
rejected because they are too bright.)

\begin{table*}
\caption{Statistical summary of the 1141~million sources in the secondary data set of Gaia DR1.
\label{tab:statSummarySecondary}}
\centering
\small
\setlength{\tabcolsep}{8pt}
\vspace{-2mm}
\begin{tabular}{lrrrrrl}
\hline\hline\noalign{\smallskip}
Quantity 
&& 10\% & 50\% & 90\% && Unit\\
\noalign{\smallskip}
\hline
\noalign{\smallskip}
Standard uncertainty in $\alpha$ ($\sigma_{\alpha*}=\sigma_{\alpha}\cos\delta$) &&  0.285 &  1.802 & 12.871 && mas \\ 
Standard uncertainty in $\delta$ ($\sigma_{\delta}$) &&  0.257 &  1.568 & 11.306 && mas \\ 
Semi-major axis of error ellipse in position ($\sigma_\text{pos,\,max}$):\\
\hspace{5mm} $G<16$ (7\% of the secondary data set) &&  0.106 &  0.255 &  4.118 && mas \\ 
\hspace{5mm} $G=16{-}17$ (7\%) &&  0.182 &  0.484 & 11.105 && mas \\ 
\hspace{5mm} $G=17{-}18$ (13\%) &&  0.284 &  0.761 & 11.534 && mas \\ 
\hspace{5mm} $G=18{-}19$ (22\%) &&  0.501 &  1.444 & 13.027 && mas \\ 
\hspace{5mm} $G=19{-}20$ (31\%) &&  0.986 &  2.816 & 16.314 && mas \\ 
\hspace{5mm} $G=20{-}21$ (20\%) &&  2.093 &  7.229 & 21.737 && mas \\ 
\hspace{5mm} all magnitudes (100\%) &&  0.349 &  2.345 & 15.699 && mas \\ 
Excess source noise ($\epsilon_i$) &&  0.000 &  0.594 &  2.375 && mas \\ 
Number of field-of-view transits input to the solution ($N$) &&    7 &   13 &   26 &&    \\ 
Number of good CCD observations AL used in the solution ($n_\text{good}$) &&   41 &   71 &  157 &&    \\ 
Fraction of bad CCD observations AL ($n_\text{bad}/(n_\text{good}+n_\text{bad})$) &&  0.0 &  0.0 &  2.0 && \% \\ 
Magnitude in Gaia's unfiltered band ($G$) &&  16.49 &  19.02 &  20.32 && mag \\
\noalign{\smallskip}
\hline
\end{tabular}
\tablefoot{
Columns headed 10\%, 50\%, and 90\% give the lower decile, median, and upper decile
of the quantities for the 1\,140\,662\,719 secondary sources.
See footnote~\ref{footnote:downweighting} for the definition of good and bad CCD observations.
}
\end{table*}

\begin{figure*}
\resizebox{\hsize}{!}{%
\includegraphics{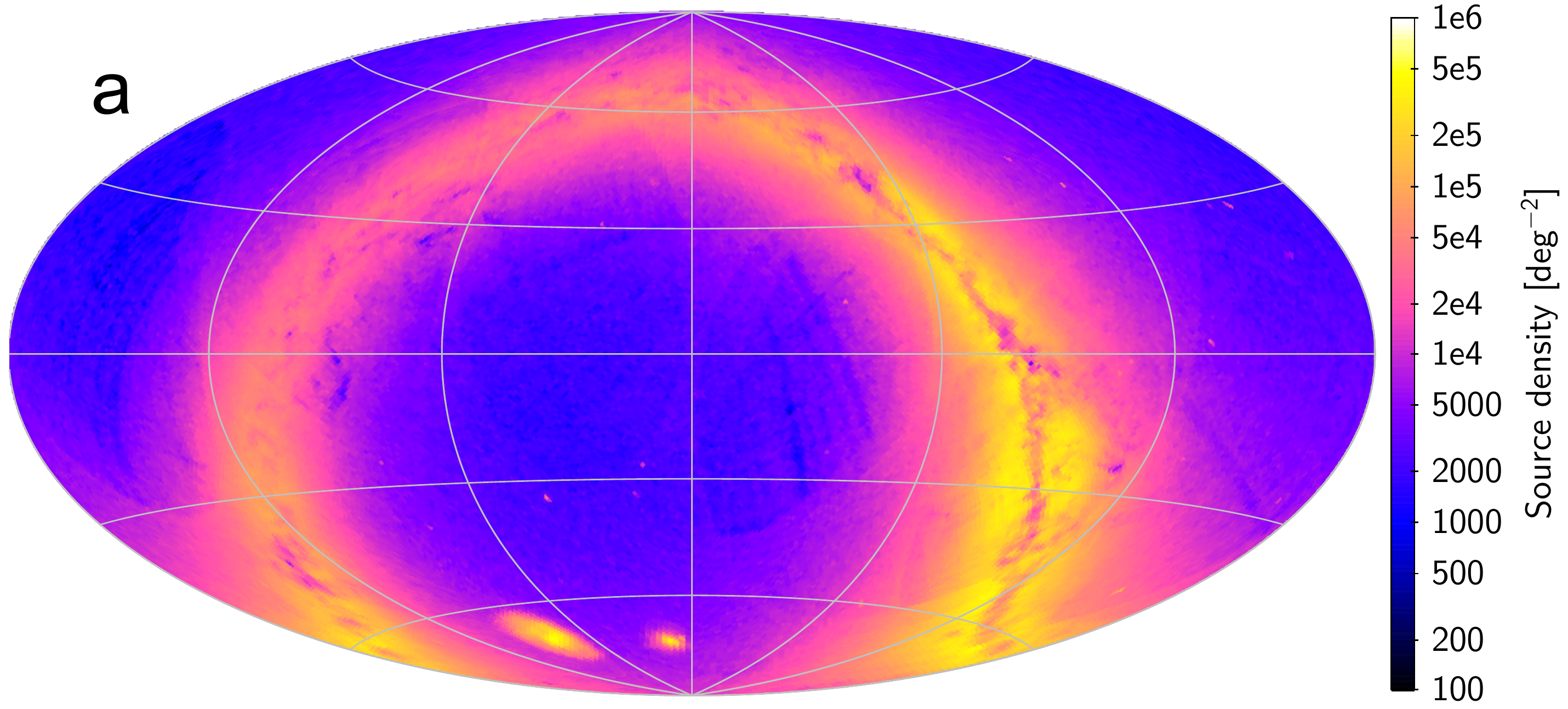}
\includegraphics{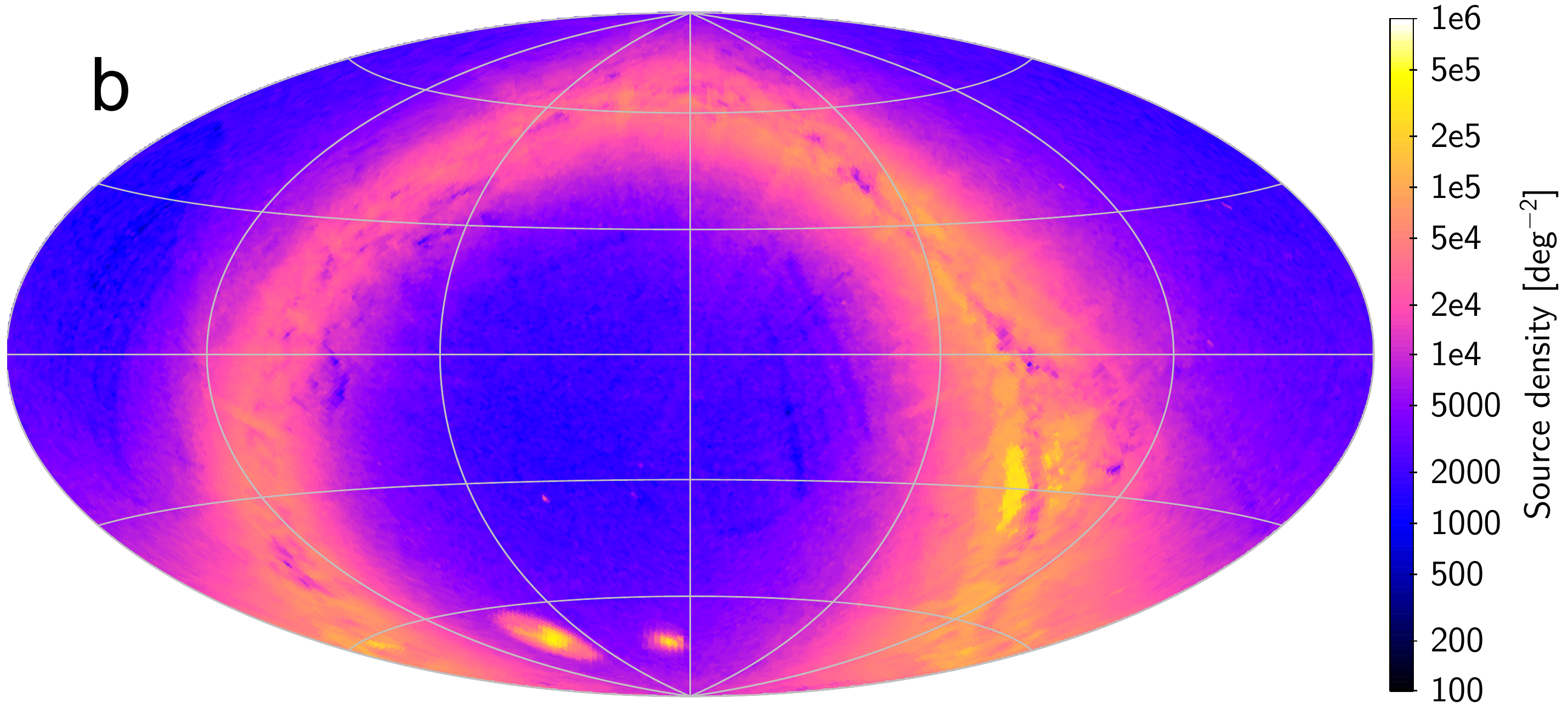}
\includegraphics{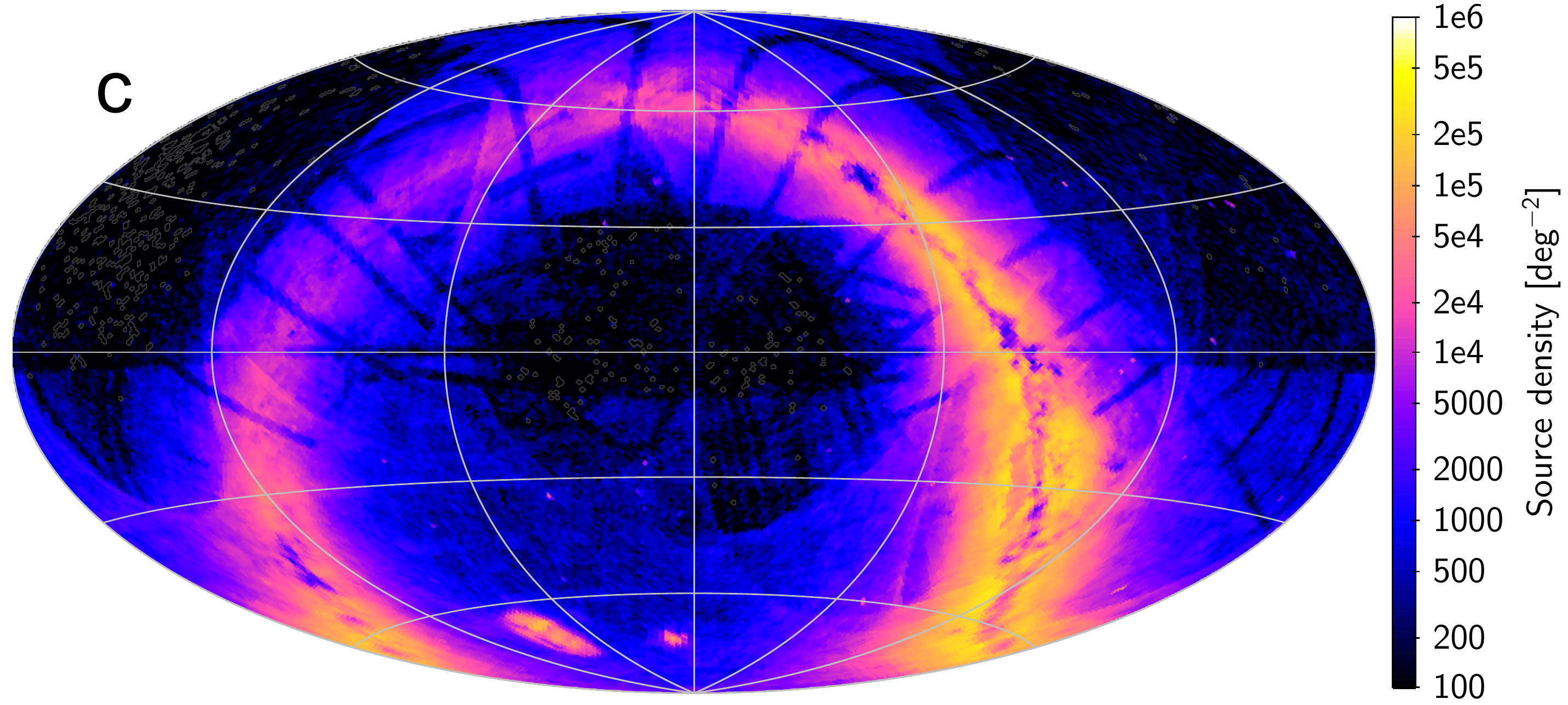}
} 
\caption{Density of sources in the secondary data set of Gaia DR1:
(\textbf{a}) all 1141~million sources in the secondary data set;
(\textbf{b}) the 685~million sources in common with the IGSL;
(\textbf{c}) the 456~million new sources. 
These maps use an Aitoff projection in equatorial (ICRS) coordinates, with origin 
$\alpha=\delta=0$ at the centre and $\alpha$ increasing from right to left. Mean densities
are shown for sources in cells of about 0.84~deg$^2$.
}

\label{fig:secondaryEqu}
\end{figure*}
\begin{figure*}
\resizebox{\hsize}{!}{%
\includegraphics{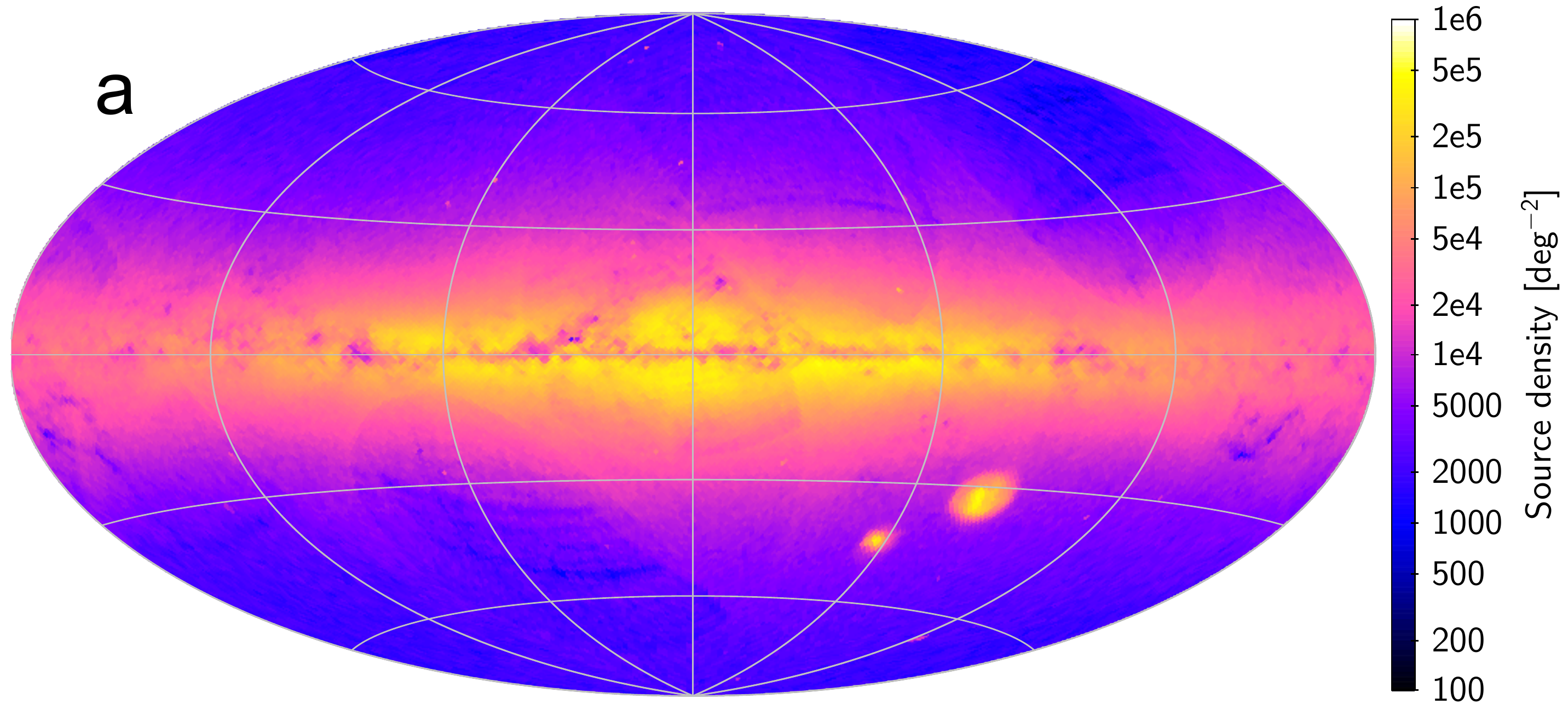}
\includegraphics{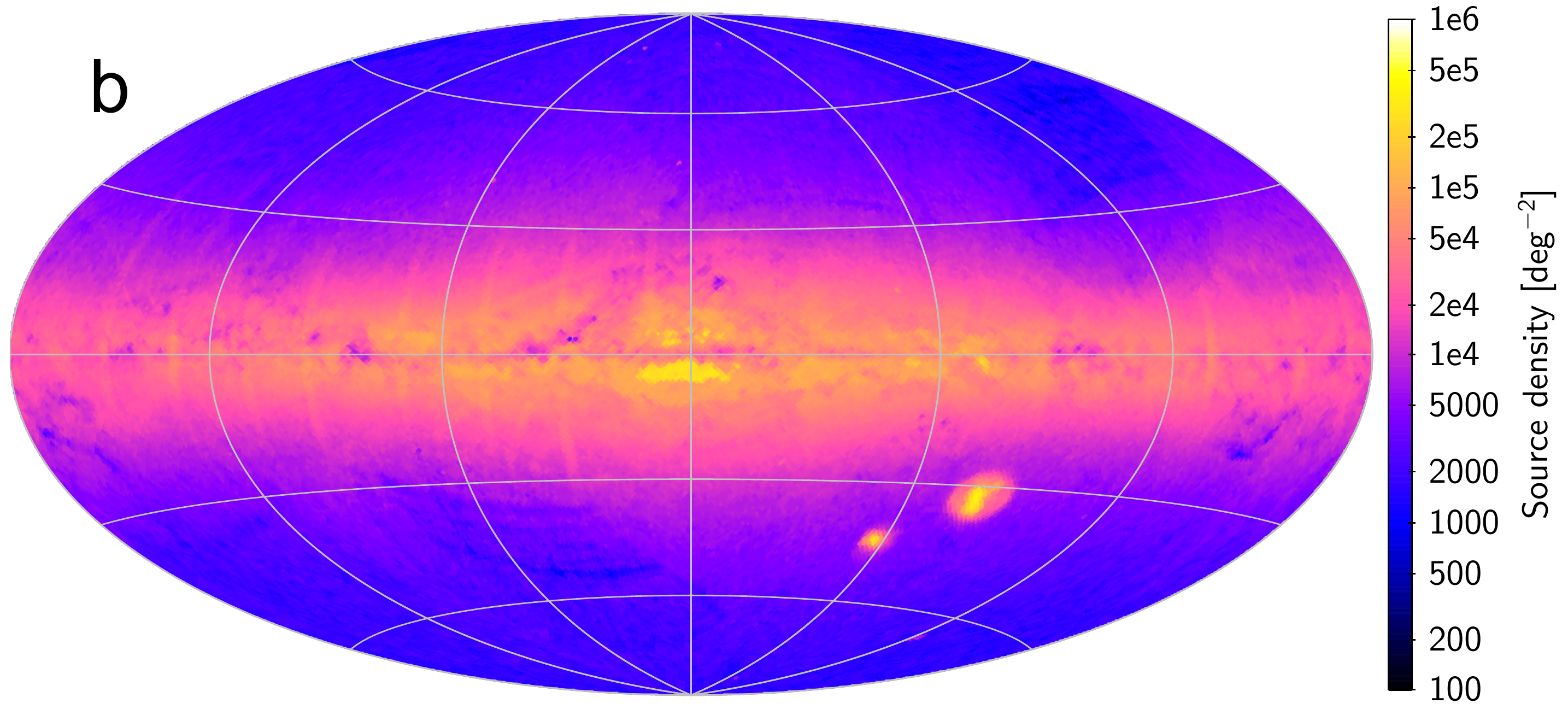}
\includegraphics{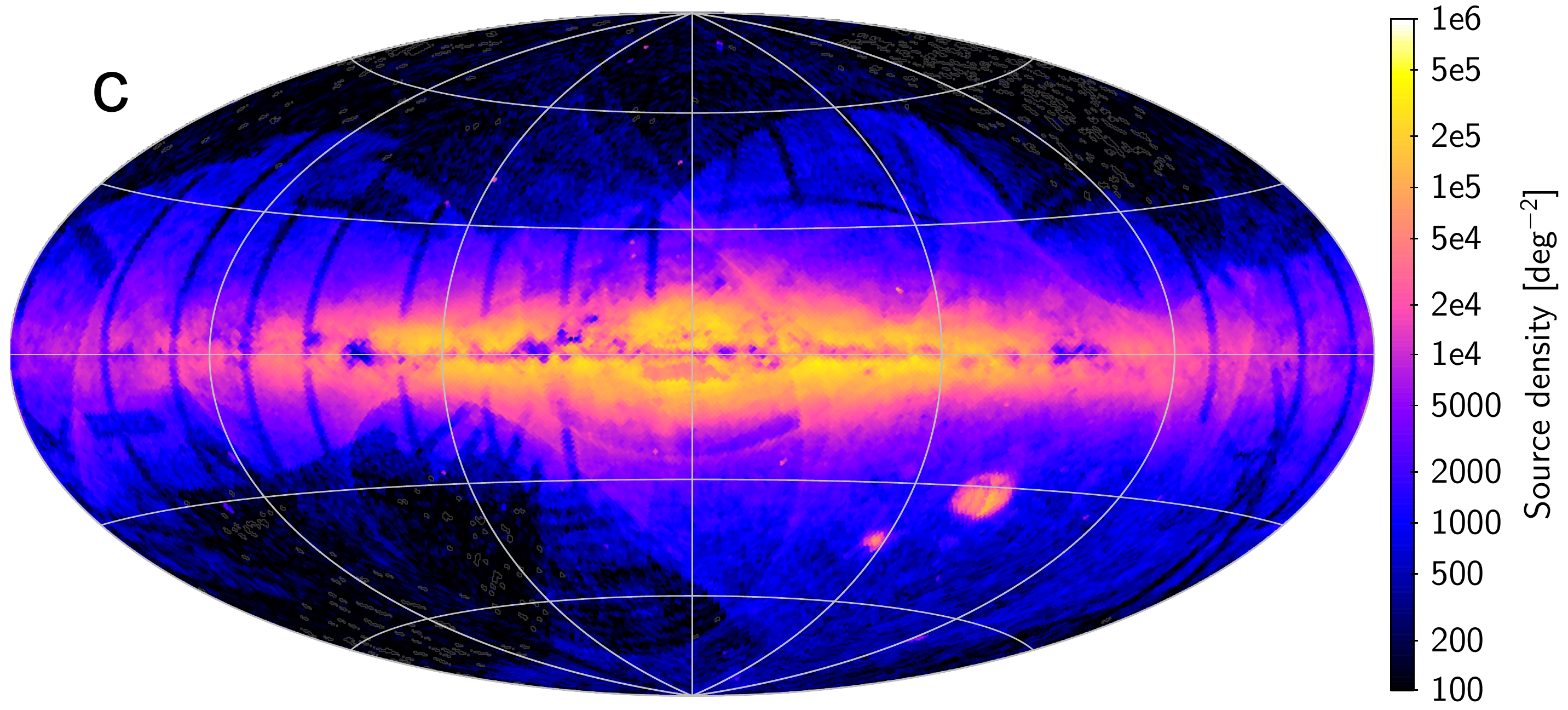}
} 
\caption{Density of sources in the secondary data set of Gaia DR1:
(\textbf{a}) all 1141~million sources in the secondary data set;
(\textbf{b}) the 685~million sources in common with the IGSL;
(\textbf{c}) the 456~million new sources. 
These maps use an Aitoff projection in Galactic coordinates, with origin 
$l=b=0$ at the centre and $l$ increasing from right to left. Mean densities
are shown for sources in cells of about 0.84~deg$^2$.
}
\label{fig:secondaryGal}
\end{figure*}

\subsection{Secondary data set} 
\label{sec:results_secondary}

The secondary solution gives approximate positions for more than 2.5~billion entries,
including more than 1.5~billion ``new sources'' created in the process of cross-matching the
Gaia detections to the source list \citep[see Sect.~4 in][]{2016GaiaF}.

Many of the new sources are spurious, and a suitable criterion had to be 
found to filter out most of the bad entries. On the other hand, for uniformity of the resulting 
catalogue, it is desirable that the very same criteria do not reject too many of the solutions
using observations cross-matched to the initial source list. By comparing the distributions of 
various quality indicators for the two kinds of sources, the following criterion was found 
to provide sensible rejection of obviously spurious sources while retaining nearly all 
solutions for sources in the initial source list:
\begin{equation}\label{eq:formal2}
N \ge 5 
\quad \text{and} \quad 
\epsilon_i < 20~\text{mas} 
\quad \text{and} \quad 
\sigma_\text{pos,max} < 100~\text{mas} 
\, .
\end{equation}
$N$ is the number of field-of-view transits used in the solution, 
$\epsilon_i$ is the excess source noise (Sect.~\ref{sec:results_primary}), 
and $\sigma_\text{pos,max}$ the semi-major axis of the error ellipse in position
at the reference epoch. The excess source noise is essentially 
a measure of the astrometric consistency of the $N$ transits. The first two conditions
therefore mean that the source should have been detected at least five times at 
positions consistent within some 20~mas. This limit is large enough to accommodate
attitude and calibration modelling errors as well as source modelling errors for
many unresolved binaries, while rejecting the much larger mismatches that are
typically found for spurious detections. The limit on the size of the error ellipse
in position removes very faint sources with large photon-noise uncertainties
and some sources with extremely elongated error ellipses.

That Eq.~(\ref{eq:formal2}) provides a reasonable selection was checked in several 
selected areas by superposing the positions of accepted and rejected sources on 
images obtained with the ESO VLT Survey Telescope (VST) for the Gaia ground based optical 
tracking (GBOT) project \citepads{2014SPIE.9149E..0PA} and, for some very high-density
areas in the Baade's window region, with the HST Advanced Camera for Surveys (ACS/WFC).
These checks indicate that the above criterion is even conservative in the sense that
very many real sources detected by Gaia are not retained in the present preliminary selection. 

Applying the selection criterion in Eq.~(\ref{eq:formal2}) results in accepted positional
solutions for 1467~million entries, of which 771~million are in the IGSL and 695~million 
are new sources. A large number of entries in the IGSL were found to be redundant,
resulting in nearly coinciding positional solutions. The secondary data set of Gaia DR1 
consists of the 1\,140\,622\,719 non-redundant entries that also have valid photometric information.
The leftmost maps in Fig.~\ref{fig:secondaryEqu} shows the total density of sources in 
the secondary data set; the other two maps show the densities of the IGSL and new sources.
Imprints of the ground-based surveys used in the construction of the IGSL are 
clearly seen in the latter two maps (as over- and under-densities in Fig.~\ref{fig:secondaryEqu}b 
and c, respectively). These are largely absent in the total density map (Fig.~\ref{fig:secondaryEqu}a), 
which however still shows features related to the scanning law of Gaia (cf.\ Fig.~\ref{fig:stats2}b).
Figure~\ref{fig:secondaryGal} shows the same densities in Galactic coordinates.  

The secondary data set contains only positions, with their estimated
uncertainties and other statistics, but no parallaxes or proper motions. Some
statistics are summarised in Table~\ref{tab:statSummarySecondary}. 
The standard uncertainties in position are calculated using the recipe
in \citetads{2015A&A...583A..68M}. This provides a conservative estimate based on
a Galactic model of the distribution of the (neglected) parallaxes and proper motions.

\section{Validation}
\label{sec:validation}

A significant effort has been devoted to examining the quality of the astrometric 
solutions contributing to Gaia DR1. This validation has been made in two steps,
by two independent groups using largely different approaches. The first step, 
carried out by the AGIS team responsible for the solutions, aimed to
characterise the solutions and design suitable filter criteria for the published
results. In the second step, carried out by a dedicated data validation team 
within the Gaia Data Processing and Analysis Consortium \citep{2016GaiaB}, 
a rigorous set of pre-defined tests were applied to the data provided
\citep{2016GaiaA}.

Only the validation tests performed by the AGIS team on the primary solution 
and on the auxiliary quasar solution are described here. They are of three kinds:
\begin{enumerate}
\item
The residuals of the astrometric least-squares solutions were analysed in order 
to verify that they behave as expected, or alternatively to expose deficiencies 
in the modelling of the data. See Appendix~\ref{app:residuals}.
\item
Special TGAS runs were made, in which the modelling of the Gaia instrument 
or attitude was modified, or different subsets of the observations were used.
These are consistency checks of the data, and could also reveal if the results are
unduly sensitive to details of the modelling. A direct comparison of the resulting
astrometric parameters (in particular the parallaxes) provides a direct quantification
of this sensitivity. See Appendix~\ref{app:validation}.
\item 
The results were compared with independent external data, such as astrometric parameters from 
the Hipparcos catalogue and expected results for specific astrophysical objects
(quasars, cepheids, etc.). See Appendices~\ref{app:precision} and~\ref{app:external}.
\end{enumerate}
The validation tests were completed before the final selection of sources had been made, 
and are therefore based on more sources than finally retained in Gaia DR1. 

The detailed results of these exercises are given in the appendices. 
In summary, the comparisons with external data (Appendix~\ref{app:precision} 
and~\ref{app:external}) show good agreement 
on a global level, with differences generally compatible with the stated precisions 
of the primary data set and of the comparison data. However, there are clear indications 
of systematic differences at the level of $\pm 0.2$~mas, mainly depending on colour 
and position on the sky. 
Such differences may extend over tens of degrees (Figs.~\ref{fig:valColour}--\ref{fig:valSplit}).
Very locally, even larger systematics are indicated, which 
would affect a small fraction of the sources. The statistical distributions of the
differences typically have Gaussian-like cores with extended tails including outliers.
The analysis of residuals (Appendix~\ref{app:residuals}) allows us to identify important 
contributors to the random and systematic errors, i.e.\ attitude modelling
errors (including micro-clanks and micrometeoroid hits) and colour-dependent
image shifts in the optical instrument (chromaticity). The special validation solutions 
(Appendix~\ref{app:validation}) confirm these findings and provide some quantification 
of the resulting errors, while pointing out directions for future improvements.

\section{Known problems: Causes and cures} 
\label{sec:problems}

The preliminary nature of the astrometric data contained in Gaia DR1 cannot be too
strongly emphasised. TGAS has allowed us to 
develop our understanding of the instrument, exercise the complex data analysis 
systems, and obtain astrophysically valuable results in a much shorter time than
originally foreseen. This has been possible thanks to a number of simplifications
and shortcuts, which inevitably weakens the solution in many respects. Additional
weaknesses have been identified during the validation process 
(Appendix~\ref{app:validation}), and more will undoubtedly be discovered by
users of the data.

Importantly, the weaknesses identified so far are either an expected consequence
of the imposed limitations of Gaia DR1, or of a character that will be remedied by the
planned future improvements of the data analysis. The most important known
weaknesses, and their remedies, are listed below.
\begin{enumerate}
\item
{Limited input data:} 
the data sets are based on a limited time interval -- less than a quarter of the nominal
mission length. The primary astrometric solution, providing the attitude and calibration 
parameters, uses less than 1\% of the data volume expected for the final astrometric 
solution. Both the length of the observed interval and the number of primary sources
used in the astrometric solution will increase with successive releases. 
\item
{Prior data:} 
the use of prior positional information from the Hipparcos and \mbox{Tycho-2} catalogues
limits the primary data set to a few million of the brightest stars ($\lesssim 11.5$~mag).
These are in many ways the most problematic ones because of CCD gating, partially saturated images,
etc. Moreover, the positional errors in these catalogues affect the resulting proper motions and
parallaxes. Future releases will not use any prior astrometric information at all,
except for aligning the reference frame.
\item
{Cyclic processing:} 
the astrometric solution is designed to be part of a bigger processing loop, including
the gradual refinement of the calibration of LSF and PSF 
versus the spectral energy distribution of the sources. 
For Gaia DR1 this loop had not been closed, and the centroiding was done against a 
bootstrap library prepared pre-launch using the limited knowledge of the instrument at 
the time. The image centroids used for the present solutions are therefore strongly
affected by chromaticity and other uncalibrated variations of the LSF and PSF.
The effect of this is clearly seen both in the residuals (Appendix~\ref{sec:chrom}) and in 
the astrometric data (Appendix~\ref{app:validation}). For the next data release the
loop will have been closed and executed once, which should drastically reduce some of 
these effects. The final astrometric solution will be based on several cyclic processing loops, 
which should almost completely eliminate the centroid errors caused by systematic variations
of the LSF and PSF, including chromaticity.
\item
{Cross-matching:} 
the cross-matching of Gaia observations to sources is far from perfect owing to
the use of crude estimates of the attitude and calibration, and an initial source list 
compiled mainly from ground-based data. The lack of stars with high proper motion
($\mu>3.5$~arcsec~yr$^{-1}$) in Gaia DR1 is one unfortunate consequence. The
astrometric solutions for subsequent releases will be based on the much improved
cross-matchings made as part of the cyclic processing loop mentioned above. 
The final list of sources detected and observed by Gaia will be independent 
of ground-based surveys.
\item
{Attitude model:}
The relatively low density of sources in the primary solution ($\sim$10~deg$^{-2}$
in large parts of the sky; see Fig.~\ref{fig:stats1}a) required the use of a longer
knot interval (30~s) for the attitude model than foreseen in the final astrometric solution
($<$\,10~s; see the AGIS paper, Sect.~7.2.3, and \citeads{2013A&A...551A..19R}). Residual
modelling errors contribute significant correlated noise in the present solution.
This will be eliminated by the vastly improved attitude modelling made possible by
a much higher density of primary sources.
\item
{Micro-clanks and micrometeoroid hits:} these are not treated at all, or only
by placing gaps around major micrometeoroid hits. Micro-clanks are much more 
frequent than expected from pre-launch estimates, and could be a major 
contribution to the attitude modelling errors even for very short knot intervals,
if not properly handled. The use of rate data (estimates of the spacecraft angular velocity 
that do not require AGIS) to detect and quantify micro-clanks 
was not foreseen before the commissioning 
of Gaia, but has emerged as an extremely efficient way to eliminate the detrimental 
effect of micro-clanks (Appendix~\ref{sec:clanks}). For future data releases this
will be implemented, and a similar technique can be used to mitigate the effects
of small micrometeoroid hits and other high-frequency attitude irregularities.
\item
{Source model:} all sources are treated as single stars, and the radial component
of their motions is ignored. Thus, all variations in proper motion due to orbital
motion in binaries or perspective effects are neglected. The proper motions given are
the mean proper motions between the Hipparcos/Tycho epoch (around J1991.25)
and the Gaia DR1 epoch (J2015.0). For resolved binaries, it could be that the
positions at the two epochs are inconsistent, e.g.\ referring to different components,
or to one of the components at one epoch, and to the photocentre at the other. 
\item
{Calibration model:} the geometric instrument calibration model used for the current 
primary solution does not include the full range of dependencies foreseen in the
final version. This concerns in particular the small-scale irregularities, i.e.\ the 
small AL displacements from one pixel column to the next, and their
dependence on the gate and time. Moreover, the large-scale calibration parameters
evolve too quickly for the currently used time resolution (see Fig.~\ref{fig:cal1}).
These issues can be resolved by better adapting the model to the observed
variations, for example by using polynomial segments or splines for their
temporal evolution.  
\item
{Basic-angle variations:} for this data release, basic-angle variations have been
corrected by simply adopting the (smoothed) variations measured by the BAM
(Appendix~\ref{sec:calibration_bam}). We know from 
simulations that a very wide range of basic-angle variations (depending on their
frequency and other characteristics) can in fact be calibrated as part of the astrometric
solution. Special validation solutions, which include the harmonic coefficients of the 
basic angle variations as unknowns, show that this is indeed possible for variations
of the kind seen in actual data. It is expected that future astrometric solutions
will have the basic-angle variations largely determined by such self-consistent calibrations
rather than relying on BAM data. The latter will still be important as an independent
check and for detecting basic-angle jumps and other high-frequency features. 
\item
{Spatially correlated systematics:} Several of the weaknesses mentioned above
combine to produce systematic errors that are strongly correlated over areas that may
extend over tens of degrees. Such errors are not much reduced by averaging over any
number of stars in a limited area, e.g.\ when calculating the mean parallax or mean  
proper motion of a stellar cluster. This will greatly improve in future releases of Gaia 
data thanks to the generally improved modelling of the instrument and attitude.  
\end{enumerate}  
With such a long list of problems and weaknesses identified in the data already before
their release, one might wonder if the release should not have been postponed until a number of 
these issues have been fixed or mitigated. However, we believe that the current results are
immensely valuable in spite of these problems, provided that the users are aware of
them. Moreover, future improvements of the data analysis can only benefit from 
experiences gained in the early astrophysical use of the data.

\section{Conclusions}
\label{sec:conclusions}

The inclusion of positional information from the Hipparcos and \mbox{Tycho-2} catalogues
in the early Gaia data processing has allowed us to derive positions, parallaxes, 
and proper motions for about 2~million sources from the first 14 months of observations
obtained in the operational phase of Gaia. This primary data set contains 
mainly stars brighter than $V\simeq 11.5$. In a secondary data set, using the attitude and geometric 
calibration of Gaia's instrument obtained in the primary solution, approximate positions have 
been derived for an additional 1141~million sources down to the faint limit of Gaia
($G\simeq 20.7$). 

All positions are given in the ICRS and refer to the 
epoch J2015.0. For the primary data set, the overall alignment of the positions with the 
extragalactic radio frame (ICRF2) is expected to be accurate to about 0.1~mas in each
axis at the reference epoch. 
The proper motion system is expected to be non-rotating with respect to the ICRF2
to within 0.03~mas~yr$^{-1}$. The positional reference frame of Gaia DR1 coincides with the 
Hipparcos reference frame at epoch J1991.25, but the Hipparcos frame is rotating with 
respect to the Gaia DR1 frame by about 0.24~mas~yr$^{-1}$ (and hence with respect to
ICRF by a similar amount). The median uncertainty of individual proper motions is
0.07~mas~yr$^{-1}$ for the Hipparcos stars and 1.4~mas~yr$^{-1}$
for non-Hipparcos \mbox{Tycho-2} stars. The derived proper motions represent the mean
motions of the stars between the two epochs J1991.25 and J2015.0, rather than their 
instantaneous proper motions at J2015.0.

The trigonometric parallaxes derived for the primary data set have a median standard
uncertainty of about 0.32~mas. This refers to the random errors. 
Systematic errors, depending mainly on position and colour, could exist at a typical level 
of $\pm 0.3$~mas. This includes a possible global offset of the parallax zero point 
by $\pm 0.1$~mas, and the regional (spatially correlated) and colour-dependent 
systematics of $\pm 0.2$~mas revealed by the special validation solutions described 
in Appendix E. These systematics cannot be much reduced by 
averaging over a number of stars in a small area, such as in a stellar cluster.

The many solutions and validation experiments leading up to the Gaia DR1 data sets
have vastly expanded our understanding of Gaia's astrometric behaviour and 
boosted our confidence that Gaia will in the end provide results of extraordinary
quality. Meanwhile, users of Gaia DR1 data should be extremely aware of the preliminary
nature of the current results, and of the various deficiencies discussed in this paper, 
as well as the potential existence of other yet undetected issues.


\begin{acknowledgements}

This work has made use of data from the ESA space mission Gaia, 
processed by the Gaia Data Processing and Analysis Consortium
(DPAC).
Funding for the DPAC has been provided by national institutions, in
particular the institutions participating in the Gaia Multilateral Agreement.
The Gaia mission website is \url{http://www.cosmos.esa.int/gaia}.

The authors are members of the Gaia DPAC.  
This work has been supported by: 

MINECO (Spanish Ministry of Economy) - FEDER through grant ESP2013-48318-C2-1-R
and ESP2014-55996-C2-1-R and MDM-2014-0369 of ICCUB (Unidad de Excelencia
'Mar\'ia de Maeztu'); 

the Netherlands Research School for Astronomy (NOVA); 

the German Aerospace Agency DLR under grants 50QG0501, 50QG1401 50QG0601, 50QG0901 and 50QG1402; 

the European Space Agency in the framework of the Gaia project; 

the Agenzia Spaziale Italiana (ASI) through grants  ASI  I/037/08/0, ASI I/058/10/0, ASI 2014-025-R.0, and ASI 2014-025-R.1.2015 and the Istituto Nazionale di AstroFisica (INAF);

the Swedish National Space Board; 

the United Kingdom Space Agency;

the Centre National d'Etudes Spatiales (CNES);

Funda\c{c}\~ao para a Ci\^{e}ncia e a Tecnologia through the contract Ci\^{e}ncia2007 and project grant PTDC/CTE-SPA/118692/2010.

We thank the Centre for Information Services and High Performance Computing (ZIH) at TU Dresden for generous allocations of computer time.

This research has profited significantly from the services of the Centre de Donn\'ees Astronomiques Strasbourg, CDS, (SIMBAD/VizieR/Aladin).

Our work was eased considerably by the use of the astronomy-oriented data handling and visualisation software TOPCAT
\citepads{2005ASPC..347...29T}.
We gratefully acknowledge its author, Mark Taylor, for providing support and implementing additional features for our needs.

In addition to the authors of this work there are many other people who have made valuable contributions to Gaia's astrometric reduction. 
Among these, we want to specifically mention Sebastian Els, Michael Perryman, Floor van Leeuwen, and the former members of the Gaia core processing team who have meanwhile moved on to other projects.
We thank the anonymous referee for constructive comments on the original version of the manuscript.
\end{acknowledgements}

\bibliographystyle{aa} 
\bibliography{refs} 

\appendix

\section{Geometric calibration of the Gaia instrument}
\label{app:calibration}

This appendix gives some details on the instrument calibration model
used in the current astrometric solutions and presents selected results 
on some key calibration parameters. It also explains how the BAM data 
were used to correct the observations.

\subsection{Calibration parameters estimated in the astrometric solution}
\label{sec:calibration_parameters}
 
The instrument calibration model is both an extension and simplification of the one described by 
Eqs.~(15)--(18) in Sect.~3.4 of the AGIS paper \citepads{2012A&A...538A..78L}.  
The model consists of a nominal part, a constant 
part, and a time-dependent part. For the AL component it can be written
\begin{equation}\label{eq:cal1}
\eta_{fngw}(\mu,\, t) = \eta^0_{ng}(\mu) + \Delta\eta_{fngw}(\mu) 
+ \Delta\eta_{fn}(\mu,\, t)\, ,
\end{equation}
where $\mu$ is the AC pixel coordinate (running from 13.5 to 1979.5 across the
CCD columns) and $t$ is time; 
$\eta^0_{ng}$ is the nominal geometry depending on the CCD (index $n$) 
and gate ($g$) used; 
$\Delta\eta_{fngw}$ is the constant part depending also on the field index ($f$) 
and window class ($w$, see footnote \ref{footnote1}); and $\Delta\eta_{fn}$ is the 
time-dependent part. The dependence on $\mu$ (within a CCD) and/or $t$ 
(within a calibration interval) is written as a linear combination of shifted 
Legendre polynomials $L^*_l(x)$, orthogonal on $[0,\,1]$ and reaching 
$\pm 1$ at the end points, i.e.\ $L^*_0(x)=1$, $L^*_1(x)=2x-1$, and 
$L^*_2(x)=6x^2-6x+1$. 

In the current AL calibration model, the constant part is decomposed as
\begin{equation}\label{eq:cal2}
\Delta\eta_{fngw}(\mu) =
\sum_{l=0}^2 \Delta\eta^\text{g}_{lfng}L^*_l(\tilde{\mu})
+ \sum_{l=0}^2 \Delta\eta^\text{w}_{lfnw}L^*_l(\tilde{\mu})
+ \sum_{l=0}^1 \Delta\eta^\text{b}_{lfnb}L^*_l(\overline{\mu}_b) \, ,
\end{equation}
where the superscripted constants are the calibration parameters and
$\tilde{\mu}=(\mu-13.5)/1966$ is the normalised AC pixel coordinate.
The dependence on CCD gate (superscript ``g") is different in the preceding 
and following field of view, caused by
the slightly different effective focal lengths; hence $\Delta\eta^\text{g}$ 
must depend on the field index $f$. 
The effect of the window class (``w") could also depend on $f$, and
similarly the third term (``b") in Eq.~(\ref{eq:cal2}), which represents the 
intermediate-scale irregularities of the CCD that cannot be modelled by a 
polynomial over the full AC extent of the CCD. In practice the medium-scale irregularities
are largely associated with the discrete stitch blocks resulting from the 
CCD manufacturing process \citep{2016GaiaP}.  
The stitch blocks 
are 250 pixel columns wide, except for the two outermost blocks which are
108 columns wide; the exact block boundaries are therefore 
$\mu=13.5$, $121.5$, $371.5$, $\dots$, $1621.5$, $1871.5$, $1979.5$. 
The intermediate-scale errors are here modelled by a separate linear polynomial 
for each stitch block, depending on the block index
$b=\lfloor (\mu+128.5)/250\rfloor$%
\footnote{$\lfloor x\rfloor$ is the floor function, i.e.\ the largest integer $\le x$.} 
and the normalised intra-block pixel coordinate 
$\overline{\mu}_b=(\mu-\mu_b)/(\mu_{b+1}-\mu_b)$.
Here, $[\mu_b,\,\mu_{b+1}]$ are the block boundaries given above
for $b=0\dots 8$.
Small-scale irregularities, which vary on a scale of one or a few CCD pixel 
columns, are clearly present but not modelled in the current solution.

\begin{figure}[t] 
\resizebox{1.00\hsize}{!}{\includegraphics{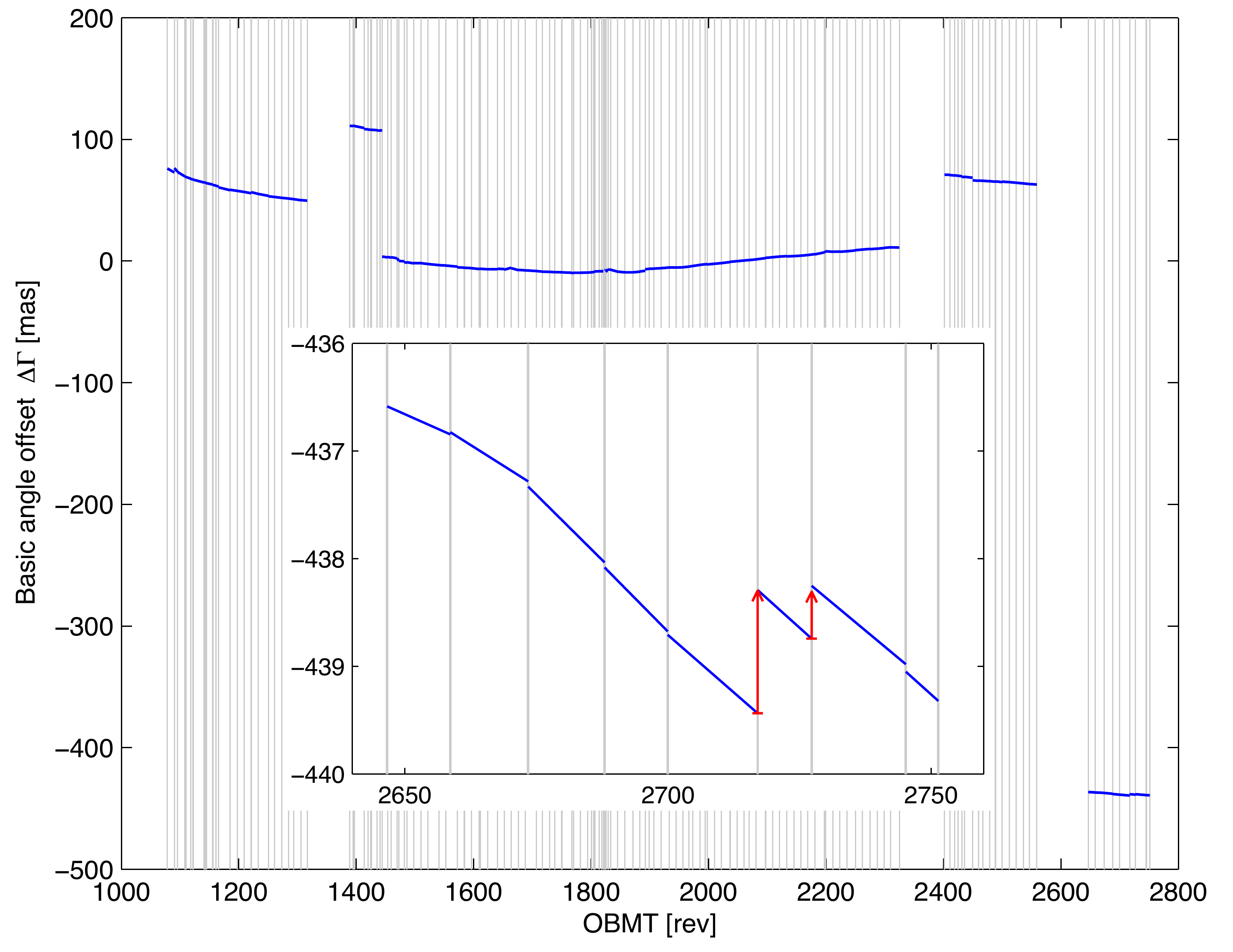}} 
\resizebox{1.00\hsize}{!}{\includegraphics{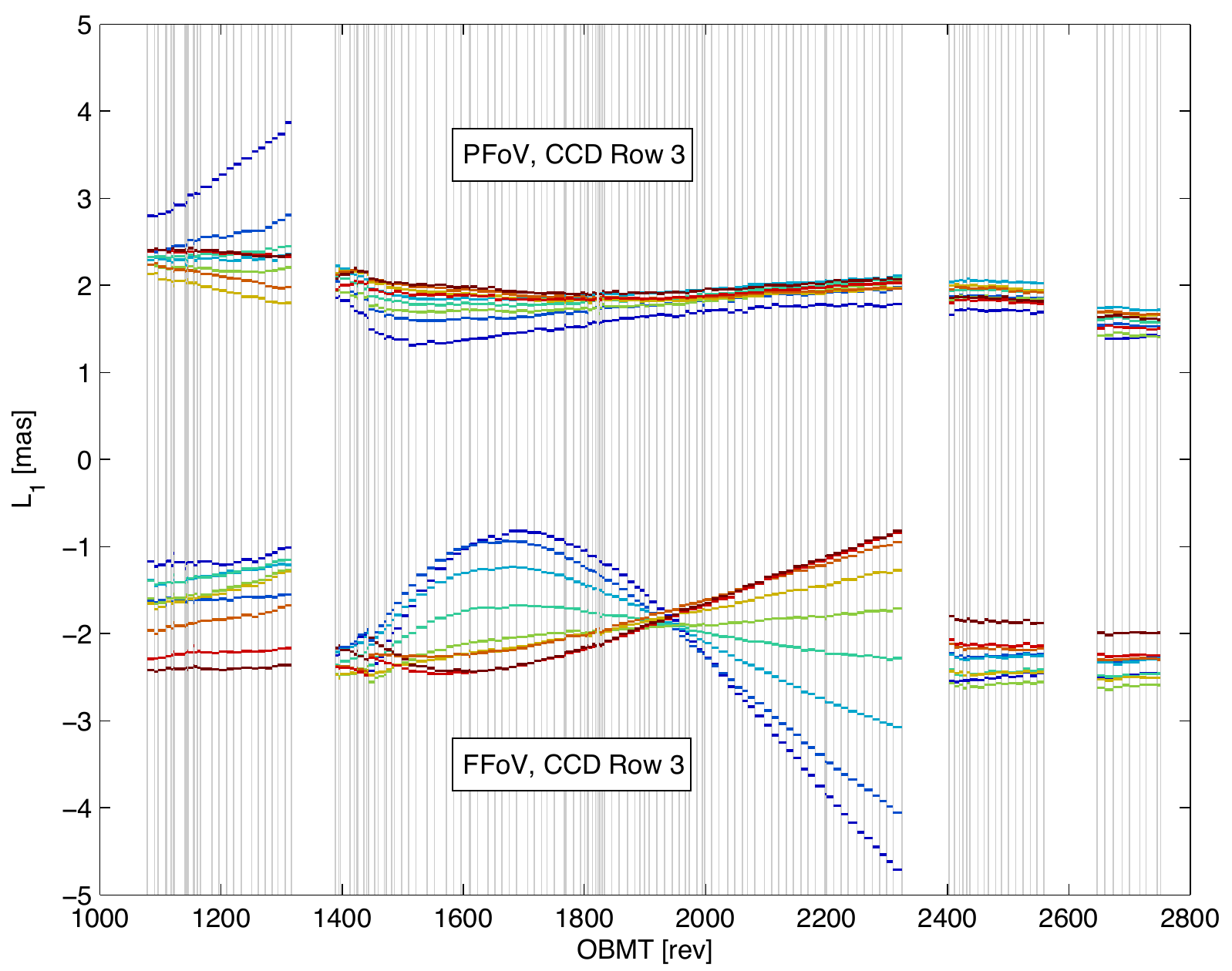}} 
\caption{Evolution of selected calibration parameters estimated in the primary solution.
Time is expressed in revolutions of the onboard mission timeline (OBMT; Sect.~\ref{sec:input}).
Vertical grey lines indicate the breakpoints $t_j$ of the calibration model.
Top: basic-angle offset, Eq.~(\ref{eq:DeltaGamma}), with a zoom to the final $\sim$100 revolutions.
Bottom: parameter $\Delta\eta^{(0)}_{1fnj}$, representing a small rotation of the CCD in
its own plane, for the nine CCDs in row 3. Colours violet to brown are used for AF1 to AF9 (see Fig.~\ref{fig:fpa}), 
respectively.
}
\label{fig:cal1}
\end{figure}

The time-dependent part of the AL calibration needs to take into account 
the joint dependence on $\mu$ and $t$, which quite generally can be expanded 
in terms of the products of one-dimensional basis functions.
With $\tilde{t}_j=(t-t_j)/(t_{j+1}-t_j)$ denoting the normalised time coordinate
in calibration interval $j$, we have 
\begin{equation}\label{eq:cal3}
\Delta\eta_{fn}(\mu,\, t) =
\sum_{l=0}^L \sum_{m=0}^{M_l} 
\Delta\eta^{(m)}_{lfnj}L^*_l(\tilde{\mu})L^*_m(\tilde{t}_j)\, ,
\end{equation}
where $L$ is the maximum degree of the polynomial in $\mu$ and
$M_l$ is the maximum degree of the polynomial in $t$ that is combined 
with a polynomial in $\mu$ of degree $l$. The current model uses $L=2$,
as for the constant part, and $M_0=1$, $M_1=M_2=0$; thus 
Eq.~(\ref{eq:cal3}) simplifies to
\begin{equation}\label{eq:cal4}
\Delta\eta_{fn}(\mu,\, t) =
\sum_{l=0}^2 \Delta\eta^{(0)}_{lfnj}L^*_l(\tilde{\mu})
+ \Delta\eta^{(1)}_{0fnj}L^*_1(\tilde{t}_j)\, .
\end{equation}
In analogy with Eq.~(19) in the AGIS paper, the basic-angle offset can be
computed from the calibration parameters as 
\begin{equation}\label{eq:DeltaGamma}
\Delta\Gamma(t) = \frac{1}{62}\sum_{n\in\text{AF}}\sum_f
\left(\Delta\eta^{(0)}_{0fnj}+ \Delta\eta^{(1)}_{0fnj}L^*_1(\tilde{t}_j)\right)f \, ,
\end{equation}
where $f=\pm 1$ for the preceding and following field of view, respectively. In the present model
this function is piecewise linear as illustrated in the top panel of Fig.~\ref{fig:cal1}. 

\begin{table}[t]
\caption{Number of parameters of different kinds in the geometric calibration model
used for Gaia DR1.\label{tab:cal}}
\small
\begin{tabular}{lrrrrrrrr}
\hline\hline
\noalign{\smallskip}
\multicolumn{1}{l}{Kind of} & \multicolumn{7}{c}{Multiplicity} & \multicolumn{1}{c}{Total}\\
\multicolumn{1}{l}{parameter} & $l$ & $f$ & \multicolumn{1}{c}{$n$} & $g$ & $b$ & 
\multicolumn{1}{c}{$w$} & \multicolumn{1}{c}{$j$/$k$} & \multicolumn{1}{c}{number} \\
\noalign{\smallskip}
\hline
\noalign{\smallskip}
$\Delta\eta^{(0)}_{lfnj}$ & 3 & 2 & 62 & 1 & 1 & 1 & 141 & 52\,452\\[3pt]
$\Delta\eta^{(1)}_{lfnj}$ & 1 & 2 & 62 & 1 & 1 & 1 & 141 & 17\,484\\[3pt]
$\Delta\eta^\text{g}_{lfng}$ & 3 & 2 & 62 & 9 & 1 & 1 & 1 & 3\,348\\[3pt]
$\Delta\eta^\text{b}_{lfnb}$ & 2 & 2 & 62 & 1 & 9 & 1 & 1 & 2\,232\\[3pt]
$\Delta\eta^\text{w}_{lnw}$ & 3 &  2 & 62 & 1 & 1 & 3 & 1 & 1\,116\\[3pt]
\noalign{\smallskip}
\hline
\noalign{\smallskip}
$\Delta\zeta^{\,(0)}_{lfnk}$ & 3 & 2 & 62 & 1 & 1 & 1 & 87 & 32\,364\\[3pt]
$\Delta\zeta^{\,(1)}_{lfnk}$ & 1 & 2 & 62 & 1 & 1 & 1 & 87 & 10\,788\\[3pt]
$\Delta\zeta^\text{\,b}_{lfng}$ & 3 & 2 & 62 & 9 & 1 & 1 & 1 & 3\,348\\
\noalign{\smallskip}
\hline
\end{tabular}
\tablefoot{The columns headed Multiplicity give the number of distinct
values for each dependency: polynomial degree ($l$), field index ($f$),
CCD index ($n$), gate ($g$), stitch block ($b$), window class ($w$), and
time interval ($j$ or $k$). The last column 
is the product of multiplicities, equal to the number of calibration parameters
of the kind.}
\end{table}

For the AC calibration we have in analogy with Eq.~(\ref{eq:cal1})
\begin{equation}\label{eq:cal5}
\zeta_{fng}(\mu,\, t) = \zeta^{\,0}_{fn}(\mu) + \Delta\zeta_{fng}(\mu) 
+ \Delta\zeta_{fn}(\mu,\, t) \, .
\end{equation}
The AC model has fewer breakpoints for the time dependence, no dependence
on window class, and no intermediate or small-scale irregularities. Thus,
\begin{align}\label{eq:cal6}
\Delta\zeta_{fng}(\mu) &= \sum_{l=0}^2 \Delta\zeta^\text{\,b}_{lfng}L^*_l(\tilde{\mu})\\[3pt]
\Delta\zeta_{fn}(\mu,\, t) &=
\sum_{l=0}^2 \Delta\zeta^{\,(0)}_{lfnk}L^*_l(\tilde{\mu})
+ \Delta\zeta^{\,(1)}_{0fnk}L^*_1(\tilde{t}_k)\, ,
\end{align}
where $\tilde{t}_k=(t-t_k)/(t_{k+1}-t_k)$ are normalised time coordinates
relative to the breakpoints $t_k$ for the AC calibration time intervals.

The calibration model does not include colour- or magnitude-dependent terms, 
although such dependencies can be expected from chromaticity and non-linear
charge transfer inefficiency in the CCDs. Chromatic effects are indeed apparent
in the residuals, and will have an effect on the astrometric results as discussed 
in Appendix~\ref{sec:valColour}.

The model as described applies to the 62 CCDs in the AF; 
for the SM the nominal calibration $\eta^0_{ng}(\mu)$,
$\eta^0_{ng}(\mu)$ is not updated in the current solution as the SM observations
are not used for the astrometric solution in this release.

Table~\ref{tab:cal} summarises the number of parameters of the different
kinds. The total number of calibration parameters is 76\,632 for the AL model 
and 46\,500 for the AC model. The calibration model as described above is
degenerate because it does not specify a unique division between the different
components. For example, the parameter $\Delta\eta^{(0)}_{0fnj}$, averaged over
all calibration intervals $j$, describes an AL offset of CCD $n$ in field $f$ that 
is independent of $\mu$; but $\Delta\eta^\text{b}_{0fnb}$ could describe exactly the 
same offset by means of a constant value for all stitch blocks $b$. In the solution
a number of constraints are imposed on the calibration parameters, which make
them non-degenerate with each other and with the attitude model. These constraints 
are essentially the same as Eqs.~(16)--(18) in the AGIS paper and not repeated here.

A few examples of calibration results are shown in Fig.~\ref{fig:cal1}. The top panel
shows the long-term evolution of the basic-angle offset $\Delta\Gamma$. Major 
discontinuities between the continuous segments are usually real; two examples
are shown in the inset diagram where the red arrows show the sizes of jumps
determined from BAM data at two of the breakpoints. Refocusing and decontamination
cause much larger jumps. The bottom panel shows the evolution of the coefficient 
of $L^*_1(\tilde{t}_j)$ in Eq.~(\ref{eq:cal4}) for selected CCDs in both fields of view. 
This parameter represents a small, apparent rotation of the CCD in its own plane, 
caused mainly by the optical distortion. Between refocusing and decontamination
events, this parameter varies smoothly over time and according to position (CCD)
in the field, but very differently in the two fields of view. Plots such as these,
showing a generally smooth development of calibration parameters from one discrete 
time interval to the next, suggest
that the adopted geometric calibration model is physically sound and adequate at
a precision level better than 0.1~mas.

\subsection{Calibration parameters derived from BAM data}
\label{sec:calibration_bam}
 
\begin{figure}[t] 
\resizebox{0.89\hsize}{!}{\includegraphics{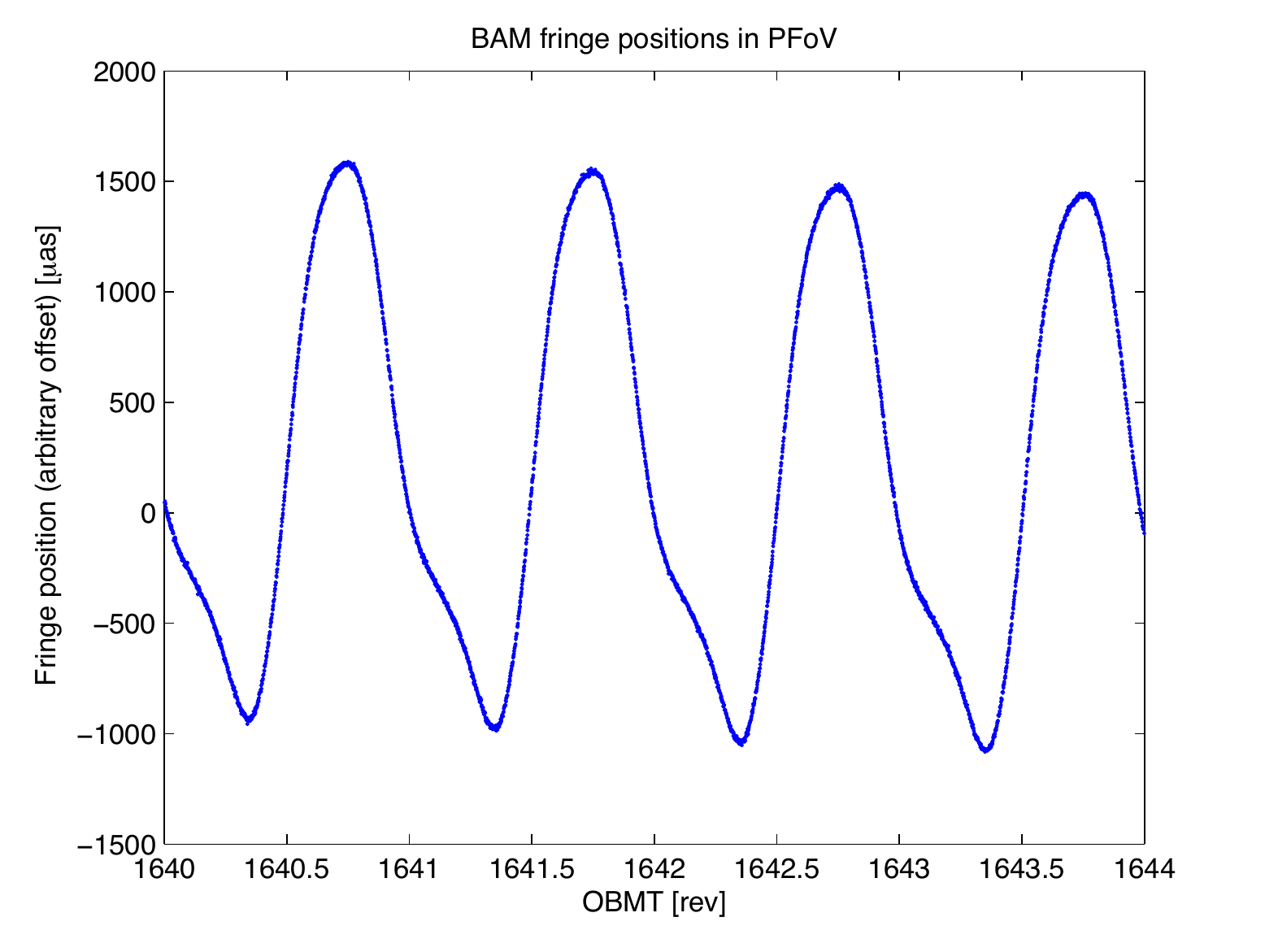}} 
\resizebox{0.89\hsize}{!}{\hspace{4mm}\includegraphics{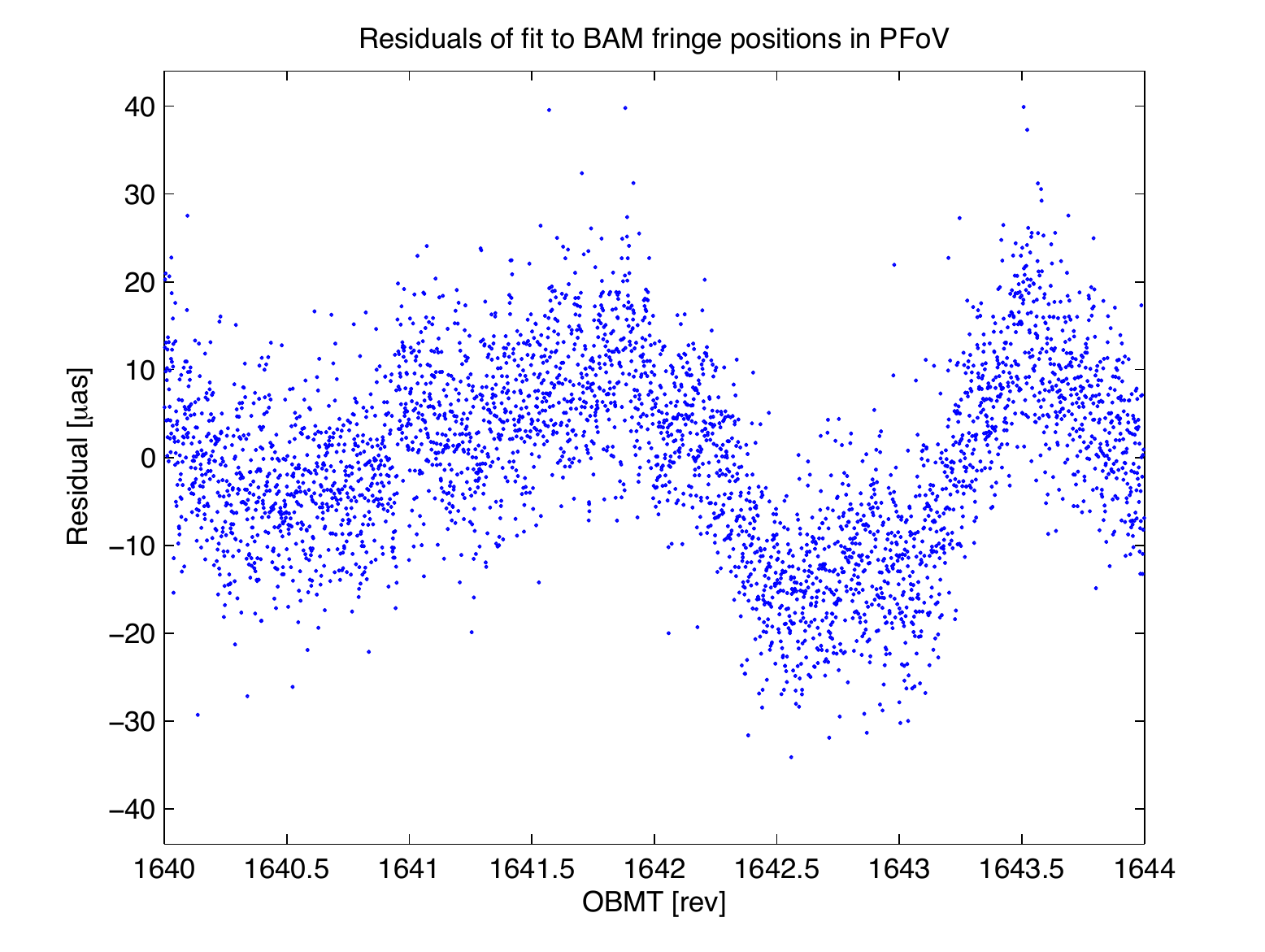}} 
\caption{Example of the BAM signal for the preceding field of view. 
Time is expressed in revolutions of the onboard mission timeline (OBMT; Sect.~\ref{sec:input}).
Top: individual fringe position measurements $\xi^\text{P}$ after removal of outliers. 
Bottom: residuals after fitting the model in Eq.~(\ref{eq:bamLOS}). 
}
\label{fig:bamRaw}
\end{figure}

The periodic variations seen in the BAM signal are strongly 
coupled to the spin phase of the satellite with respect to the Sun. The heliotropic spin 
phase $\Omega$ increases by $360^\circ$ for each $\sim$6~hr spin period and is zero 
when the direction to the apparent Sun is symmetrically located between the two fields 
of view (see Fig.~1 in \citeads{2016A&A...586A..26M}). 
Figure~\ref{fig:bamRaw} shows an example of the line-of-sight variations in the 
preceding field of view during a one-day interval (four successive spin periods).
In such a time interval, and for a given field of view (P or F), the following model was 
usually found to provide a reasonable fit to the line-of-sight variations, as represented 
by the location $\xi$ (expressed as an angle) of the central fringe on the BAM CCD:
\begin{equation}\label{eq:bamLOS}
\xi^\text{P}(t) = C^\text{P}_{0} + (t-t_0){C'_0}^\text{P}
+ \sum_{k=1}^8\Bigl[ C_k^\text{P}\cos k\Omega(t) + S_k^\text{P}\sin k\Omega(t) \Bigr]\, , 
\end{equation}
with a similar expression for $\xi^\text{F}(t)$ in the other field of view. 
Here $t_0$ is the mid-time of the 
interval and $C^\text{P}_0$, ${C'}^\text{P}_0$,  $C^\text{P}_k$, and $S^\text{P}_k$
($k=1,\,\dots,\,8$) are constants in the interval. Detected discontinuities were 
subtracted before fitting this model. Residuals of the fit are typically on
the level of a few tens of $\mu$as and contain systematic patterns (e.g.\ as seen 
in the lower panel of Fig.~\ref{fig:bamRaw}) that correlate with spacecraft activities 
such as changes in the telemetry rates. The constant and linear coefficients 
$C^\text{P}_0$ and ${C'_0}^\text{P}$ are not further used in the analysis. 

Fits using Eq.~(\ref{eq:bamLOS}) were made independently for the preceding (P) and 
following (F) fields of view, resulting in two sets of harmonic coefficients 
for each fitted time interval. The differences between these, in the sense F minus P,
provide a corresponding harmonic representation of the basic-angle variations:
\begin{equation}\label{eq:bam}
\Delta\Gamma(t) = \sum_{k=1}^8\Bigl[ C_k\cos k\Omega(t) + S_k\sin k\Omega(t) \Bigr]\, , 
\end{equation}
with a separate estimate of $C_k=C^\text{F}_k-C^\text{P}_k$ and 
$S_k=S^\text{F}_k-S^\text{P}_k$ ($k=1\dots 8$) obtained for each one-day interval.

The sizes of the harmonic coefficients $C_k$, $S_k$ decrease rapidly with increasing 
order $k$ (Table~\ref{tab:valVbacF}). The harmonic coefficients are only 
approximately constant over the investigated 14 months of BAM data. At least 
three different kinds of variations can be 
distinguished: (i) an annual periodic variation; (ii) a secular trend; and (iii) seemingly
more irregular, rapid variations on timescales of weeks to months.

The annual and secular variations are well fitted by the following analytical model,
in which each coefficient is approximated as a linear function modulated by the 
expected inverse-square dependence on solar distance:
\begin{equation}\label{eq:bam1}
C_k(t) = \Bigl[ C_{k,0} + (t-t_\text{ref})C_{k,1}\Bigr] \times d(t)^{-2}\, , \quad k=1\dots 8\, 
\end{equation}
(and similarly for $S_k$). Here $t_\text{ref}=\text{J2015.0}$ and $d(t)$ is Gaia's 
heliocentric distance in au. This analytical fit was used to correct the observations for the 
basic angle variations in the astrometric solutions for Gaia DR1.

\begin{figure}[t] 
\resizebox{0.94\hsize}{!}{\includegraphics{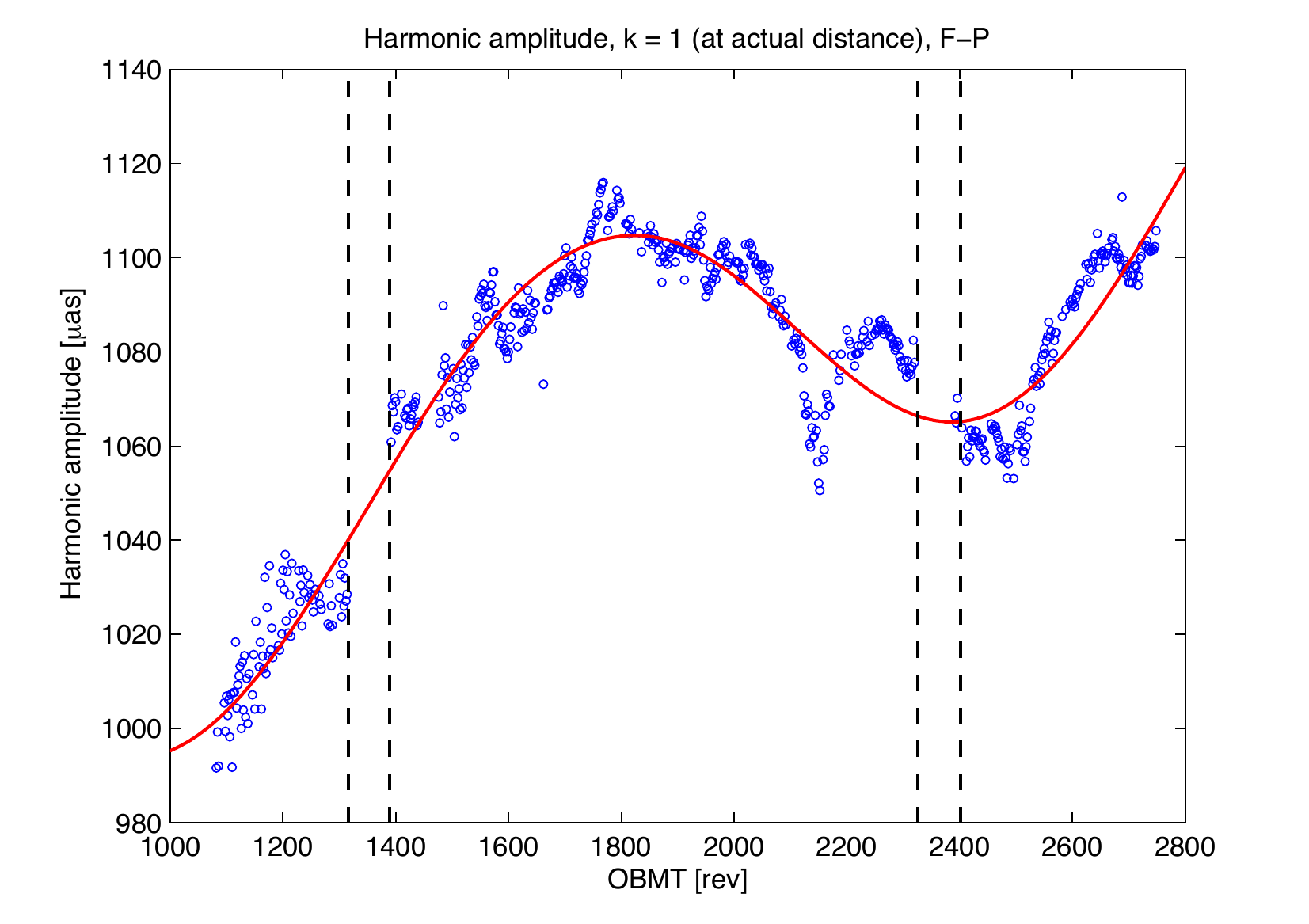}} 
\resizebox{0.94\hsize}{!}{\hspace{4mm}\includegraphics{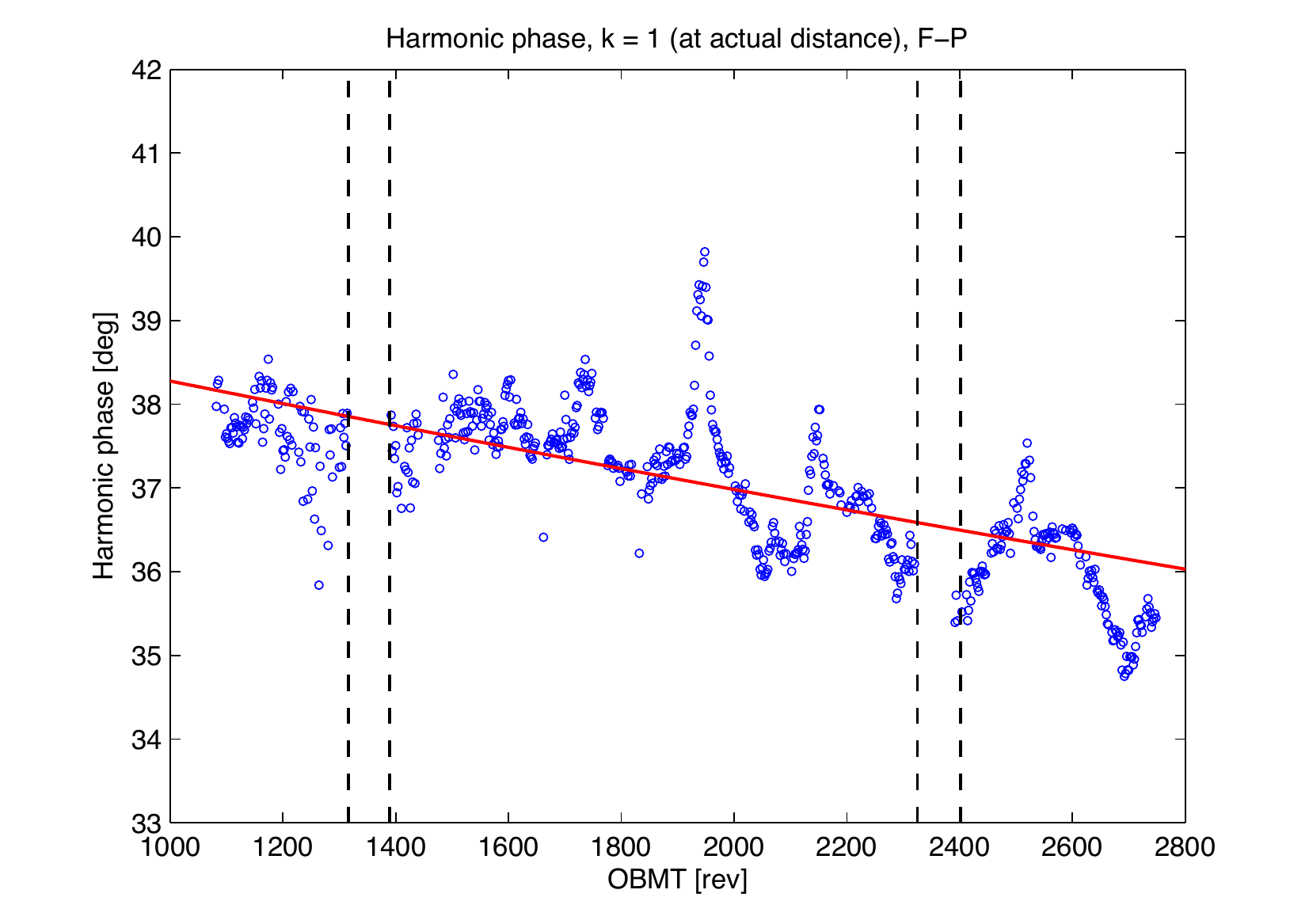}} 
\caption{Amplitude ($A_1$) and phase ($\phi_1$) of the first harmonic in Eq.~(\ref{eq:bam}) 
fitted to the BAM signal. 
Time is expressed in revolutions of the onboard mission timeline (OBMT; Sect.~\ref{sec:input}).
Circles are the values for individual one-day intervals;
the solid curve is the global model used to correct the observations in Gaia DR1. 
The vertical dashed lines mark the two major data gaps caused by decontamination
procedures.}
\label{fig:bamCoeff}
\end{figure}

The temporal evolution of the dominant first order
($k=1$) is shown in Fig.~\ref{fig:bamCoeff}, where the Fourier coefficients 
have been transformed to amplitude $A_1$ and phase $\phi_1$ such that
$C_1=A_1\cos\phi_1$ and $S_1=A_1\sin\phi_1$.
The fitted Eq.~(\ref{eq:bam1}) is shown by the red solid curves. In addition to the
annual variation of $\pm 3.3$\% in amplitude, caused by the eccentricity of Gaia's 
heliocentric orbit, the plots show secular trends in both amplitude and phase 
at the level of several percent, as well as systematic deviations from the model 
in Eq.~(\ref{eq:bam1}).
At least some of these deviations are related to the mean rate of observations.
At the time of writing it is not clear if they represent actual changes in the basic-angle
variations, or if they are merely an artefact of the BAM. Until this has been established, 
the smoothed model in Eq.~(\ref{eq:bam1}) is used to correct the observations.

\section{Estimating the precision of parallaxes from a comparison with Hipparcos data}
\label{app:precision}

In this appendix we describe how the external uncertainties of the TGAS parallaxes were
estimated based on a comparison with Hipparcos data. These estimates were used to
calculate the inflation factor in Eq.~(\ref{eq:infl}) applied to all formal uncertainties in
the primary data set of Gaia DR1. 

In AGIS the least-squares estimates of the astrometric parameters are 
rigorously computed in the iterative solution, but the associated uncertainties 
are only approximately estimated, using a number of simplifications. For a given source,
the formal standard errors (uncertainties) of the five astrometric parameters are computed 
as described in Sect.~6.3 of the AGIS paper, i.e.\ from the diagonal elements of the 
inverse of the corresponding $5\times 5$ part of the normal matrix. As discussed by
\citetads{2012A&A...543A..14H} this neglects the statistical correlations 
introduced by the attitude and calibration models, which couple the observation equations 
of different sources to each other. This will cause the actual uncertainties to be  
underestimated. In Gaia DR1 the underestimation may be particularly severe because of the large 
modelling errors and relatively low redundancy of observations. It is therefore important 
to investigate the relation between the formal standard uncertainties computed from the 
least-squares solution, here denoted by $\varsigma$, and the actual standard uncertainties, 
denoted by $\sigma$.
(The word standard here signifies that the quantities represent standard deviations.
It does not imply that the errors follow, or even are assumed to follow, the normal distribution.
For the subsequent derivation it is sufficient to assume that the errors have finite variance.) 

A comparison of the Hipparcos parallaxes with the corresponding values from the
current primary (TGAS) solution 
offers an interesting possibility to investigate this relation, thanks to the following 
circumstances: (i) the parallax errors in the two data sets are uncorrelated, since the 
Hipparcos parallaxes were not used in the solution \citepads{2015A&A...574A.115M}; 
(ii) the standard uncertainties do not differ too much between the two data sets; and 
(iii) the number of common stars is large enough for accurate statistics. 

For the comparison we use the parallaxes (and their uncertainties) from the new 
reduction of the Hipparcos data \citepads{2007ASSL..350.....V}. The primary solution,
after application of the filter in Eq.~(\ref{eq:formal1}), contains data for 101\,106 
Hipparcos stars that were used for the present study, although not all of them are 
retained in Gaia DR1. The two sets of parallax values are
here distinguished by subscript H (for Hipparcos) and T (for TGAS). 
For the stars in common the median formal uncertainty is 
$\varsigma_\text{H}\simeq 0.9$~mas for the Hipparcos parallaxes and 
$\varsigma_\text{T}\simeq 0.15$~mas for the TGAS parallaxes.   

The non-correlation between the two sets of parallaxes implies that the variance 
of $\Delta\varpi=\varpi_\text{T}-\varpi_\text{H}$ equals the sum of the actual 
mean variances,
\begin{equation}\label{eq:app1}
\text{Var}(\Delta\varpi)=\langle\sigma_\text{T}^2\rangle + \langle\sigma_\text{H}^2\rangle \, .
\end{equation}
The angular brackets denote averages over the stars, which is necessary in order to 
take into account the non-uniformity (heteroscedasticity) of the data sets. $\text{Var}(\Delta\varpi)$ is 
readily estimated, e.g.\ as the sample variance of the parallax differences, and thus
provides a firm estimate of the combined mean variances of the data sets. This
should be compared with the combined formal variances,
\begin{equation}\label{eq:app2}
\langle\varsigma_{\Delta\varpi}^2\rangle= \langle\varsigma_\text{T}^2\rangle + 
\langle\varsigma_\text{H}^2\rangle \, . 
\end{equation}

Consider for example the $\simeq\,$86\,000 stars with formal parallax standard 
uncertainties $\varsigma_\text{T}\le 0.7$~mas and $\varsigma_\text{H}\le 1.5$~mas. 
The rms formal standard uncertainties are 
$\langle\varsigma_\text{T}^2\rangle^{1/2}=0.226$~mas and   
$\langle\varsigma_\text{H}^2\rangle^{1/2}=0.915$~mas, giving a combined  
standard deviation $\langle\varsigma_{\Delta\varpi}^2\rangle^{1/2}=0.942$~mas. 
However, the sample standard deviation of $\Delta\varpi$ is 1.218~mas
(excluding nine stars for which $|\Delta\varpi|>10$~mas).
From this we conclude that $\varsigma_\text{T}$ and/or $\varsigma_\text{H}$ 
significantly underestimate the true errors. This analysis can be repeated 
for various selections of formal uncertainties, providing in each case an estimate 
of the combined uncertainties.  

However, as shown below, it is also possible to estimate the relative contributions 
of the data set to the combined variance, and hence the variance of each data set
separately. The 
method depends on the practical circumstance that the probability density function 
of the true parallaxes has a steep edge towards small values. 

Let $\varpi\ge 0$ denote the true parallax of a star and 
$e_\text{H}=\varpi_\text{H}-\varpi$,
$e_\text{T}=\varpi_\text{T}-\varpi$ the measurement errors in the two data sets. 
Let us first assume that the measurements are unbiased, 
$\text{E}(e_\text{T})=\text{E}(e_\text{H})=0$, where $\text{E}$ is the expectation 
or mean value. The non-correlation assumption is 
\begin{equation}\label{eq:app3}
\text{E}(e_\text{T}e_\text{H}) = 0 \, ,
\end{equation}
which results in 
\begin{equation}\label{eq:app4}
\sigma_{\Delta\varpi}^2 = 
\text{E}[(e_\text{T}-e_\text{H})^2] = 
\text{E}(e_\text{T}^2) + \text{E}(e_\text{H}^2) = 
\langle\sigma_\text{T}^2\rangle + \langle\sigma_\text{H}^2\rangle \, ,
\end{equation}
which is Eq.~(\ref{eq:app1}). 
Consider now the weighted mean parallax,
\begin{equation}\label{eq:app5}
\varpi_x = (1-x)\varpi_\text{T} + x\varpi_\text{H} \, 
\end{equation}
for $0\le x\le 1$. The error of $\varpi_x$ is $e_x = (1-x)e_\text{T} + xe_\text{H}$,
and its covariance with $\Delta\varpi$ is
\begin{equation}\label{eq:app6}
\text{Cov}(e_x,\,\Delta\varpi)=
\text{E}[e_x(e_\text{T}-e_\text{H})]
= (1-x)\langle\sigma_\text{T}^2\rangle - x\langle\sigma_\text{H}^2\rangle \, .
\end{equation}
This is clearly zero if\,%
\footnote{It is worth noting that this $x$ also minimises the 
variance of $e_x$ and equals the weight ratio, 
$x=\langle\sigma_\text{H}^2\rangle^{-1}/\left(
\langle\sigma_\text{T}^2\rangle^{-1} + \langle\sigma_\text{H}^2\rangle^{-1}\right)$.}
\begin{equation}\label{eq:app7}
x = \frac{\langle\sigma_\text{T}^2\rangle}{\langle\sigma_\text{T}^2\rangle 
+ \langle\sigma_\text{H}^2\rangle} \, .
\end{equation}
Since this holds for any value of the true parallax $\varpi$, it follows that
also the covariance between $\varpi_x$ and $\Delta\varpi$ is zero for this 
value of $x$, provided that the errors are not correlated with $\varpi$,
which is a reasonable assumption based on how parallaxes are computed.

If therefore $\Delta\varpi=\varpi_\text{T}-\varpi_\text{H}$ is plotted against 
$\varpi_x$, and $x$ is adjusted for zero correlation between the plotted 
quantities, the mean variances of the data sets can be calculated as
\begin{equation}\label{eq:app8}
\langle\sigma_\text{T}^2\rangle = x_0\sigma^2_{\Delta\varpi}\, , \quad 
\langle\sigma_\text{H}^2\rangle = (1-x_0)\sigma^2_{\Delta\varpi}\, , 
\end{equation}
where $x_0$ is the value of $x$ for which the correlation is zero.
In practice this procedure only works for small enough parallaxes because the
correlation is only apparent when the errors cause the measured parallaxes 
to be scattered into negative values.

Equations~(\ref{eq:app1})--(\ref{eq:app8}) were derived under the assumption 
that $\varpi_\text{T}$ and $\varpi_\text{H}$ are unbiased. However, it is easily
verified that the same relations hold when they are biased, provided
that the bias is not a function of $\varpi$. While the difference in bias can
be estimated as the mean value of $\Delta\varpi$, it is not possible to
separate out the bias of each data set with this method.

\begin{figure} 
\resizebox{\hsize}{!}{\includegraphics{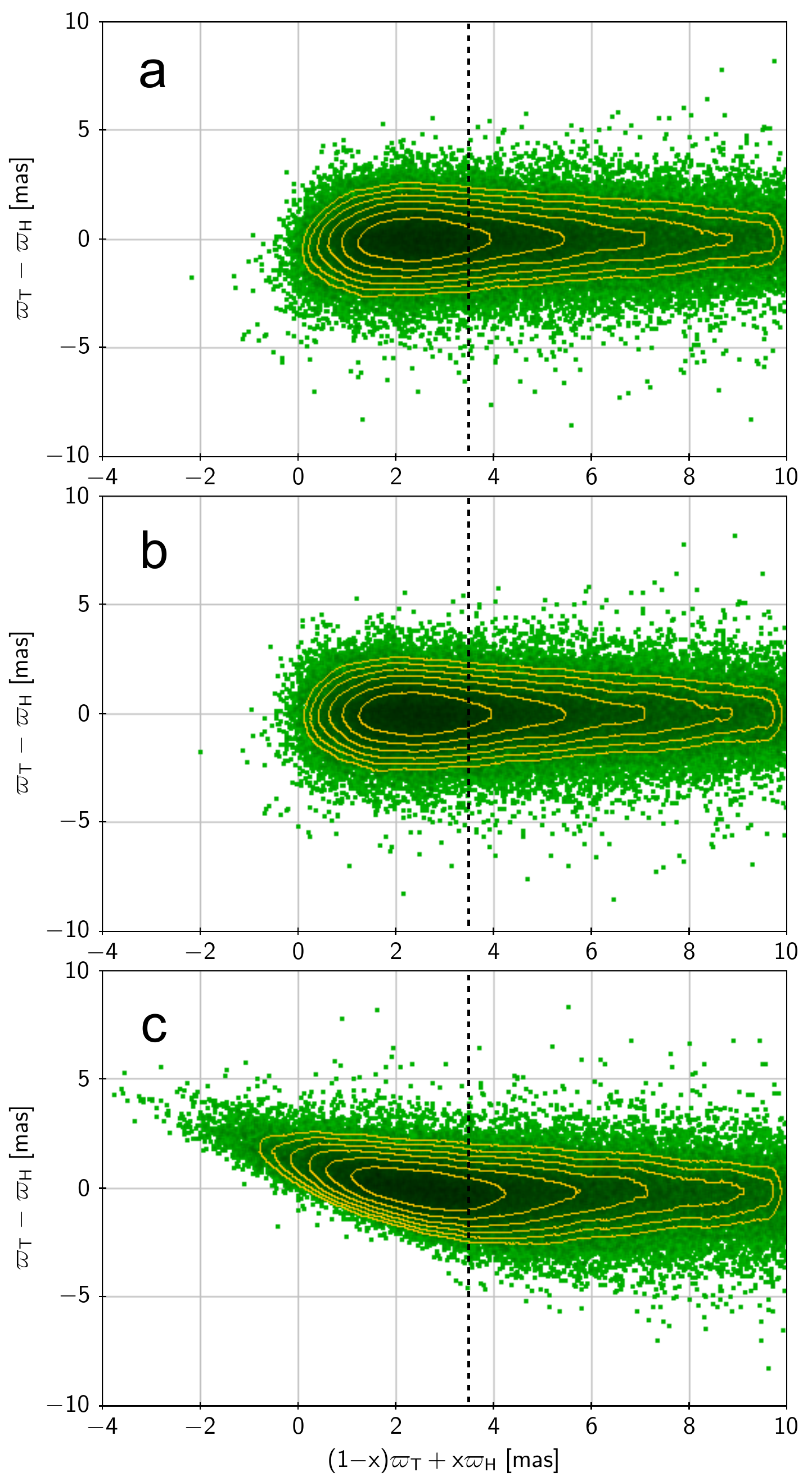}}  
\caption{Parallax difference between TGAS and Hipparcos plotted against the
weighted mean parallax (Eq.~\ref{eq:app5}) for three different weight factors $x$: 
({\bf a}) $x=0$, i.e.\ the abscissa is the TGAS parallax; 
({\bf b}) $x=0.1$; and
({\bf c}) $x=1$, i.e.\ the abscissa is the Hipparcos parallax. 
See text for further explanation.}
\label{fig:app1}
\end{figure}

Figure~\ref{fig:app1} illustrates the application of the method to the previously
mentioned selection, $\varsigma_\text{T}\le 0.7$~mas and 
$\varsigma_\text{H}\le 1.5$~mas. $\Delta\varpi$ is here plotted versus 
$\varpi_x$ for $x=0.0$, 0.1, and 1.0. (Only the $\simeq\,$73\,000 points with   
$\varpi_x<10$~mas are shown.) In the top panel ({\bf a}), the case $x=0$ exhibits 
a weak positive correlation most clearly seen from the slightly asymmetric 
distribution of $\Delta\varpi$ for the smallest parallaxes.
In the bottom panel ({\bf c}), the case $x=1$ shows a very strong negative correlation.
For $x=0.1$, shown in the middle panel ({\bf b}), the correlation virtually disappears.
Thus we conclude that $x_0\simeq 0.1$.
With $\sigma_{\Delta\varpi}=1.218$~mas from the sample standard deviation, 
Eq.~(\ref{eq:app8}) gives $\langle\sigma_\text{T}^2\rangle^{1/2}=0.385$~mas. 
Comparing with the rms formal uncertainty,  
$\langle\varsigma_\text{T}^2\rangle^{1/2}=0.226$~mas, 
we conclude that the formal parallax uncertainties for this particular sample 
on the average need to be increased roughly by the inflation factor $F\simeq 1.7$.

In this example $x_0$ was estimated by visual inspection of a sequence of 
$(\varpi_x,\,\Delta\varpi)$-plots for different values of $x$. It is not difficult 
to devise an objective and more precise criterion to estimate $x_0$ and hence $F$. 
Let $\rho(\varpi_x,\Delta\varpi\,|\,x,c)$ denote the sample correlation coefficient 
between $\varpi_x$ and $\Delta\varpi$ calculated for a given value of $x$,
using only points with $\varpi_x \le c$, where $c$ is some positive constant. 
While this sample correlation coefficient in general depends on $c$, we clearly expect 
$\rho(\varpi_x,\Delta\varpi\,|\,x_0,c)=0$ to hold for any value of 
$c$. Thus, $x_0$ can in principle be obtained by solving this equation for arbitrary 
$c$. In practice we should choose $c$ to minimise the statistical uncertainty 
of $x_0$. Using bootstrap resampling \citep{efron1994introduction} to estimate the
uncertainty, it appears that $c=3.5$~mas 
(dashed line in Fig.~\ref{fig:app1})
is close to optimal, and we find for the 
three cases in Fig.~\ref{fig:app1}, respectively, 
$\rho(\varpi_x,\Delta\varpi\,|\,x,c)=+0.077$, $-0.005$, and $-0.522$
(excluding 13 points for which $|\Delta\varpi|>10$~mas).
Estimating $x_0$ by bisection we obtain $x_0=0.095 \pm 0.006$, from which 
$\langle\sigma_\text{T}^2\rangle=0.141\pm 0.009$~mas$^2$ or
$F=1.66\pm 0.05$.

\begin{figure} 
\resizebox{\hsize}{!}{\includegraphics{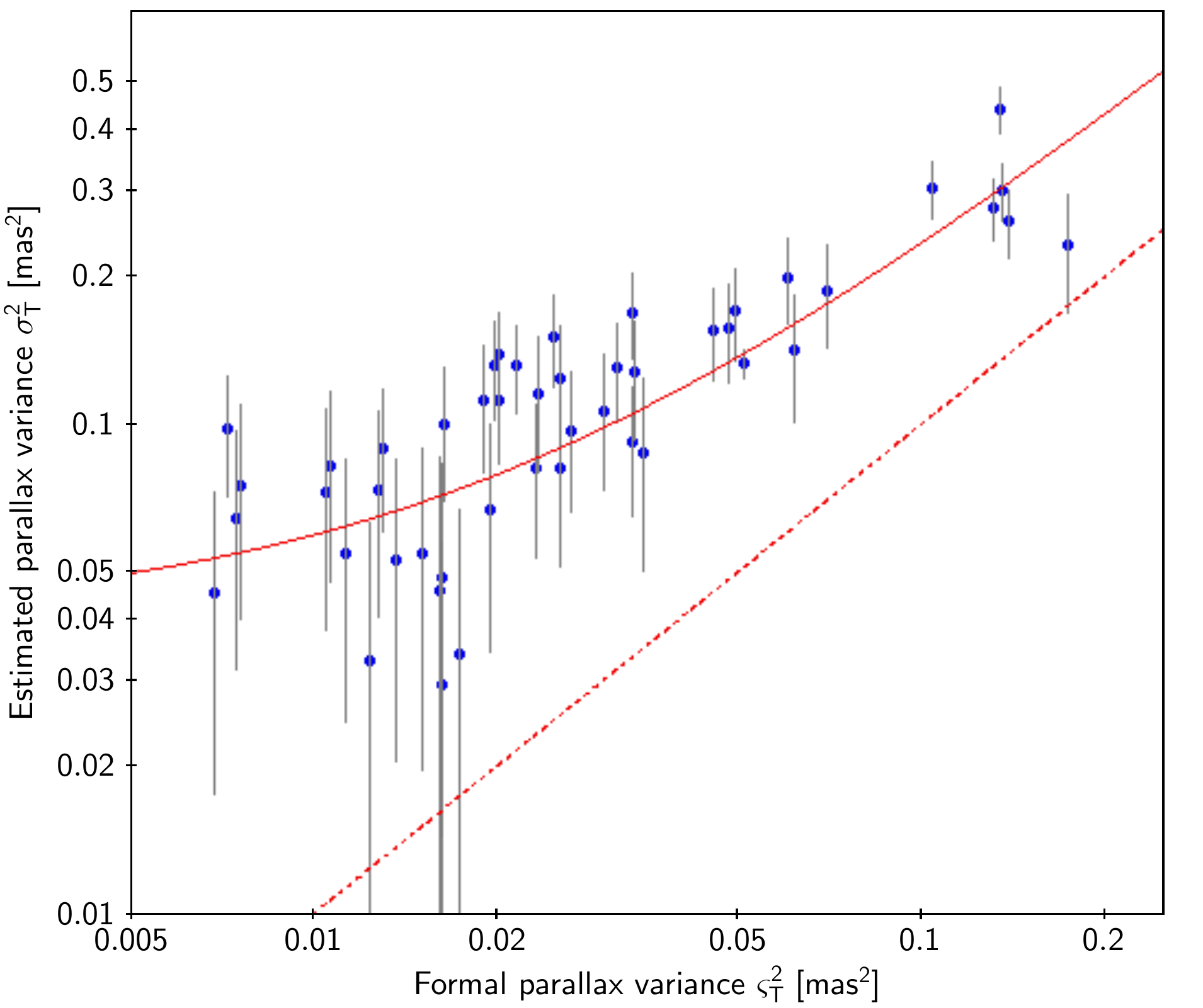}}  
\caption{Statistical relation between the formal parallax variances of Hipparcos
stars in the primary (TGAS) solution, and the actual variances estimated as described
in the text. The solid line is the fitted relation in Eq.~(\ref{eq:app9}); 
the dashed line is the 1:1 relation.}
\label{fig:app2}
\end{figure}

It is not expected that the inflation factor $F$ should be the same for all
sources, independent of $\varsigma_\text{T}$. To investigate this, the method 
described above was applied to different subsamples of the data sets, selected according 
to their formal uncertainties. This makes it possible to trace out the statistical relation 
between the formal and actual uncertainties. Figure~\ref{fig:app2} shows the
result of such an analysis of the parallaxes in the current primary solution.
The estimated mean actual variances $\langle\sigma_\text{T}^2\rangle$, with 
68\% confidence limits obtained by bootstrapping, are plotted against the mean 
formal variances $\langle\varsigma_\text{T}^2\rangle$ for 49 different subsamples
using $c=3.5$~mas and removing points with $|\Delta\varpi|>10$~mas. 
The solid curve is the relation
\begin{equation}\label{eq:app9}
\langle\sigma_\text{T}^2\rangle \simeq 
a^2\langle\varsigma_\text{T}^2\rangle + b^2 \, 
\end{equation}
for $a=1.4$ and $b=0.2$~mas, obtained by a weighted least-squares fitting 
(with some rounding). The adopted inflation factors in Eq.~(\ref{eq:infl}) 
correspond to this curve. The linear form of this relation is mainly empirical, 
but not without theoretical foundation: neglected correlations tend to give a 
multiplicative factor to the variance ($a^2$), while unmodelled uncorrelated errors 
add a constant variance ($b^2$).

\section{Comparison with external data}
\label{app:external}

In this appendix we compare astrometric parameters in the primary (TGAS) solution
with some external data of comparable accuracy. The main purpose 
is to check for possible systematic errors in the primary data set and, if possible,
characterise them in terms of their size and dependence on position, colour, etc.

In order to summarise key properties in a few numbers, we generally use robust statistics
such as the median for the location of a distribution, and the RSE 
(see footnote~\ref{footnote:RSE}) for the scale or dispersion of the distribution. 
For brevity, the median and RSE of quantity $x$ are denoted by $\text{med}(x)$ and 
$\text{RSE}(x)$. Where relevant, the standard uncertainty of the median is estimated 
using bootstrapping \citep{efron1994introduction}.

The results of a dedicated validation procedure applied to the Gaia DR1 data are given
by \citet{2016GaiaA}.

\begin{figure*} 
\resizebox{\hsize}{!}{%
\includegraphics{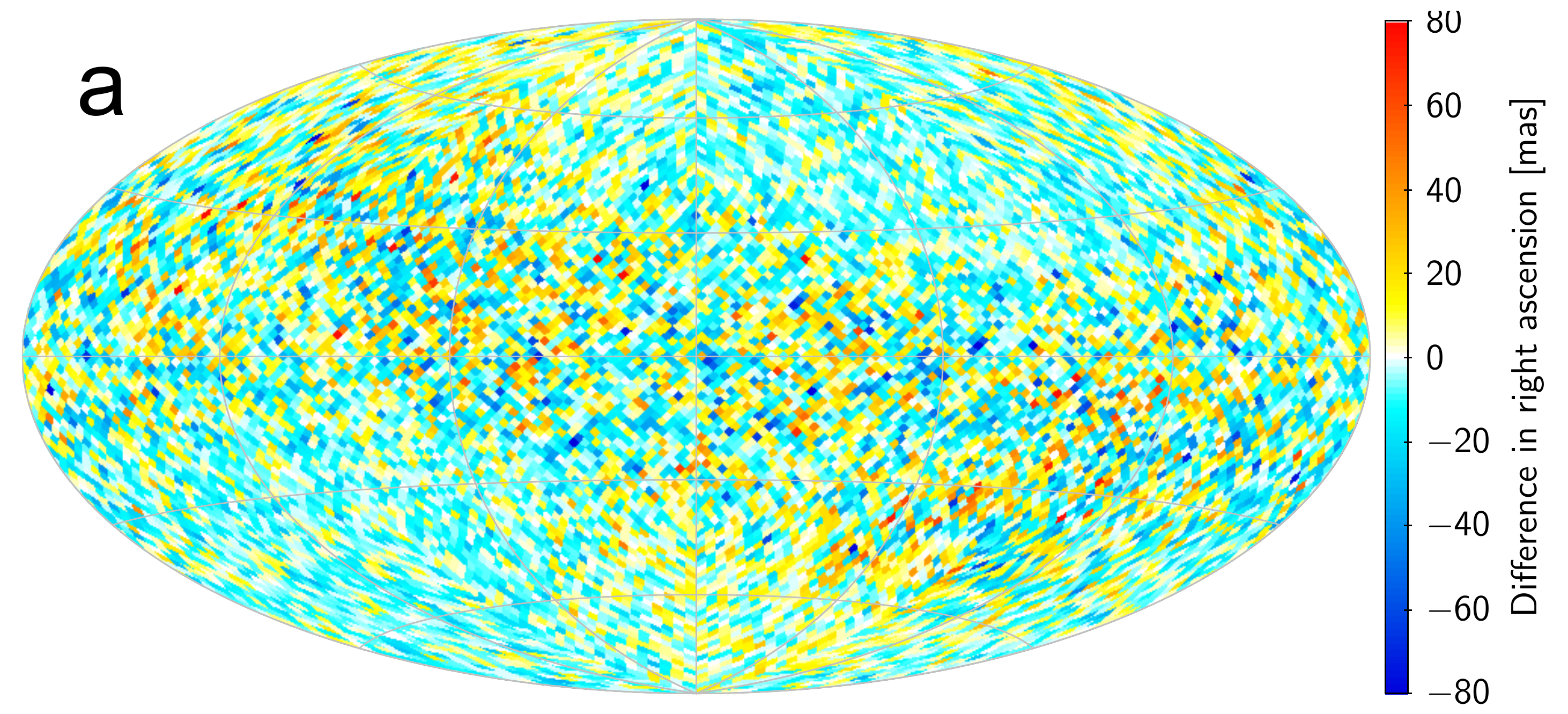}  
\includegraphics{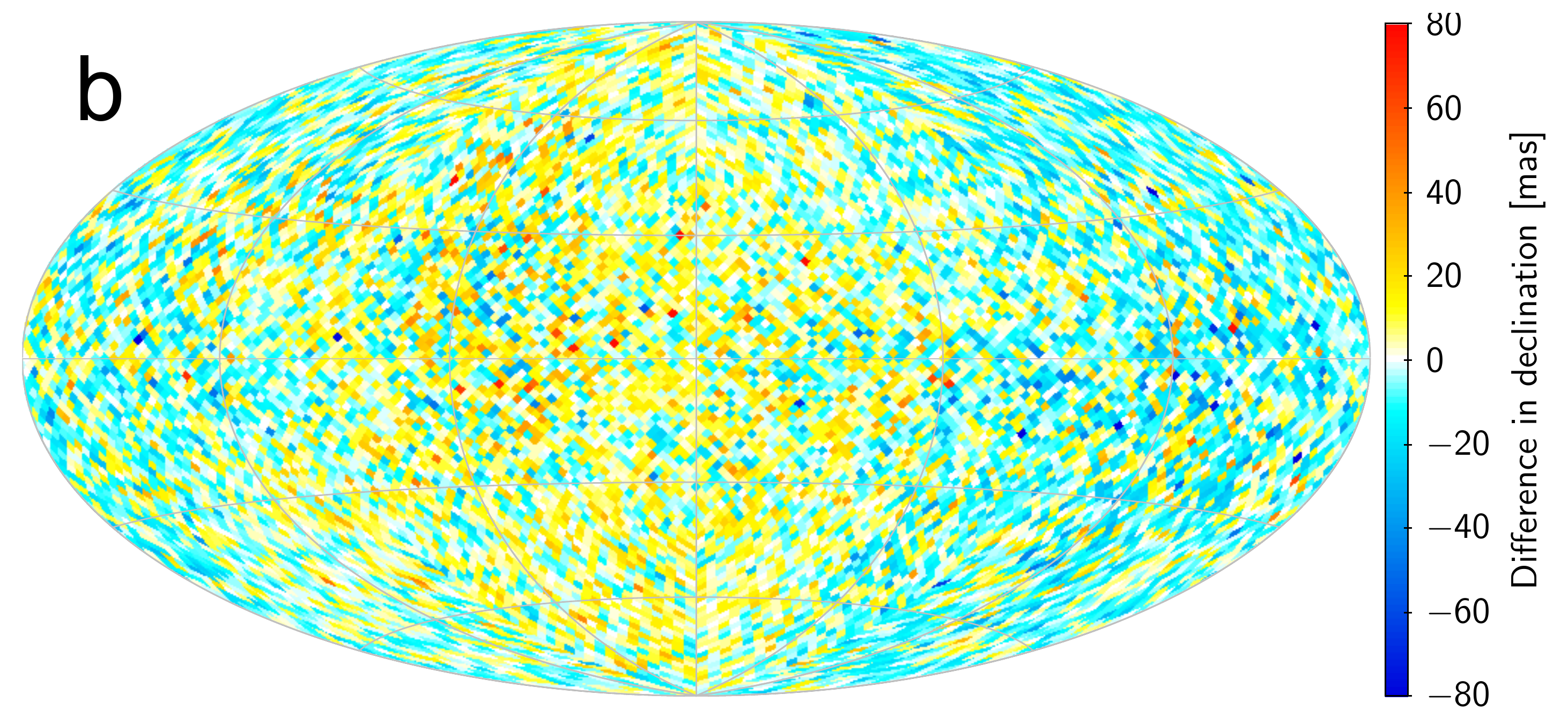}  
\includegraphics{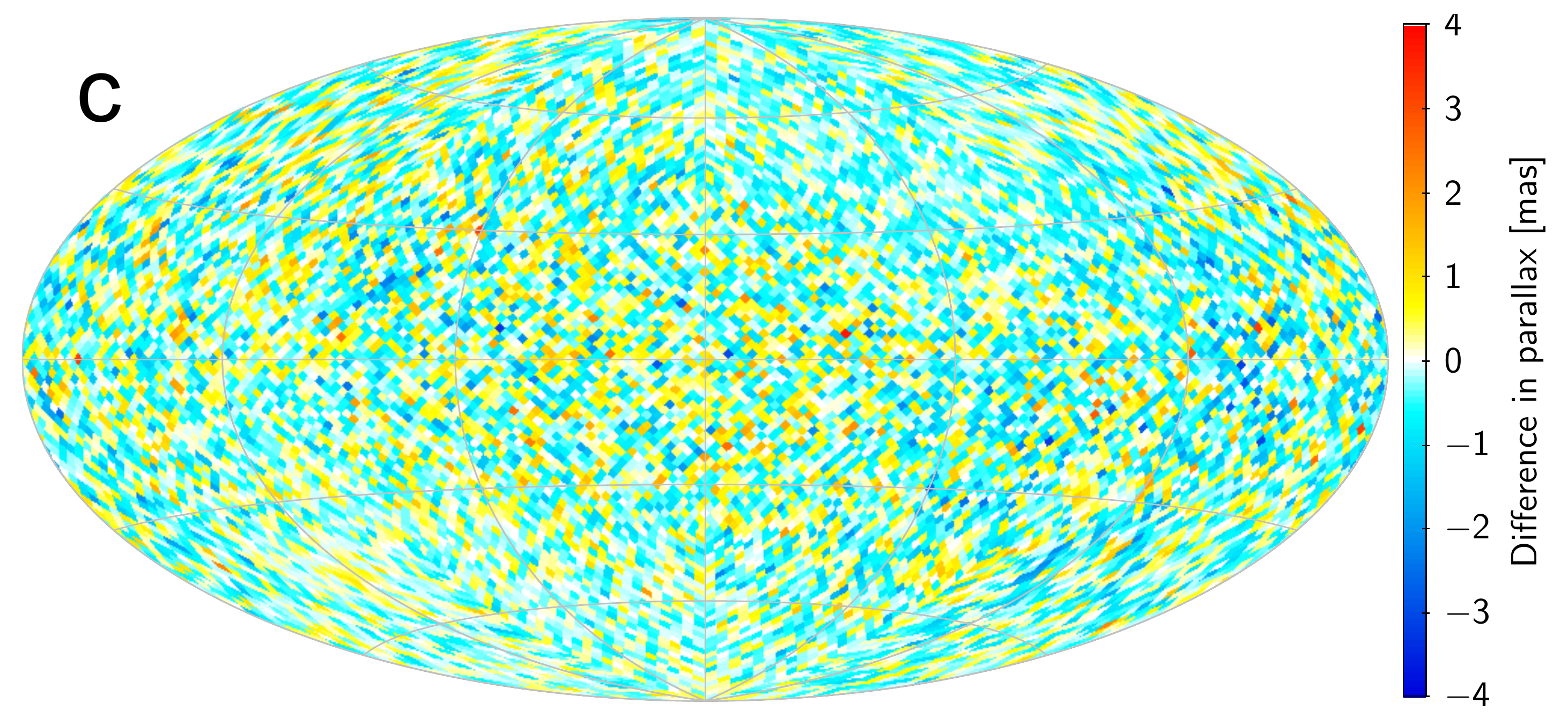}  
} 
\caption{Differences in position and parallax between the primary data set (TGAS) and the 
Hipparcos catalogue for 86\,928 sources: 
({\bf a}) difference in right ascension, $(\alpha_\text{T}-\alpha_\text{H})\cos\delta$; 
({\bf b}) difference in declination, $\delta_\text{T}-\delta_\text{H}$;
({\bf c}) difference in parallax, $\varpi_\text{T}-\varpi_\text{H}$.
Median differences at epoch J2015.0 are shown in cells of about 3.36~deg$^2$.
The position differences have not been corrected for the orientation difference 
between the Hipparcos reference frame and the reference frame of Gaia DR1.
The maps use an Aitoff projection in equatorial (ICRS) coordinates, with origin 
$\alpha=\delta=0$ at the centre and $\alpha$ increasing from right to left.}   
\label{fig:valHip1a}
\end{figure*}

\begin{figure*} 
\resizebox{\hsize}{!}{%
\includegraphics{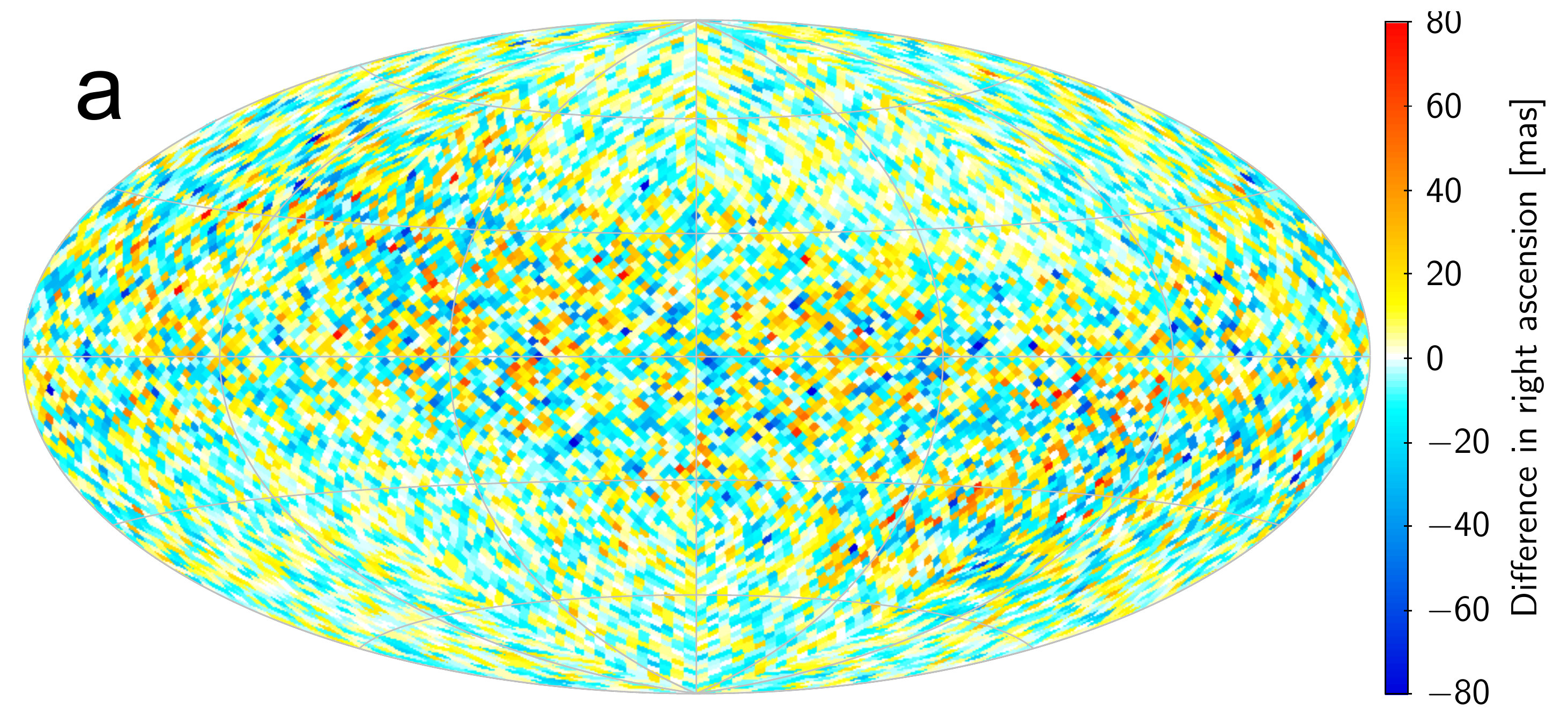}  
\includegraphics{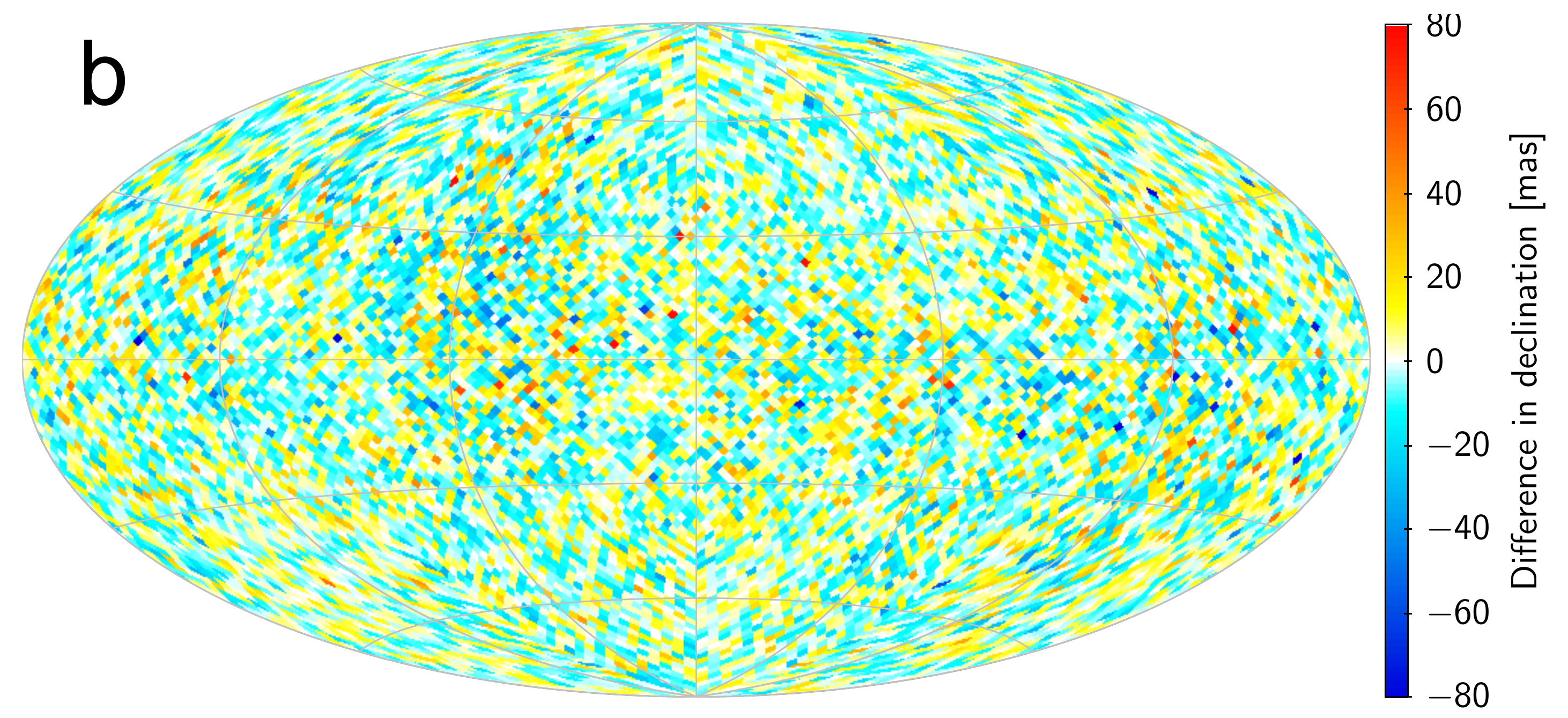}  
\includegraphics{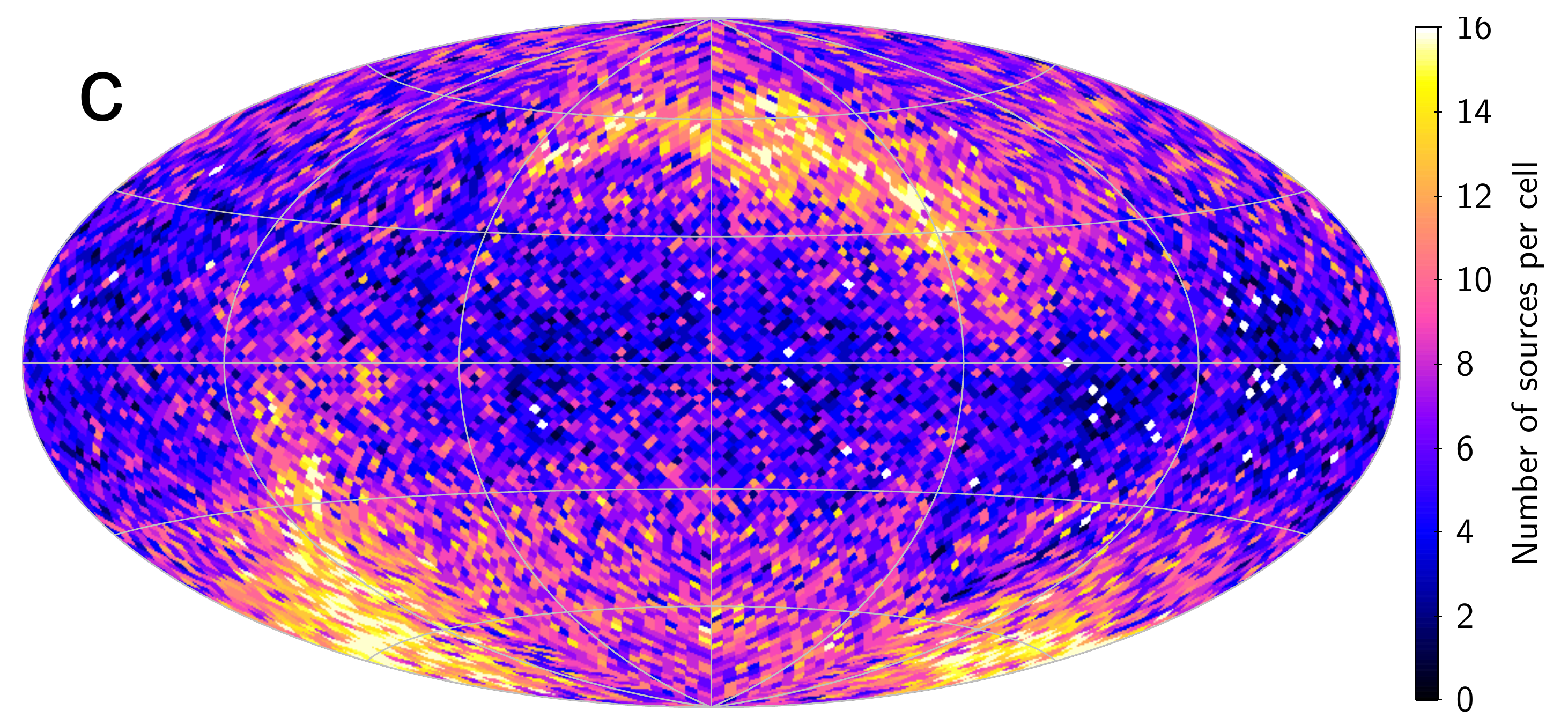}  
} 
\caption{Differences in position between the primary data set (TGAS) and the Hipparcos catalogue
 for 86\,928 sources: 
({\bf a}) difference in right ascension, $(\alpha_\text{T}-\alpha_\text{H})\cos\delta$; 
({\bf b}) difference in declination, $\delta_\text{T}-\delta_\text{H}$.
Median differences at epoch J2015.0 are shown in cells of about 3.36~deg$^2$.
({\bf c}) Number of sources per cell used to compute the median differences here and 
in Fig.~\ref{fig:valHip1a}. Some empty cells are shown in white.
The position differences have been corrected for the orientation difference 
between the Hipparcos reference frame and the reference frame of Gaia DR1.
The maps use an Aitoff projection in equatorial (ICRS) coordinates, with origin 
$\alpha=\delta=0$ at the centre and $\alpha$ increasing from right to left.}
\label{fig:valHip1b}
\end{figure*}

\subsection{Comparison with the Hipparcos and Tycho-2 catalogues}
\label{sec:hipparcos}

The following comparisons are based on the 2\,086\,766 sources from the
primary solution that satisfy Eq.~(\ref{eq:formal1}), even though not all of them
are retained in Gaia DR1. For 101\,106 sources with Hipparcos identifiers
we compare with the re-reduction of the raw Hipparcos data by 
\citetads{2007ASSL..350.....V} as retrieved from the CDS. The Hipparcos astrometric data
were propagated to epoch J2015.0 using rigorous formulae 
\citepads{2014A&A...570A..62B}, but neglecting light-time and perspective effects by 
assuming zero radial velocity for all stars. The perspective effect is only relevant for
a small number of stars with high proper motion, most of which are missing in Gaia DR1.
Unless otherwise specified, 
the comparison of positions and proper motions is made after rotating the Hipparcos 
data to the Gaia DR1 frame as explained in Sect.~\ref{sec:alignment}. Only
entries with parallax uncertainty $\le 1.5$~mas in the Hipparcos catalogue are
used for the comparison below, consisting of 86\,928 entries in the primary data set.
Values from the Hipparcos catalogue are denoted with subscript H,
those from the primary (TGAS) data set by T.

In all comparisons we first consider the global differences, i.e.\ including all sources
irrespective of their position, colour, and other characteristics. It should be kept in mind
that the resulting statistics are indeed only valid on a global level. The data are in general
very inhomogeneous, and as soon as they are broken down according to position, colour, 
etc., a much more complex picture emerges with sometimes much stronger systematic
differences and locally higher dispersions. In this section we focus on the dependence on
position (i.e.\ regional systematics) and, to some extent, on colour.

\paragraph{Hipparcos positions}
 
The global statistics of the positional differences at J2015.0 are
$\text{med}(\Delta\alpha*)=-0.073\pm0.101$~mas,   
$\text{med}(\Delta\delta)=+0.154\pm0.089$~mas,  
$\text{RSE}(\Delta\alpha*)=27.8$~mas, and  
$\text{RSE}(\Delta\delta)=24.0$~mas,  
where $\Delta\alpha*=(\alpha_\text{T}-\alpha_\text{H})\cos\delta$
and $\Delta\delta=\delta_\text{T}-\delta_\text{H}$ are the
position differences in right ascension and declination.
The large RSE values are mainly attributable to the Hipparcos errors 
propagated to J2015.0, where the Hipparcos positions have rms 
uncertainties of 21.7~mas ($\sigma_{\alpha*}$) and 18.3~mas ($\sigma_\delta$), not    
accounting for possible non-linear motions caused by binarity, etc.

Panels ({\bf a}) and ({\bf b}) in Figs.~\ref{fig:valHip1a}--\ref{fig:valHip1b} 
show the median differences in $\alpha$ and $\delta$ broken down 
according to celestial position. In Fig.~\ref{fig:valHip1a} the position 
differences are shown as calculated from the catalogue values; 
in Fig.~\ref{fig:valHip1b} they have been corrected for the orientation
difference ($\vec{\varepsilon}$) according to Eq.~(\ref{eq:align1}).
The tessellation uses a Healpix scheme with 12\,285 pixels, giving 
a pixel size of 3.36~deg$^2$. The mean number of sources per pixel  
is thus eight, but the local number varies significantly as shown in  
Fig.~\ref{fig:valHip1b}c. The smaller density of stars 
in the ecliptic region $\left|\,\beta\,\right|\lesssim 45^\circ$) is partly inherent in 
the Hipparcos catalogue, but enhanced by our selection 
$\sigma_{\varpi\text{H}}\le 1.5$~mas.

The positional differences in Fig.~\ref{fig:valHip1a}a--b show a clear
signature of the $\simeq\,$5.6~mas orientation difference between the
Gaia DR1 reference frame and the Hipparcos reference frame at J2015.0.
This signature is not visible in Fig.~\ref{fig:valHip1b}a--b, where the
Hipparcos positions have been rotated by $\vec{\varepsilon}$.

The median differences, especially in right ascension, show a markedly 
larger scatter in the ecliptic region than in other parts of the sky. This
is partly explained by the lower number of sources per pixel in
that region, but mainly reflects the variation of Hipparcos proper motion 
uncertainties with ecliptic latitude. The propagated Hipparcos positions
are clearly not good enough to validate the TGAS positions on a small
scale, but do not indicate any large systematics on a semi-global scale.
For example, the median differences computed separately for 
octants of the celestial sphere differ from the global value by at most
1~mas in $\Delta\alpha*$ and 0.6~mas in $\Delta\delta$. A stricter validation of 
the TGAS positions is possible by means of VLBI data 
(Sect.~\ref{sec:validation_radiostars}).

\paragraph{Hipparcos proper motions}
 
The global statistics of the proper motion differences, after
correcting the Hipparcos values to the Gaia DR1 reference frame, are
$\text{med}(\Delta\mu_{\alpha*})=-0.003\pm0.004$~mas~yr$^{-1}$,   
$\text{med}(\Delta\mu_\delta)=+0.006\pm0.004$~mas~yr$^{-1}$,   
$\text{RSE}(\Delta\mu_{\alpha*})=1.17$~mas~yr$^{-1}$, and   
$\text{RSE}(\Delta\mu_\delta)=1.01$~mas~yr$^{-1}$,   
where $\Delta\mu_{\alpha*}=\mu_{\alpha*\text{T}}-\mu_{\alpha*\text{H}}$
and $\Delta\mu_\delta=\mu_{\delta\text{T}}-\mu_{\delta\text{H}}$ are the
proper motion differences in right ascension and declination. 
As expected, these values are almost exactly equal to the corresponding position
differences divided by the epoch difference of 23.75~yr. The maps of
the median proper motion differences are not given here, as they 
are virtually indistinguishable from the corresponding maps of position 
differences, if the colour 
scales of the latter are interpreted as proper motion scales in the range
$[-3.4,\, +3.4]$~mas~yr$^{-1}$ ($\pm 80~\text{mas}/23.75~\text{yr}$).

\begin{figure} 
\resizebox{\hsize}{!}{\includegraphics{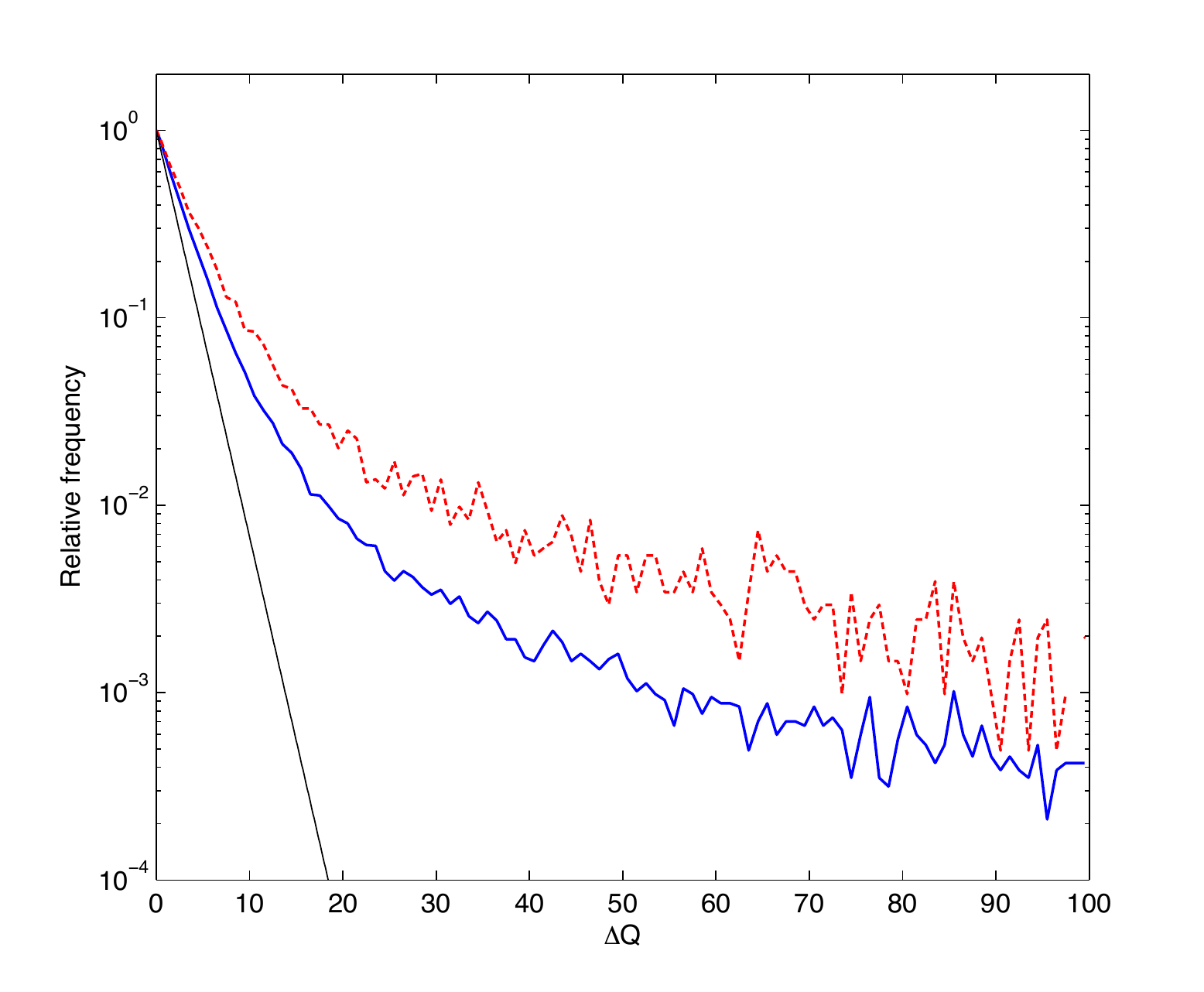}} 
\caption{Relative frequencies of the statistic $\Delta Q$ for two selections
of stars in the Hipparcos subset of the primary solution: 91\,939 bona fide single stars
(solid blue curve) and 9167 other stars (dashed red). The black line is the
theoretically expected distribution.}
\label{fig:DeltaQ}
\end{figure}

A related comparison is provided by the statistic $\Delta Q$ defined by 
Eq.~(\ref{eq:DeltaQ}).  $\Delta Q$ measures the proper motion difference
between the primary data set (TGAS) and the Hipparcos catalogue, normalised by the 
covariances provided in the two catalogues. For genuinely single stars,
$\Delta Q$ is expected to have an exponential distribution. 
Figure~\ref{fig:DeltaQ} shows the relative frequencies of $\Delta Q$ for 
two samples of the Hipparcos entries: the solid blue curve shows bona fide 
single stars (91\,939 entries without any indication of duplicity in the analysis 
by \citealt{2007ASSL..350.....V}, i.e.\ of solution type $\text{Sn}=5$), while
the dashed red curve shows the remaining stars (9167 entries with 
$\text{Sn}\ne 5$). The latter include known binaries, acceleration 
solutions, etc. For comparison, the black line shows the expected
exponential distribution. Both samples show an approximately 
exponential distribution for small values of $\Delta Q$, albeit with 
a smaller slope than theoretically expected. This could be an effect
of underestimated formal uncertainties in either or both catalogues,
or as a real cosmic scatter caused by the fact that most stars are
actually non-single. The higher relative frequency of large $\Delta Q$
among sources with $\text{Sn}\ne 5$ confirms the expected sensitivity of
$\Delta Q$ to duplicity. The sample of bona fide single stars 
contains some 50 entries with $\Delta Q>1000$, 
$\sim$1000 with $\Delta Q>100$, and 
$\sim$10\,000 with $\Delta Q>10$. These are clearly candidates for
further investigation.

\paragraph{Hipparcos parallaxes}
 
\begin{figure} 
\resizebox{0.9\hsize}{!}{\includegraphics{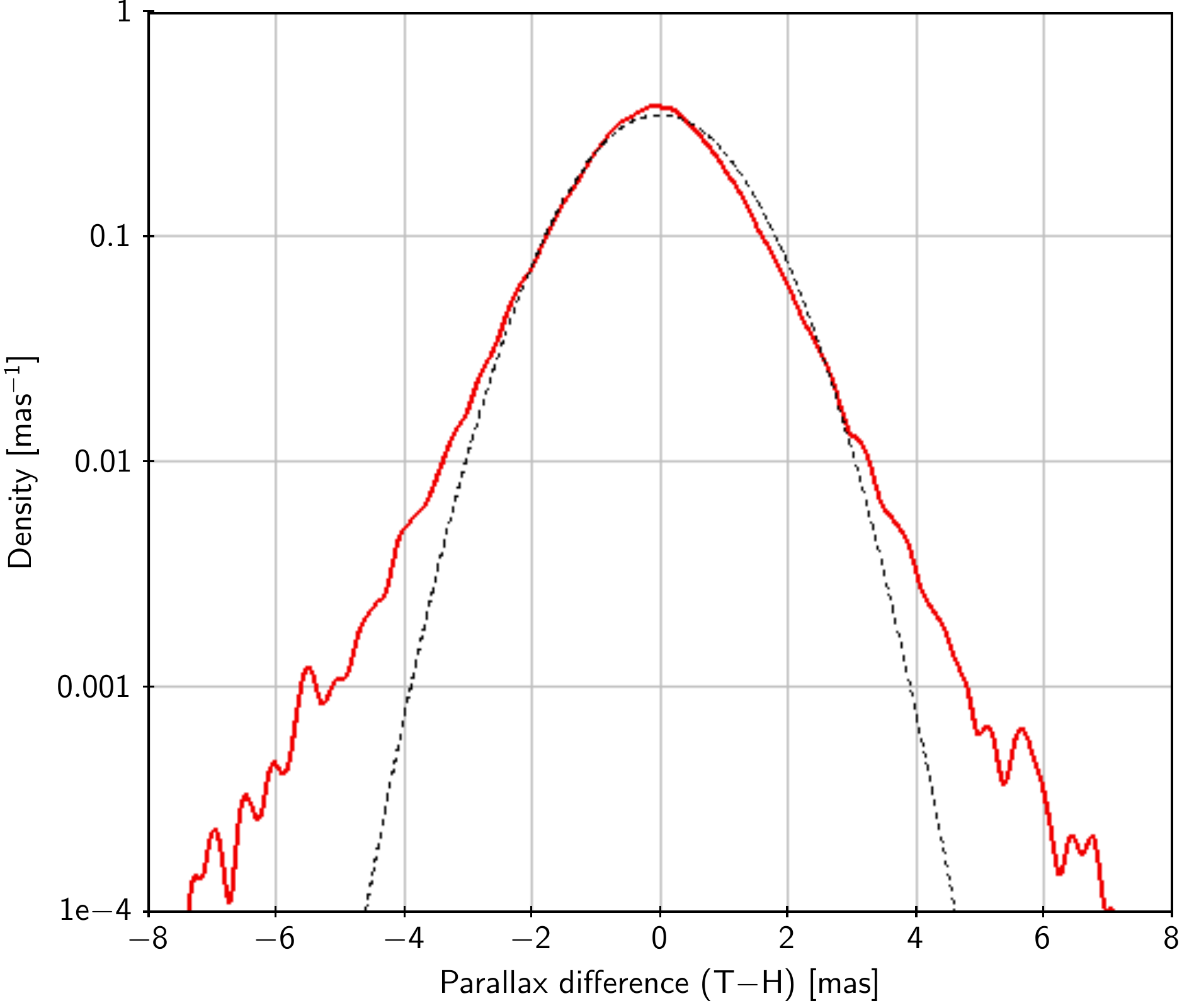}}  
\resizebox{0.9\hsize}{!}{\includegraphics{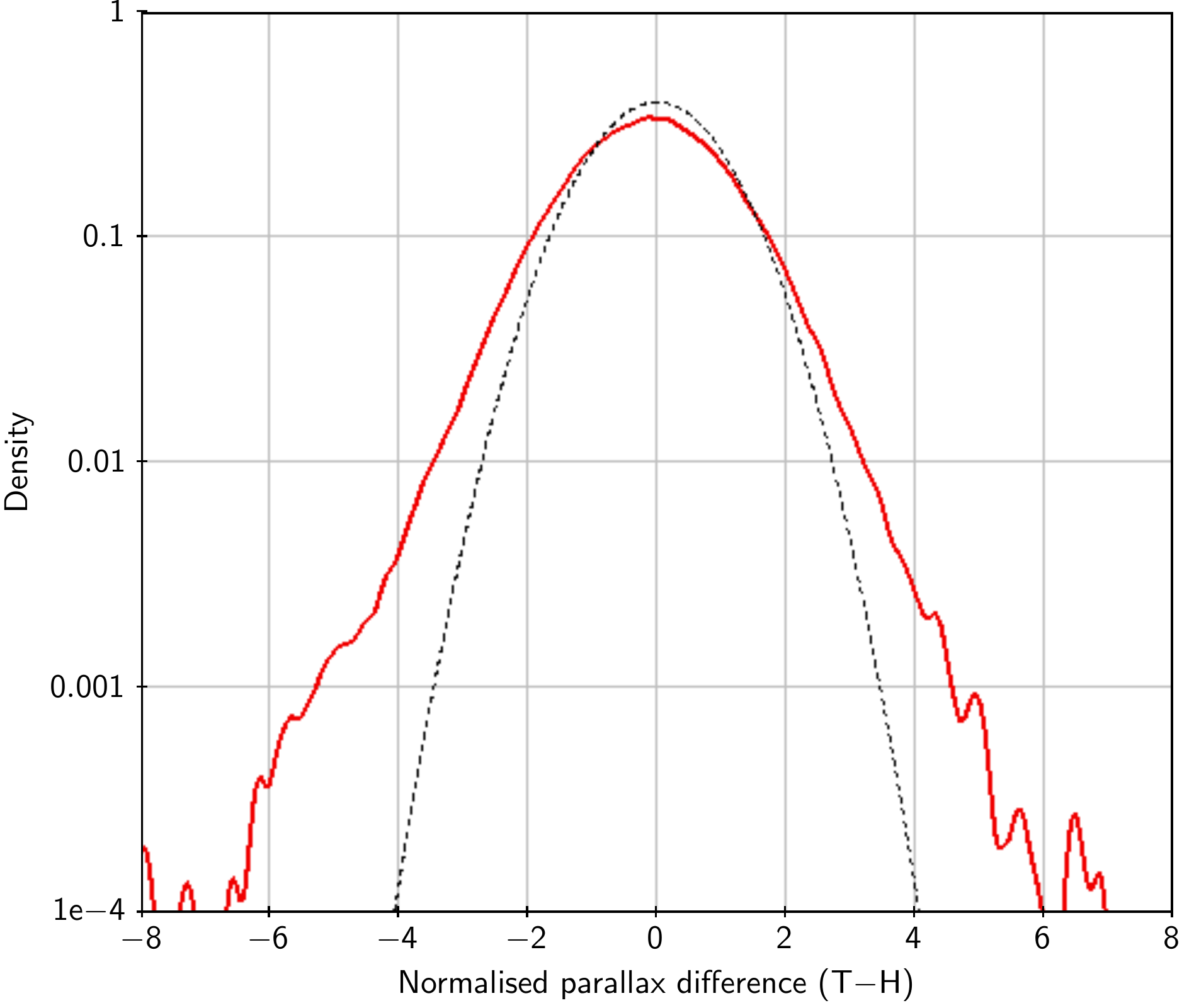}}  
\caption{Probability density plots (see footnote~\ref{footnote:KDE}) of parallax differences,
taken in the sense TGAS minus Hipparcos, for a common subset of 86\,928    
sources. Top: empirical probability density of $\Delta\varpi$ (solid), 
and for comparison a normal probability density function with standard deviation 
1.14~mas (dashed), equal to the RSE of the differences.       
Bottom: probability density of the normalised parallax
differences (solid), and for comparison the unit normal probability density 
function (see text for details).}
\label{fig:valHip3}
\end{figure}

The global statistics of the parallax differences are
$\text{med}(\Delta\varpi)=-0.089\pm0.006$~mas and     
$\text{RSE}(\Delta\varpi)=1.14$~mas, where                     
$\Delta\varpi=\varpi_\text{T}-\varpi_\text{H}$.
The slightly negative median difference is statistically
significant and is clearly seen in a probability density plot%
\footnote{Figures~\ref{fig:valHip3}, \ref{fig:icrfResNorm}, \ref{fig:valQsoPlx},
and \ref{fig:statResHist} were produced using kernel density estimation 
(KDE; e.g.\ \citeads{2012msma.book.....F}) with a Gaussian kernel having
a standard width of about 0.1 times the RSE of the distribution.\label{footnote:KDE}}
of the differences (Fig.~\ref{fig:valHip3}). The bottom diagram in
Fig.~\ref{fig:valHip3} shows the distribution of normalised differences
$\Delta\varpi/(\sigma_{\varpi\text{T}}^2+\sigma_{\varpi\text{H}}^2)^{1/2}$,
using the inflated standard uncertainties $\sigma_{\varpi\text{T}}$ from
Eq.~(\ref{eq:infl}) and $\sigma_{\varpi\text{H}}$ as given in the Hipparcos catalogue.
The distribution is slightly wider than the expected unit normal distribution
(the RSE of the normalised parallax differences is 1.22), suggesting that the 
standard uncertainties are slightly underestimated in one or both data sets.
It also displays the non-Gaussian, almost exponential tails often seen in 
empirical error distributions.

\begin{figure} 
\resizebox{\hsize}{!}{\includegraphics{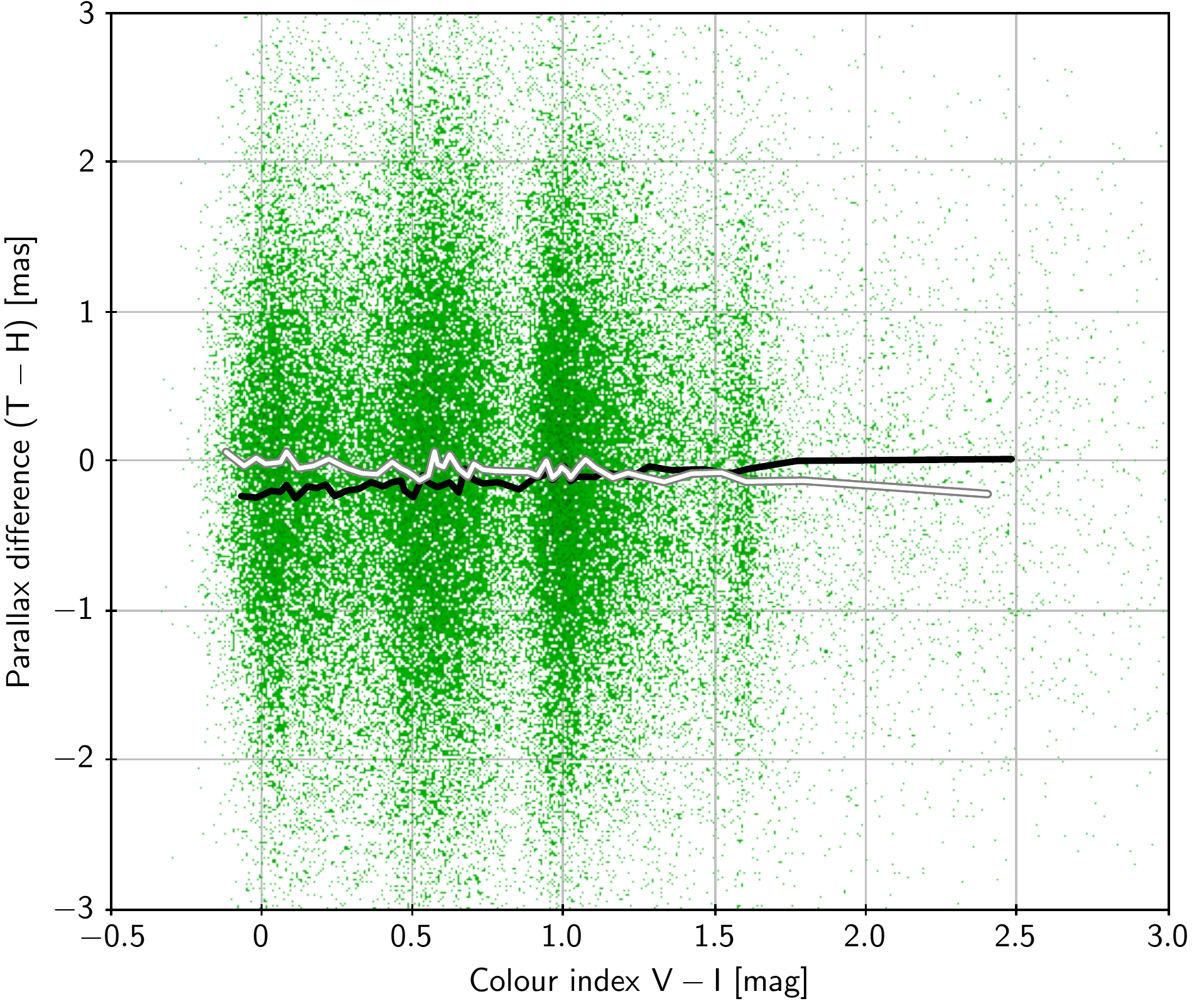}}    
\caption{Parallax differences (TGAS minus Hipparcos) for 86\,928 sources,   
plotted against colour index. 
The black line is for northern ecliptic latitudes ($\beta>0$), the grey-white 
line for southern ($\beta<0$). The lines connect median values calculated in 
50 bins subdividing the data according to $V-I$. Each bin contains about 
900 data points per hemisphere.}
\label{fig:valHip4}
\end{figure}

The parallax difference map (Fig.~\ref{fig:valHip1a}c) has many
interesting features but we will only comment on a few. The larger scatter
in the ecliptic region is obvious, as is the patchiness of the visible structures,
suggesting strong spatial correlations on a scale of a few degrees. Both features
are expected to be present, to some extent, in both data sets, and it is not 
possible to conclude from this comparison if they are (mainly) a feature in 
one or the other data set. Another conspicuous feature is that the northern 
ecliptic region ($\beta>45^\circ$, where $\beta$ is the ecliptic latitude)
is on the whole slightly more negative (blue) than the southern 
($\beta<-45^\circ$). This is confirmed by partitioning the differences 
according to ecliptic latitude:
\begin{equation}\label{eq:hip1}
\text{med}(\Delta\varpi) = 
\begin{cases} -0.130\pm 0.006~\text{mas} &\text{for $\beta>0$}\, , \\     
-0.053\pm 0.006~\text{mas} &\text{for $\beta<0$}\, . \end{cases}            
\end{equation}
Further analysis reveals that the north--south asymmetry in $\Delta\varpi$
depends on the colour of the star. Subdividing the data according to colour 
index $V-I$, taken from the Hipparcos catalogue, shows approximately linear 
trends with opposite signs (Fig.~\ref{fig:valHip4}) in the two hemispheres. 
Over the investigated range
of colours, the total amplitude of the effect is $\pm 0.1$~mas. While it cannot
be excluded that this effect, at least partly, originates from the Hipparcos data,
there are strong indications that it is caused by the -- as yet uncalibrated -- 
chromaticity of the Gaia instrument (see Appendices~\ref{sec:validation_quasars},
\ref{sec:chrom}, and \ref{sec:valColour}).

\begin{figure} 
\resizebox{\hsize}{!}{\includegraphics{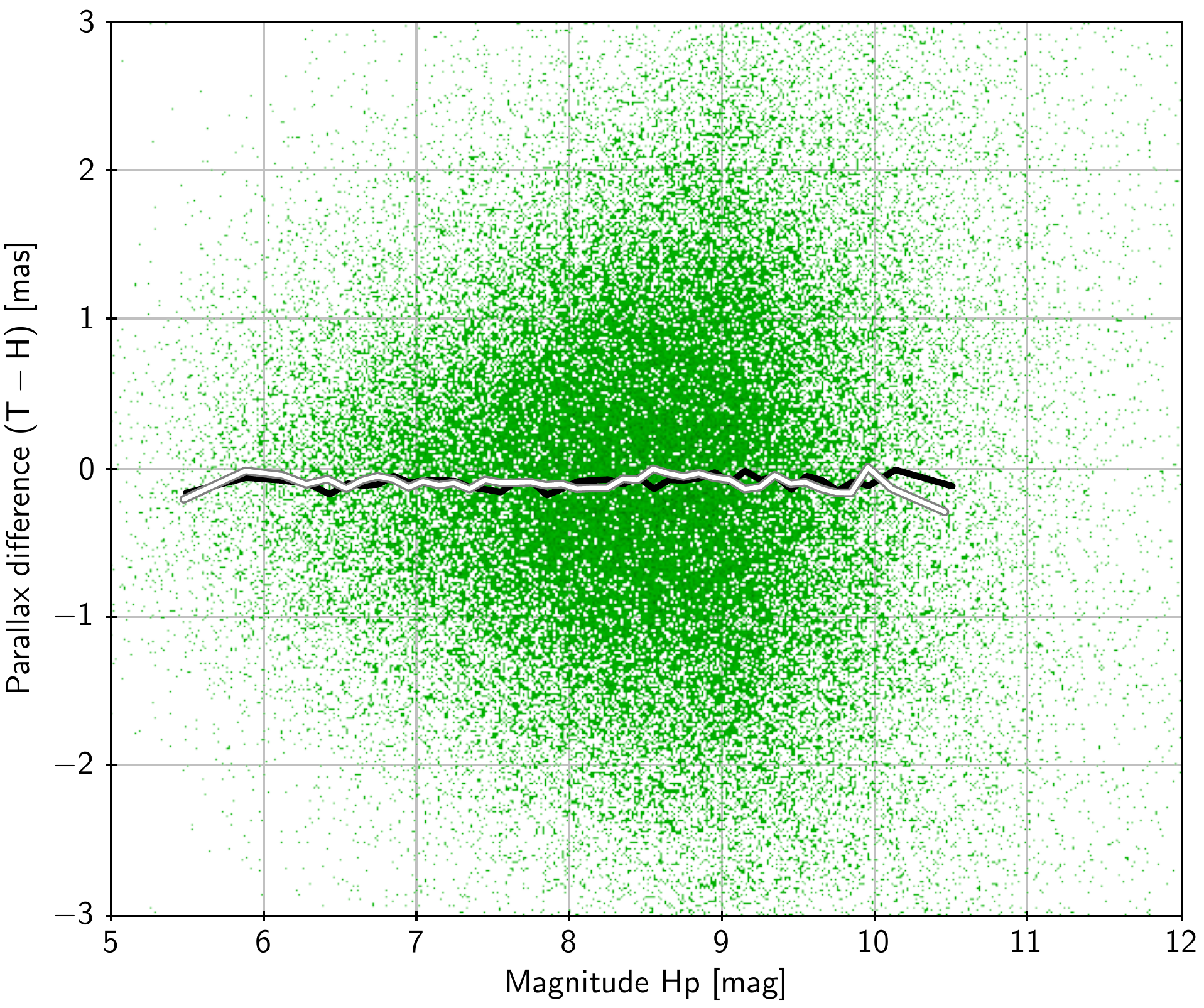}}  
\caption{Parallax differences (TGAS minus Hipparcos) for 86\,928 sources,
plotted against magnitude. 
The black line is for northern ecliptic latitudes ($\beta>0$), the grey-white 
line for southern ($\beta<0$). The lines connect median values calculated in 
50 bins subdividing the data according to the Hipparcos magnitude $H\!p$. 
Each bin contains about 900 data points per hemisphere.}
\label{fig:valHip5}
\end{figure}

If the same data are instead subdivided according to magnitude, using $H\!p$ from 
the Hipparcos catalogue (Fig.~\ref{fig:valHip5}), there is no clear systematic trend
in either hemisphere.

\paragraph{Tycho-2 proper motions}

The proper motions in the \mbox{Tycho-2} catalogue \citepads{2000A&A...355L..27H}
were derived by combining the positions obtained from the Hipparcos star mappers,
here called \mbox{Tycho-2} positions, with positions from earlier transit circle 
and photographic programs \citepads{2000A&A...357..367H}, including in particular 
the Astrographic Catalogue at a mean epoch around 1907 \citepads{1998AJ....115.1212U}.
Although a big effort was made to put the old positions on the Hipparcos reference 
frame, systematic errors remain which are then reflected in the \mbox{Tycho-2} proper
motions. For this reason, only the \mbox{Tycho-2} positions (at the effective epoch of
observation around 1991--92) have been used as prior in TGAS, but not the 
\mbox{Tycho-2} proper motions. 

\begin{figure*} 
\resizebox{\hsize}{!}{
\includegraphics{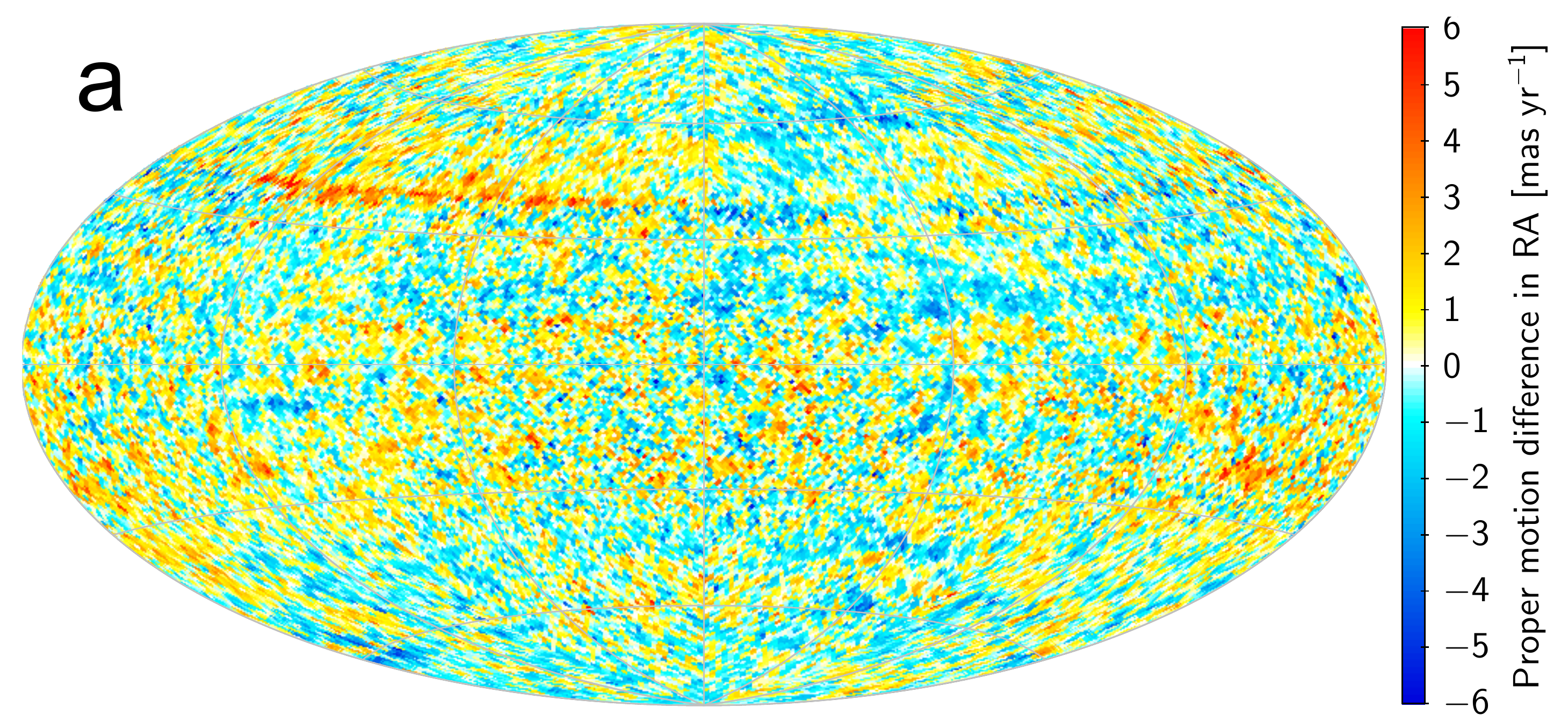}
\includegraphics{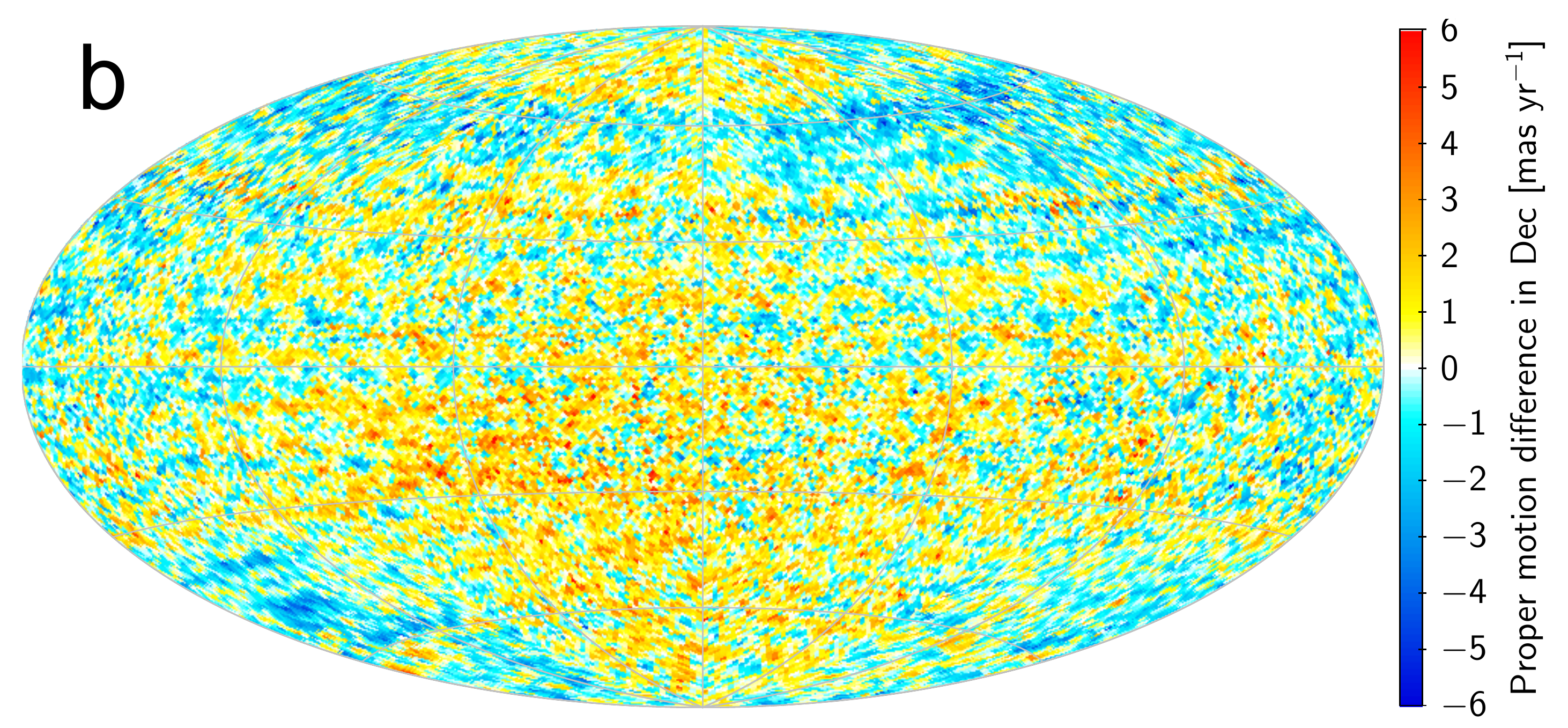}
\includegraphics{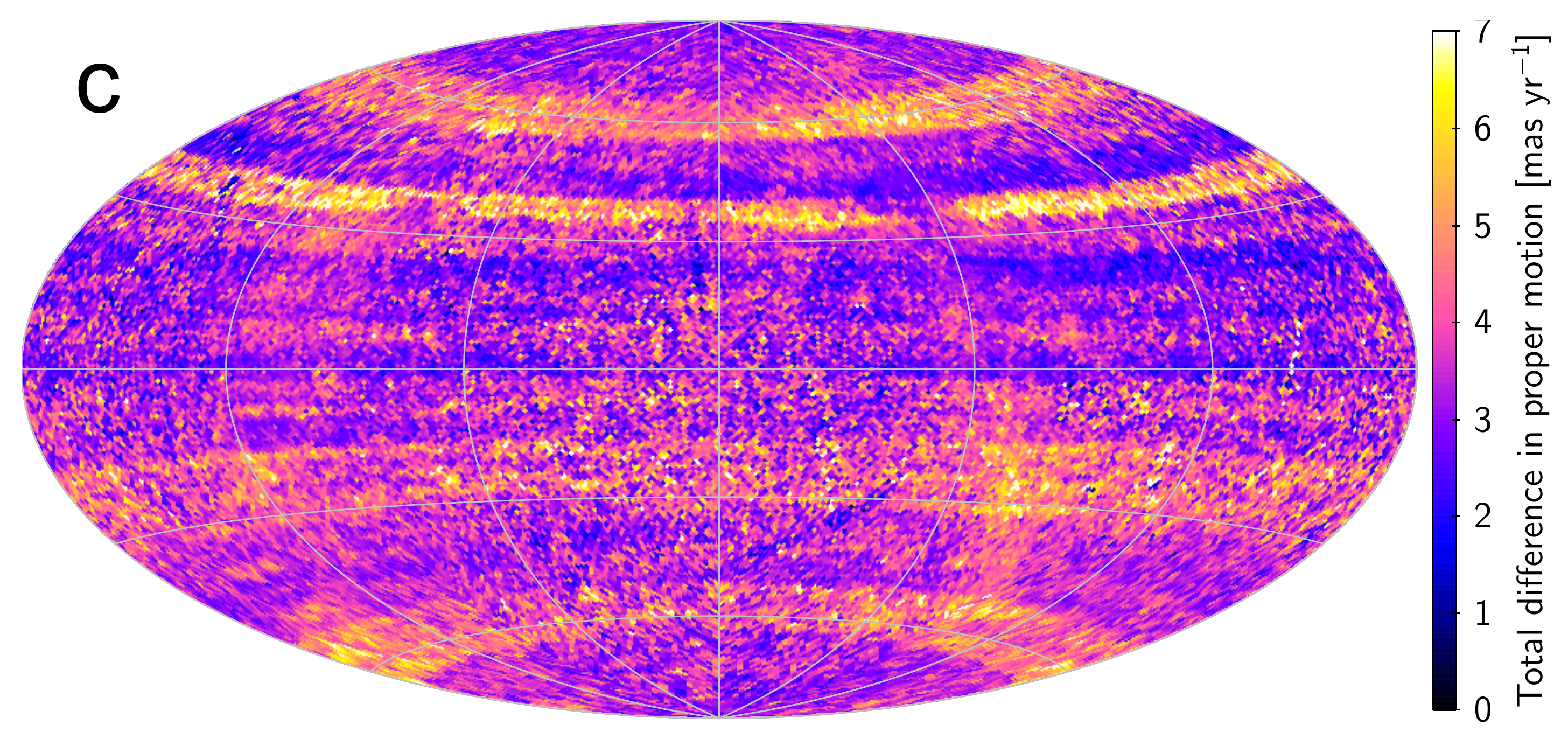} 
} 
\caption{Differences in proper motion between the primary (TGAS) solution
and the \mbox{Tycho-2} catalogue for 1\,997\,003 sources:
({\bf a}) differences in $\mu_{\alpha *}$; 
({\bf b}) differences in $\mu_\delta$; and 
({\bf c}) total differences $(\Delta\mu_{\alpha}^2 + \Delta\mu_\delta)^{1/2}$. 
Differences are taken in the sense TGAS minus \mbox{Tycho-2}, after rotation 
of the latter to the Gaia DR1 reference frame.
Median differences are shown in cells of about 0.84~deg$^2$.
Some empty cells are shown in white.
The maps use an Aitoff projection in equatorial (ICRS) coordinates, with origin 
$\alpha=\delta=0$ at the centre and $\alpha$ increasing from right to left.}
\label{fig:valTyc}
\end{figure*}

A comparison of TGAS proper motions with \mbox{Tycho-2} proper motions will
therefore mainly show the errors in the century-old positional catalogues, and
is therefore of limited value as a validation of TGAS. Nevertheless, a comparison
has been made after rotating the \mbox{Tycho-2} proper motions to the Gaia DR1 reference
frame, using Eq.~(\ref{eq:align3}). The global statistics for the proper motion differences
(TGAS minus \mbox{Tycho-2}) are 
$\text{med}(\Delta\mu_{\alpha *}) =+0.07$~mas~yr$^{-1}$, 
$\text{med}(\Delta\mu_\delta) =+0.20$~mas~yr$^{-1}$, 
$\text{RSE}(\Delta\mu_{\alpha *})= 3.6$~mas~yr$^{-1}$,
$\text{RSE}(\Delta\mu_\delta)= 3.3$~mas~yr$^{-1}$.
Maps of median differences are shown in 
Fig.~\ref{fig:valTyc}. The maps show significant systematic errors, mainly
in zones of constant declination. The alignment of these features with the
equatorial coordinate system very clearly points to the old ground-based
positions as the main source of systematics.

\subsection{Quasars}
\label{sec:validation_quasars}

\begin{figure} 
\resizebox{\hsize}{!}{\includegraphics{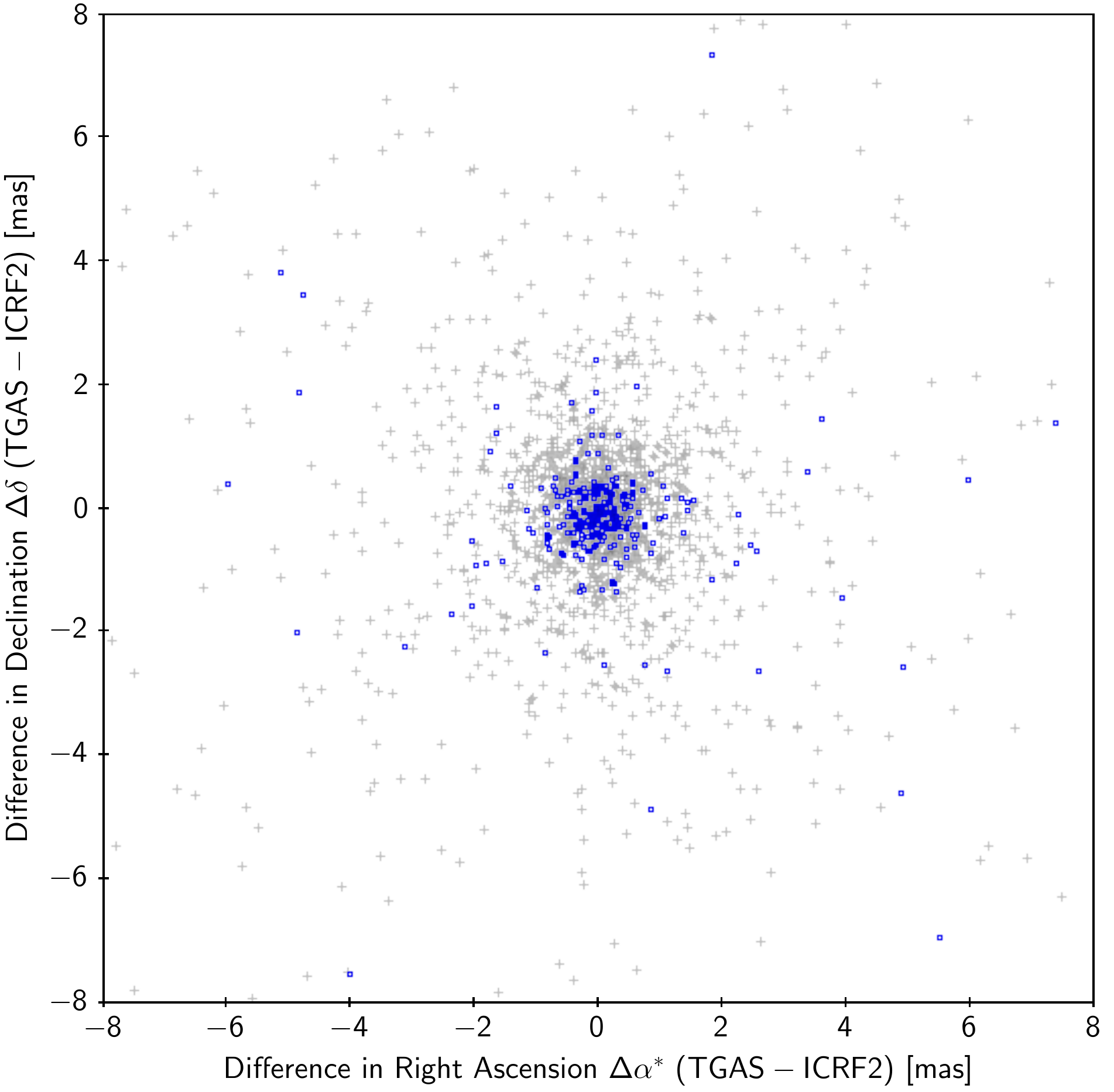}} 
\caption{Positional offsets of the optical sources matched to the VLBI positions
of ICRF2 sources. The blue circles are defining sources in ICRF2, the grey crosses
non-defining sources. 2035 sources are inside the displayed area, 156 are 
outside.}
\label{fig:icrfRes}
\end{figure}

\begin{figure} 
\resizebox{\hsize}{!}{\includegraphics{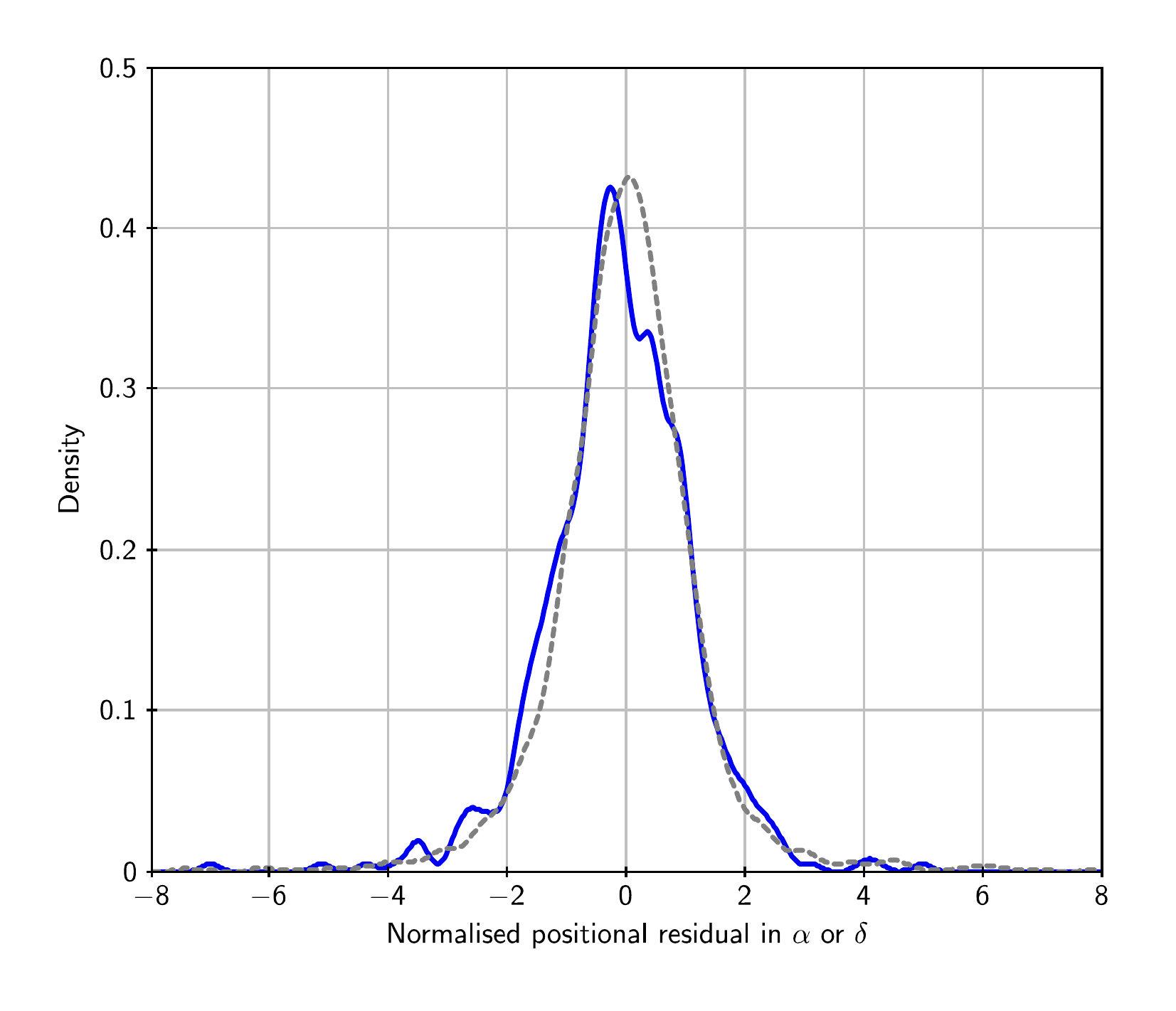}} 
\caption{Probability density plots (see footnote~\ref{footnote:KDE}) of normalised position 
differences for ICRF2 sources. The solid blue curve is for 262 defining sources, the
dashed grey for 1929 non-defining sources. $\Delta\alpha{*}/\sigma_{\Delta\alpha*}$ 
and $\Delta\delta/\sigma_{\Delta\delta}$ are considered together, as their 
distributions are not markedly different.}
\label{fig:icrfResNorm}
\end{figure}

\paragraph{Quasar positions}
The auxiliary quasar solution (Sect.~\ref{sec:quasars}) gave precise positions
and parallaxes for more than $10^5$ extragalactic sources, including 2191 that were
matched to ICRF2 sources with accurate VLBI positions. The defining subset of ICRF2 was used
to align the positional reference frame of Gaia DR1 with ICRF2 at epoch J2015.0 as described in
Sect.~\ref{sec:alignment}. Figure~\ref{fig:icrfRes} shows the optical offsets for both 
defining and non-defining sources after the alignment. Lumping $\Delta\alpha*$ and $\Delta\delta$
together, the RSE coordinate difference is 0.70~mas for the 262 matched defining sources,
and 1.82~mas for the 1929 non-defining sources. Figure~\ref{fig:icrfResNorm}
shows the distribution of the normalised position differences, 
$\Delta\alpha{*}/\sigma_{\Delta\alpha*}$, etc.,
where $\sigma_{\Delta\alpha*}$ is the quadratically combined standard 
uncertainties in TGAS (auxiliary quasar solution), using the inflated uncertainties, and 
ICRF2. The RSE of the normalised position differences is 1.08 for the
defining sources and 1.02 for the non-defining.
The overall agreement is remarkably good, especially considering
that no allowance has been made in the error budget for possible 
radio-optical offsets.

\begin{figure} 
\resizebox{\hsize}{!}{\includegraphics{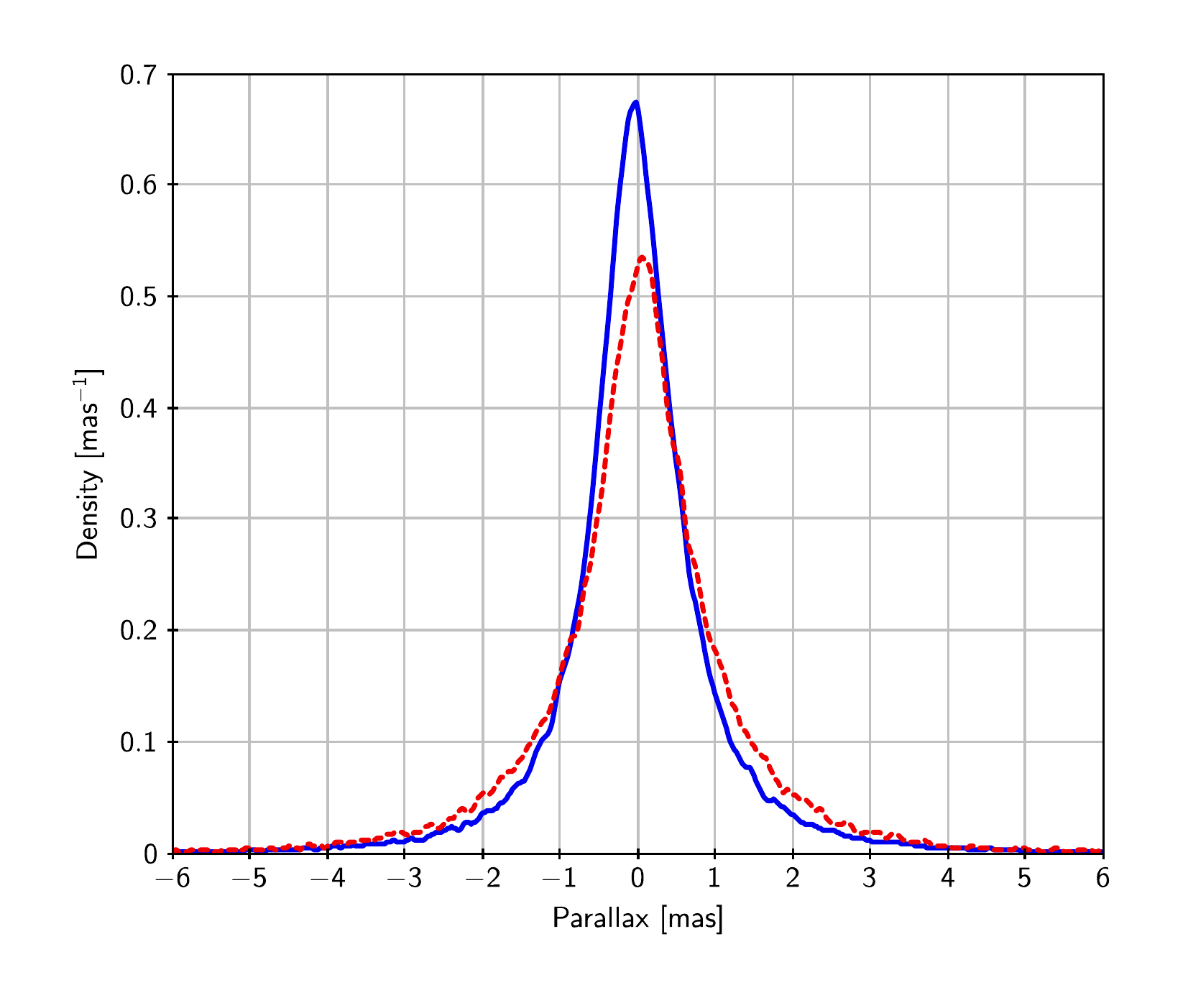}} 
\caption{Probability density plots  (see footnote~\ref{footnote:KDE}) of the measured parallaxes of 
quasars, as obtained in the auxiliary quasar solution. Blue solid curve is for
88\,641 sources at northern ecliptic latitudes ($\beta>0$), the red dashed curve
for 32\,713 sources at southern ecliptic latitudes ($\beta<0$).}
\label{fig:valQsoPlx}
\end{figure}

\paragraph{Quasar parallaxes}
The true parallaxes of quasars are negligibly small in the present context. The 
measured values therefore give an immediate impression of the dispersion 
of parallax errors and possible biases, although a detailed interpretation
will be complicated by factors that are peculiar to these objects (optical 
structure, spectral energy distribution, faintness, sky distribution, etc.).
The distribution of measured parallaxes for quasars in the primary solution
is given in Figure~\ref{fig:valQsoPlx}, where separate curves are shown for 
the northern (88\,641 sources) and for the southern 
ecliptic hemisphere (32\,713 sources). The statistics are:
\begin{equation}\label{eq:qso1}
\text{med}(\varpi) = 
\begin{cases} -0.073\pm 0.002~\text{mas} &\text{for $\beta>0$}\, , \\
+0.074\pm 0.005~\text{mas} &\text{for $\beta<0$}\, . \end{cases}
\end{equation}
The RSE is 0.85~mas (north) and 1.11~mas (south).
The north-south asymmetry in Eq.~(\ref{eq:qso1}) 
is stronger than was found in the comparison with Hipparcos data,
Eq.~(\ref{eq:hip1}). However, great caution should be exercised when interpreting 
the quasar results in view of the many complications mentioned above. 
Especially the patchy sky coverage of the GIQC is problematic, since local 
deviations could have a big impact on the global statistics.

\begin{figure}
\resizebox{\hsize}{!}{\includegraphics{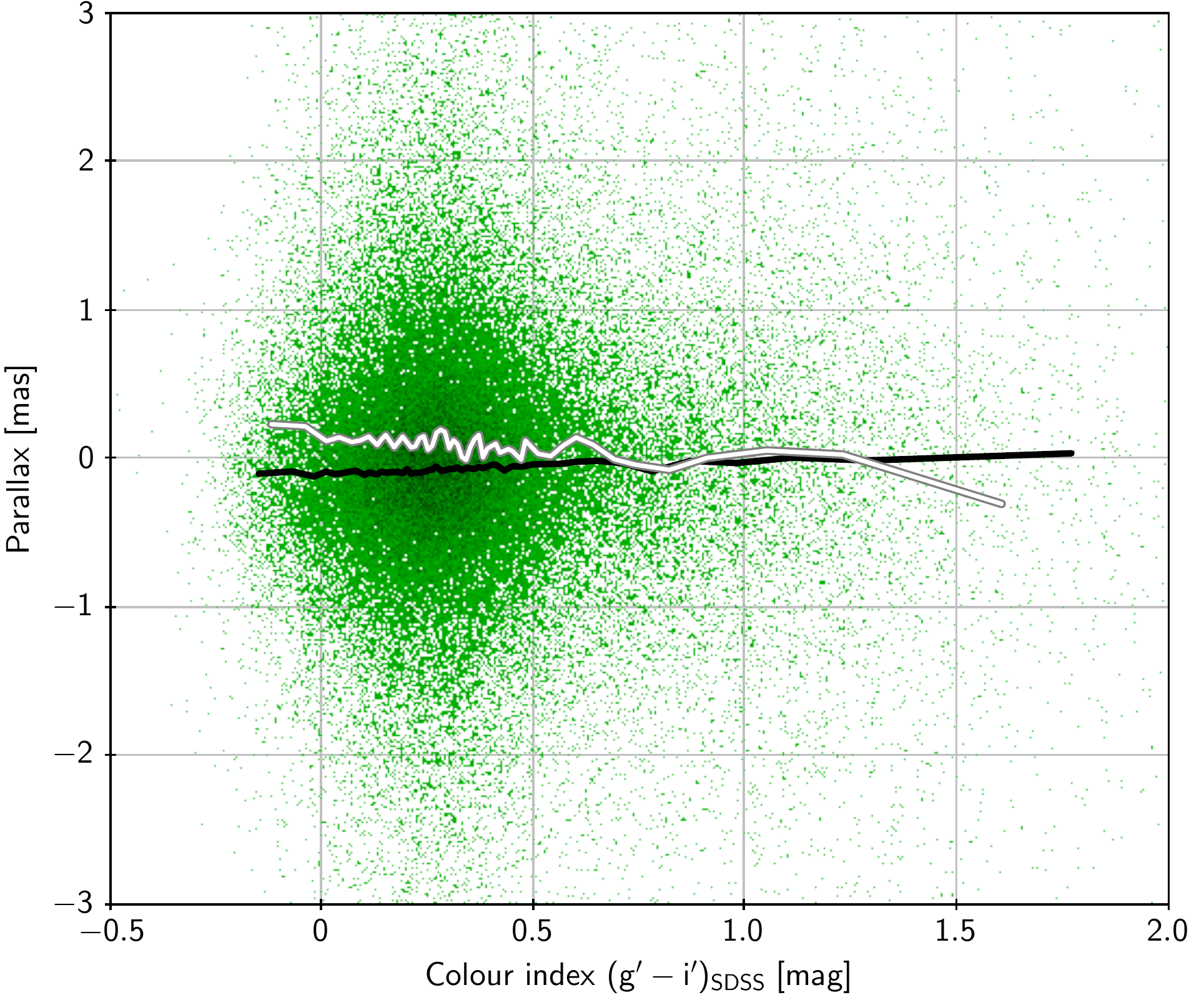}} 
\caption{Measured parallaxes for 94\,795 quasars plotted against SDSS colour index
$g'-i'$. 
The black line is for northern ($\beta>0$), the grey-white 
line for southern ecliptic latitudes ($\beta<0$). The lines connect median values calculated in 
50 bins subdividing the data according to $g'-i'$. Each bin contains about 
1600 data points for the northern and 300 for the southern latitudes.}
\label{fig:valQsoPlxCol}
\end{figure}

A further breakdown of the quasar parallaxes according to colour is then
highly interesting. Most of the quasars in GIQC have multicolour 
photometry from the Sloan Digital Sky Survey (SDSS; \citeads{2000AJ....120.1579Y}). 
Figure~\ref{fig:valQsoPlxCol} shows the results of an analysis of nearly 
95\,000 sources with SDSS colours $g'-i'$ \citepads{2002AJ....123.2121S}. 
The trends are the same as in Fig.~\ref{fig:valHip4}, comparing with the
Hipparcos parallaxes: a positive trend with increasing colour index 
for the northern hemisphere, and a negative trend for the southern hemisphere. 

\subsection{Galactic cepheids}
\label{sec:validation_cepheids}

\begin{figure} 
\resizebox{\hsize}{!}{\includegraphics{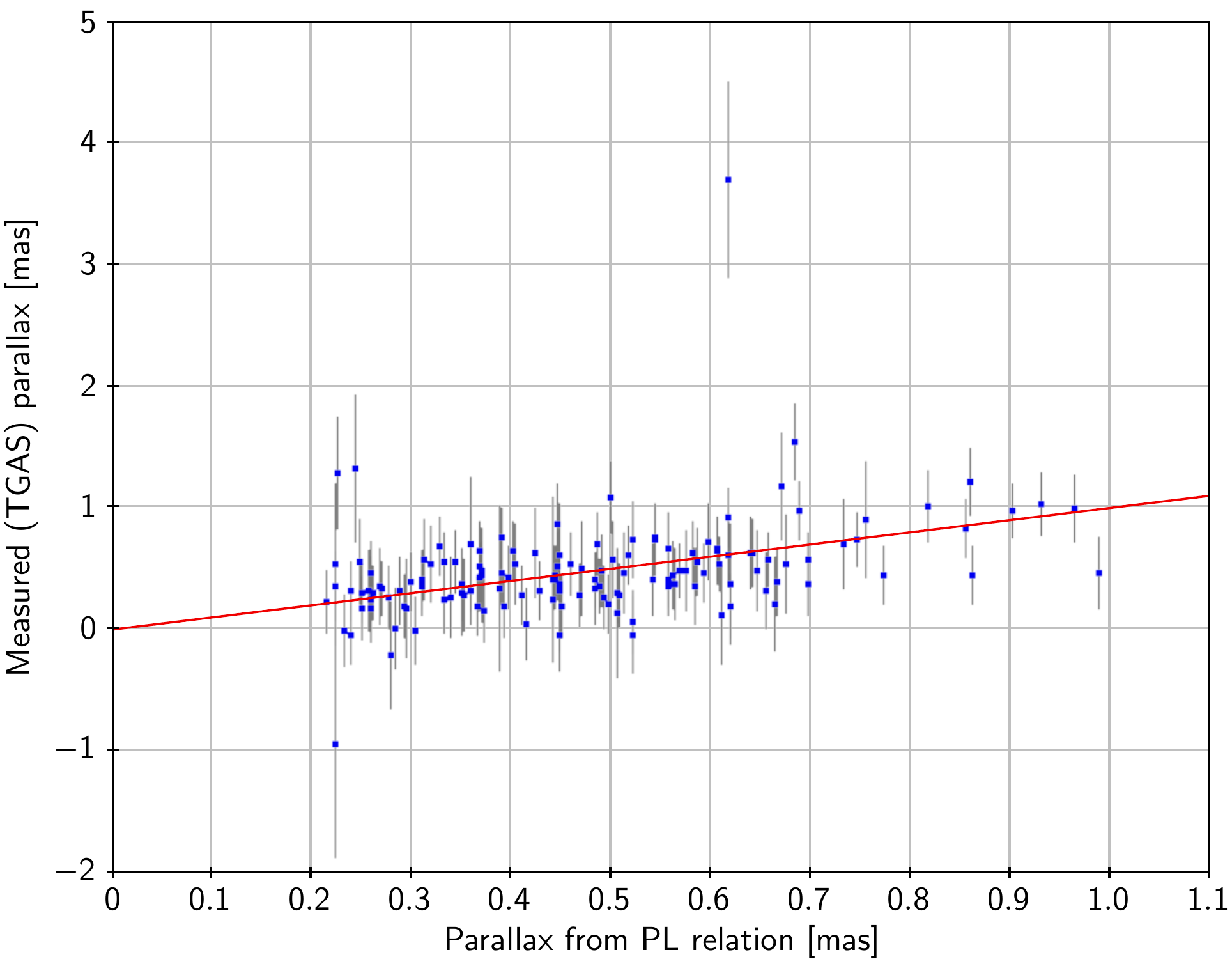}}   
\caption{Measured (TGAS) parallaxes of Galactic cepheids plotted against their
parallaxes computed from photometric data and the $V$ band period-luminosity relation
by \citetads{2007A&A...476...73F}. Error bars are the $1\sigma$ uncertainties in the
TGAS parallaxes. The inclined line is the expected 1:1 relation. Only data for the 141 cepheids with 
$\varpi_\text{PL}<1$~mas are shown.}
\label{fig:valCep}
\end{figure}

For distant cepheids the error in the parallax computed
from photometric data and a period--luminosity (PL) relation will be small 
compared with the parallax uncertainty in the current TGAS data. They could therefore 
provide an independent check of the zero point of the Gaia parallaxes.
From the catalogue by \citetads{2003A&A...404..423T} we retrieved periods and 
photometric data (mean magnitude $V$, colour excess $E_{B-V}$) for 
169 Galactic fundamental-mode pulsators with    
TGAS parallaxes satisfying Eq.~(\ref{eq:formal1}). From the PL relation, 
their parallaxes were computed as
\begin{equation}
\varpi_\text{PL} = (100~\text{mas})\times 10^{0.2(a\log P+b-V+R_VE_{B-V})} \, ,
\end{equation}
with $a=-2.678$, $b=-1.275$, and $R_V=3.23$ taken from \citetads{2007A&A...476...73F}.
The analysis of differences $\Delta\varpi=\varpi_\text{T}-\varpi_\text{PL}$ was restricted to the 
141 cepheids with $\varpi_\text{PL}<1$~mas in order to limit possible biases 
due to errors in the adopted PL relation, extinction, etc. (For example, a 0.1~mag systematic
error in $b$ or in the total extinction translates to a 5\% error in $\varpi_\text{PL}$,
or $<0.05$~mas if $\varpi_\text{PL}<1$~mas.) 
This gave $\text{med}(\Delta\varpi)=-0.016\pm 0.023$~mas and   
$\text{RSE}(\Delta\varpi)=0.25$~mas. The normalised differences   
$\Delta\varpi/\sigma_\varpi$ have an RSE of 0.86 and a standard deviation of 0.90.   
A graphical comparison is given in Fig.~\ref{fig:valCep}. 
The north-south asymmetry is insignificant:
$\text{med}(\Delta\varpi_\text{N})-\text{med}(\Delta\varpi_\text{S})=+0.001 \pm 0.046$~mas.   
Indeed, given the median colour index $V-I\simeq 1.3$~mag of the cepheids, very little
asymmetry is expected according to Fig.~\ref{fig:valHip4}. The small number of objects and their
limited spread in $V-I$ do not permit a further breakdown according to colour.

\begin{table*}[t]
\caption{Comparison of the astrometric parameters for radio sources in TGAS with
the corresponding VLBI results.\label{tab:vlbi}}
\centering
\small\setlength{\tabcolsep}{3pt}
\begin{tabular}{lrrrrrrl}
\hline\hline
\noalign{\smallskip}
Name & 
\multicolumn{1}{r}{HIP/TYC} & 
\multicolumn{1}{c}{$\Delta\alpha *$ [mas]} & 
\multicolumn{1}{c}{$\Delta\delta$ [mas]} & 
\multicolumn{1}{c}{$\Delta\varpi$ [mas]} & 
\multicolumn{1}{c}{$\Delta\mu_{\alpha*}$ [mas~yr$^{-1}$]} & 
\multicolumn{1}{c}{$\Delta\mu_\delta$ [mas~yr$^{-1}$]} &
\multicolumn{1}{l}{Reference (VLBI)} \\
\noalign{\smallskip}
\hline
\noalign{\smallskip}
S Persei & $11093$ & $+4.969 \pm 8.474$ & $-10.626 \pm 8.497$ & $-0.001 \pm 0.720$ & $-0.153 \pm 0.242$ & $-0.225 \pm 0.211$ & \citetads{2010ApJ...721..267A} \\
LS I +61 303 & $12469$ & $-28.882 \pm 6.726$ & $+22.324 \pm 7.611$ & $+0.188 \pm 0.650$ & $-1.322 \pm 0.373$ & $+1.133 \pm 0.383$ & \citetads{1999AA...344.1014L} \\
HII 174 & 1803-0008-1 &  &  & $-0.024 \pm 0.302$ & $-0.130 \pm 0.918$ & $-0.395 \pm 0.492$ & \citetads{2014Sci...345.1029M} \\
HD 283447 & $19762$ & $+8.447 \pm 0.759$ & $-36.971 \pm 0.482$ & $-0.460 \pm 0.495$ & $-8.998 \pm 0.170$ & $+0.011 \pm 0.111$ & \citetads{2012ApJ...747...18T} \\
T Tau & $20390$ & $+231.577 \pm 1.280$ & $+514.096 \pm 0.545$ & $+0.291 \pm 0.309$ & $+7.600 \pm 0.169$ & $-12.339 \pm 0.095$ & \citetads{2007ApJ...671..546L} \\
T Lep & $23636$ & & & $-0.843 \pm 0.759$ & $-4.395 \pm 0.502$ & $+1.668 \pm 0.791$ & \citetads{2014PASJ...66..101N} \\
3C273 & $60936$ & $+0.912 \pm 1.474$ & $+0.819 \pm 1.961$ & $-0.140 \pm 0.377$ & $-0.384 \pm 0.443$ & $+0.111 \pm 0.288$ & \citetads{2009ITN....35....1M} \\
$\sigma^2$ CrB & $79607$ & $-12.772 \pm 1.142$ & $-5.524 \pm 1.491$ & $+0.104 \pm 0.913$ & $-0.471 \pm 0.065$ & $-0.009 \pm 0.086$ & \citetads{1999AA...344.1014L} \\
Cyg X-1 & $98298$ & $-0.378 \pm 0.444$ & $-0.648 \pm 0.732$ & $-0.310 \pm 0.250$ & $+0.015\pm 0.086$ & $+0.034 \pm 0.139$ & \citetads{2011ApJ...742...83R} \\
HD 199178 & $103144$ & $-4.246 \pm 8.859$ & $+6.620 \pm 9.330$ & $+0.358 \pm 0.474$ & $-0.170 \pm 0.409$ & $+0.349 \pm 0.431$ & \citetads{1999AA...344.1014L} \\
AR Lac & $109303$ & $-4.324 \pm 2.945$ & $-0.975 \pm 4.398$ & $-0.332 \pm 0.510$ & $-0.220 \pm 0.127$ & $-0.082 \pm 0.191$ & \citetads{1999AA...344.1014L} \\
IM Peg & $112997$ & $+0.503 \pm 1.029$ & $-0.014 \pm 1.046$ & $-0.039 \pm 0.372$ & $+0.001 \pm 0.093$ & $-0.110 \pm 0.096$ & \citetads{2015CQGra..32v4021B} \\
PZ Cas & $117078$ & $-29.855 \pm 3.349$ & $-9.348 \pm 4.012$ & $+0.438 \pm 0.558$ & $-0.275 \pm 0.205$ & $-0.236 \pm 0.303$ & \citetads{2013ApJ...774..107K} \\
\noalign{\smallskip}
\hline
\end{tabular}
\tablefoot{$\Delta\alpha * = \Delta\alpha\cos\delta$. 
Differences are computed as TGAS minus VLBI at epoch J2015.0.
The uncertainties are the quadratically combined standard errors of the TGAS and VLBI quantities.  
The quoted references do not give the VLBI positions for HII~174 and T~Lep. 
}
\end{table*}

\subsection{Stars observed by VLBI}
\label{sec:validation_radiostars}

TGAS results for a small number of sources observed by VLBI are summarised in
Table~\ref{tab:vlbi}. The table includes 12 Galactic sources and one quasar (3C273)
for which the quality criteria in Eq.~(\ref{eq:formal1}) were satisfied.
The VLBI data were propagated from the original epoch 
of the published data to J2015.0, using rigorous formulae for uniform space motion
\citepads{2014A&A...570A..62B}. Radial velocities needed for the propagation were
taken from the SIMBAD database \citepads{2000A&AS..143....9W}. The table gives 
differences in the astrometric parameters computed in the sense TGAS value 
minus propagated VLBI value.  
The quoted uncertainties ($\pm 1\sigma)$ are the quadratically combined
uncertainties from TGAS and VLBI.

The parallax differences are less than two standard deviations in all cases, and
less than one standard deviation for 11 out of the 13 sources. The weighted mean
difference for all 13 sources is $\Delta\varpi=-0.060\pm 0.116$~mas.

In position or proper motion there are significant differences (exceeding two standard 
deviations) for 6 out of the 13 sources. At least four of the objects, namely the young
stellar systems T~Tau and HD 283447 (V773~Tau), and the RS CVn binary 
$\sigma^2$~CrB, are known to have distant tertiary components causing 
non-linear proper motions of the inner binaries that contain the radio source 
(\citeads{2006A&A...457L...9D}, \citeads{2012ApJ...747...18T}, 
\citeads{1999AA...344.1014L}, \citeads{2011ApJ...737..104P}).
This orbital motion can likely explain the discrepant proper motions for these objects 
and the large differences between their TGAS positions and the linearly extrapolated 
VLBI positions. A similar explanation may exist for the X-ray binary LS~I~+61~303.
For the Mira star T~Lep and the red supergiant PZ~Cas, VLBI observations show multiple 
maser spots at separations up to $\sim$100~mas and internal kinematics between the
spots of a few mas~yr$^{-1}$ (\citeads{2014PASJ...66..101N}, \citeads{2013ApJ...774..107K}). 
These features could explain the position and proper motion differences 
seen in Table~\ref{tab:vlbi} for these two objects.

\section{Residual statistics}
\label{app:residuals}

In this appendix we quantify the overall scatter of the AL residuals of the primary
astrometric solution, and discuss some specific contributions to the scatter, 
i.e.\ chromaticity, high-frequency attitude noise, and micro-clanks.

\subsection{Overall scatter}
\label{sec:scatter}

\begin{figure}[t] 
\resizebox{\hsize}{!}{\includegraphics{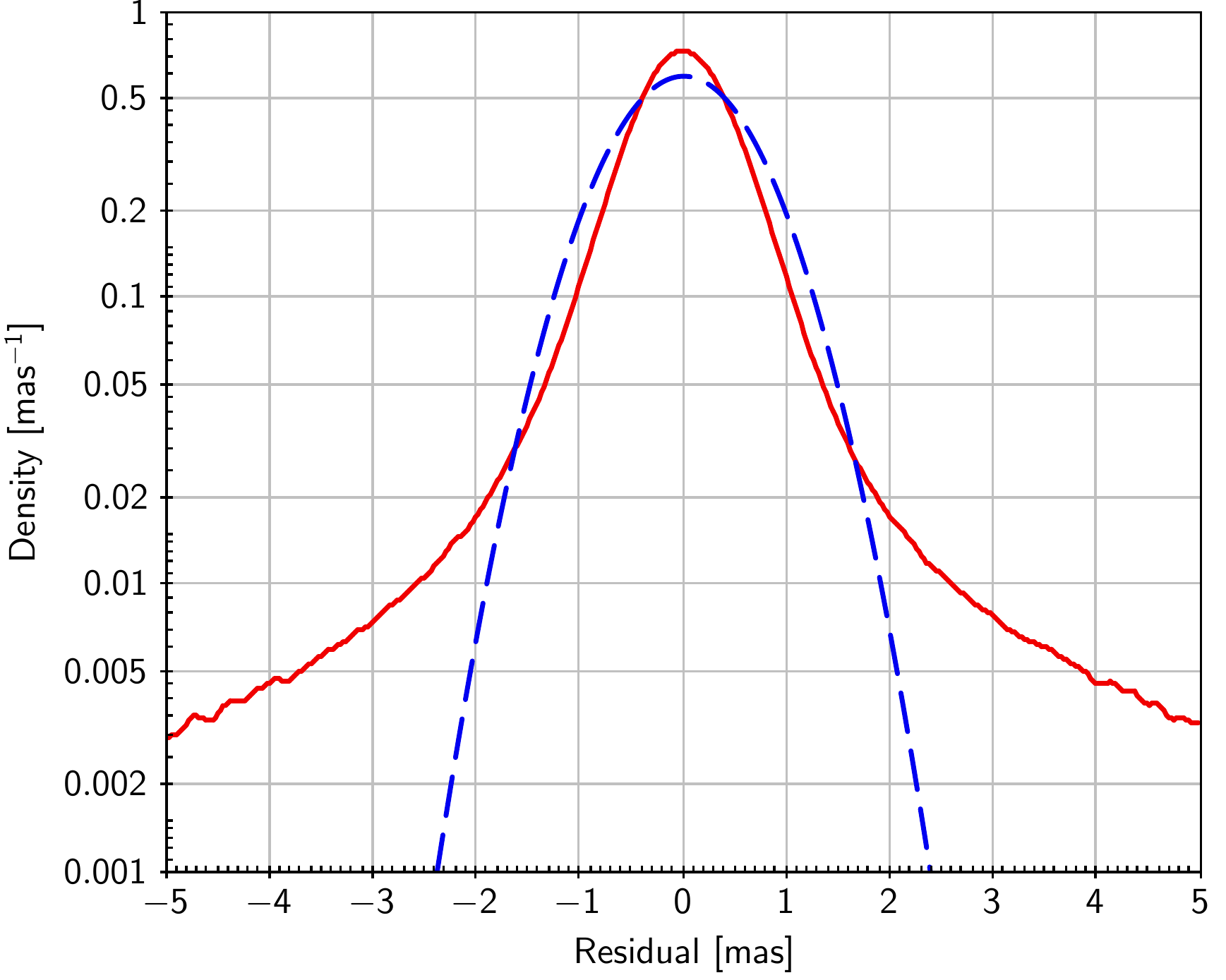}} 
\caption{Solid red curve: the probability density (see footnote~\ref{footnote:KDE}) of the AL 
residuals for individual CCD observations in the primary solution. Only good observations
(see footnote~\ref{footnote:downweighting})
in a representative five-day interval are included. Dashed blue curve: the normal probability density 
function with standard deviation 0.667~mas, equal to the RSE of the residuals.}
\label{fig:statResHist}
\end{figure}

Figure~\ref{fig:statResHist} shows the overall distribution of AL residuals in the 
primary (TGAS) solution. The width, as measured by the RSE, is 0.667~mas. The distribution 
has a Gaussian-like core with very broad wings. Although only the residuals of good 
observations (see footnote~\ref{footnote:downweighting}) were used to construct the diagram, 
some large residuals are included because they have a large excess source noise or excess 
attitude noise. This explains the presence of the broad wings in Fig.~\ref{fig:statResHist}.

\begin{figure}[t] 
\resizebox{\hsize}{!}{\includegraphics{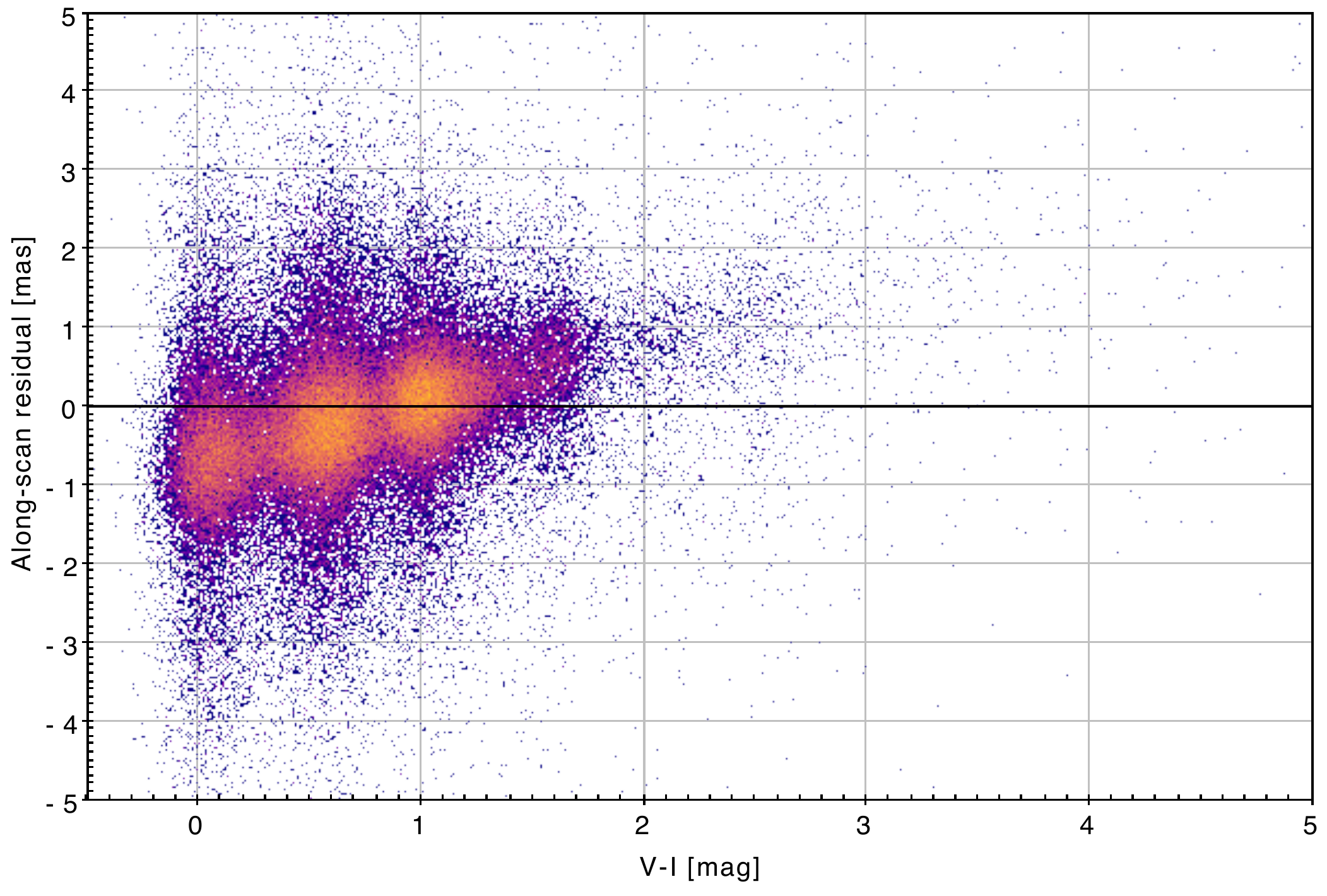}} 
\caption{Along-scan residuals for individual CCD observations in the primary solution
as function of colour index $V-I$. Only good observations of Hipparcos stars, made
at one particular CCD, are included (row 1 of AF1 in the following field of view; 
see Fig.~\ref{fig:fpa}). Colour indices are taken from the Hipparcos catalogue.}
\label{fig:statResVmI}
\end{figure}

\subsection{Chromaticity}
\label{sec:chrom}

Although the AF of the Gaia telescope does not use any refractive optics, 
the precise location of the centroid of an unresolved stellar image depends on the spectral 
energy distribution of the star. This phenomenon, known as chromaticity, is the result of
a complex interaction of the wavelength-dependent diffraction with asymmetric optical
aberrations, pixel geometry, and the centroiding algorithm. Pre-launch simulations, 
assuming realistic wavefront errors, predicted differential shifts of several mas for a typical 
range of stellar spectral classes \citepads{2006A&A...449..827B}. 
To a first approximation the shift is predicted to be a linear 
function of the effective wavenumber $\nu_\text{eff}$ \citep{2016GaiaF},
which in turn mainly depends on the overall spectral energy distribution in the optical, 
as given e.g.\ by the $V-I$ colour index. The chromaticity $\chi$, measured by the shift
in mas per magnitude of $V-I$, is expected to vary across the field of view, and to be 
different in the preceding and following fields. To the extent that the optical aberrations
vary with time, chromaticity will also be a function of time.

A plot of the AL astrometric residuals versus $V-I$ for the Hipparcos subset (using
colour indices from the Hipparcos catalogue) reveals significant chromaticity, as exemplified
by Fig.~\ref{fig:statResVmI}. Different CCD/field-of-view combinations give slopes roughly
in the range $\left|\,\chi\,\right|\lesssim 1$~mas~mag$^{-1}$. Some fraction of this shift propagates into 
the astrometric parameters of a source, depending on the number and geometry of the scans
across the source. Simulated TGAS runs show that the resulting shift in parallax is of
the order of $\pm 0.2$~mas~mag$^{-1}$, with a strong dependence on position. This
effect is more directly studied by introducing colour-dependent calibration terms,
as described in Appendix~\ref{sec:valColour}.

\begin{figure} 
\resizebox{\hsize}{!}{\includegraphics{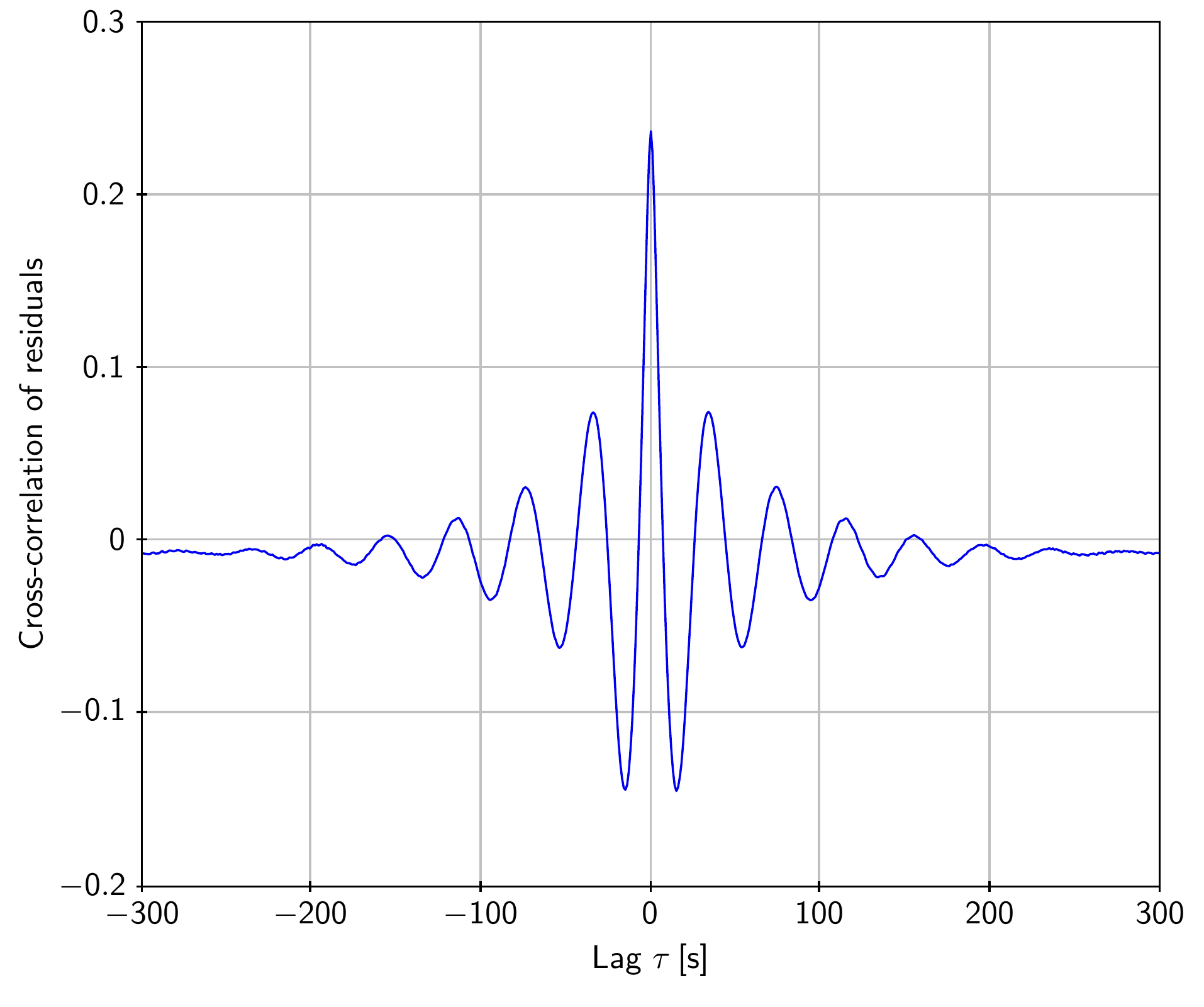}} 
\caption{Cross-correlation of the astrometric residuals in the preceding and following
fields of view.}
\label{fig:statCCF}
\end{figure}

\subsection{Correlations}
\label{sec:correlations}

The top panel of Fig.~\ref{fig:statResTime} shows, for a short stretch of observations,
the AL residuals in the baseline primary solution.
The wiggles, having an amplitude of $\simeq 0.5$~mas, 
are representative for the overall quality of the attitude fit. 
This indicates that much of the residual variance seen in Fig.~\ref{fig:statResHist} comes 
from AL attitude irregularities that are too rapid to be modelled by the attitude spline with
a 30~s knot interval (Appendix~\ref{sec:valKnots}). The resulting modelling errors
introduce temporal correlations in the observations on timescales up to several minutes,
which in turn propagate into spatial correlations among the astrometric parameters.

An analogous situation for the Hipparcos data was analysed by \citetads{2007A&A...474..653V},
who demonstrated how a careful modelling of the attitude can reduce not only the total 
size of the modelling errors but also the temporal (and hence spatial) correlations by a
large factor. For Gaia, this will be remedied in future data releases. In the meantime, it
is important to characterise the correlations that exist in Gaia DR1.

Figure~\ref{fig:statCCF} is a plot of the cross-correlation coefficient between the AL residuals 
in the two fields of view, calculated as
\begin{equation}\label{eq:statCCF}
\rho_{pf}(\tau)=\frac{\overline{p_i f_j}}{\sigma_p\sigma_f} \, ,
\end{equation}
where $p_i$ and $f_j$ are the residuals of observations in the preceding and following field
of view, respectively, and the average is taken over all residual pairs for which the time
difference $t_i-t_j$ is $\tau$ (to the nearest second). The normalisation factor is 
$\sigma_p\sigma_f=0.38$~mas$^2$, from which the cross-covariance can be recovered.%
\footnote{The quantity $\overline{p_i f_j}$ equals the cross-covariance of the residuals since 
the average residual in each field of view is practically zero. Robust estimates of this and
the denominator of Eq.~(\ref{eq:statCCF}) were obtained by binning the residuals of 
good observations (see footnote~\ref{footnote:downweighting}) in 1~s bins and 
rejecting bins for which the average residual exceeded 10~mas. The averages in 
$\overline{p_i f_j}$ were taken over the accepted bins. The residual variances 
$\sigma_p^2$ and $\sigma_f^2$ were computed from the sums of the squared residuals 
in the accepted bins, and therefore represent the dispersion of individual residuals, not 
of the mean residual per bin. 
-- Using the cross-correlation between the two fields of view, rather than the autocorrelation 
in either field, eliminates the many strong spikes caused by the highly correlated errors 
of a given star crossing the nine successive CCDs in the AF. These spikes, separated by 4.85~s 
(the time between successive CCD observations), form a triangular comb function for lags 
up to 38.8~s. They have other causes than the attitude modelling errors (e.g.\ source and 
calibration modelling errors), and it is therefore reasonable to disregard them in this analysis. 
On the other hand, the AL attitude error is practically the same in the two fields 
of view and therefore contributes to the cross-correlation.}

The cross-correlation function exhibits the characteristic pattern expected from
modelling errors in the attitude spline. Given that the knot separation is 30~s,
it may seem surprising that the zero-crossings have a typical separation of
only about 20~s; however, this is expected for an attitude spline fit using 
a small value of the regularisation parameter $\lambda$ 
(Sect.~\ref{sec:attitude}; see \citeads{2012A&A...543A..15H}).
The height of the central peak suggests that at least a quarter of the total
residual variance comes from attitude modelling errors. The actual fraction
may be higher (see below).

For lags of several minutes, the cross-correlation function in Fig.~\ref{fig:statCCF} 
settles at a slightly negative value, corresponding to a cross-covariance of
$-2600~\mu$as$^2$. This is caused by basic-angle variations that have
not been corrected based on the BAM data, nor accounted for in the calibration. 
Since the AL attitude is defined by the mean pointing of the two viewing
directions, residuals caused by basic-angle variations are anti-correlated 
between the fields. The rms amplitude of these (as yet) uncalibrated 
basic-angle variations is $(2\times 2600)^{1/2}\simeq 72~\mu$as. 
(For much bigger lags of several hours 
the cross-covariance gradually goes to zero, except around values related
to the spin period and basic angle.)

The temporal correlations shown in Fig.~\ref{fig:statCCF} are significant for
delays up to $\sim$2~min, corresponding to $2^\circ$ on the sky. Thus,
spatial correlations in the astrometric parameters can be expected for stars
that are separated by angles up to a few degrees. The extent to which the
temporal correlations propagate into spatial correlations depends in a 
complex way on the geometry of the scans and how much overlap there
is between the scans of the different stars. A rough indication is given by 
the ``coincidence fraction'' introduced by van~Leeuwen
(\citeyearads{1999ASPC..167...52V}, \citeyearads{2007A&A...474..653V}) in the
context of Hipparcos data. For Hipparcos observations the coincidence 
fraction drops rapidly from close to 1 at very small separations to
0.5 at separations of ${\sim}1^\circ$, and then more slowly. The same 
should be the case for Gaia observations, as the scanning laws and field 
sizes of the two missions are rather similar. Thus we conclude that the
spatial correlations in the Gaia DR1 astrometry, e.g.\ for parallaxes in a 
stellar cluster, may be significant (perhaps $\sim$0.25) at separations
up to ${\sim}1^\circ$, but much smaller on longer scales. In this context
it should be remembered that Fig.~\ref{fig:statCCF}, being derived from the 
residuals of the solution, inevitably underestimates the cross-correlation of the actual 
attitude modelling errors, which to some extent are absorbed by the source 
parameters. Other modelling errors may create astrometric errors that are
correlated over much longer angular scales (Appendix~\ref{app:validation}).

\begin{figure*} 
\resizebox{\hsize}{!}{%
\hspace{12pt}\includegraphics{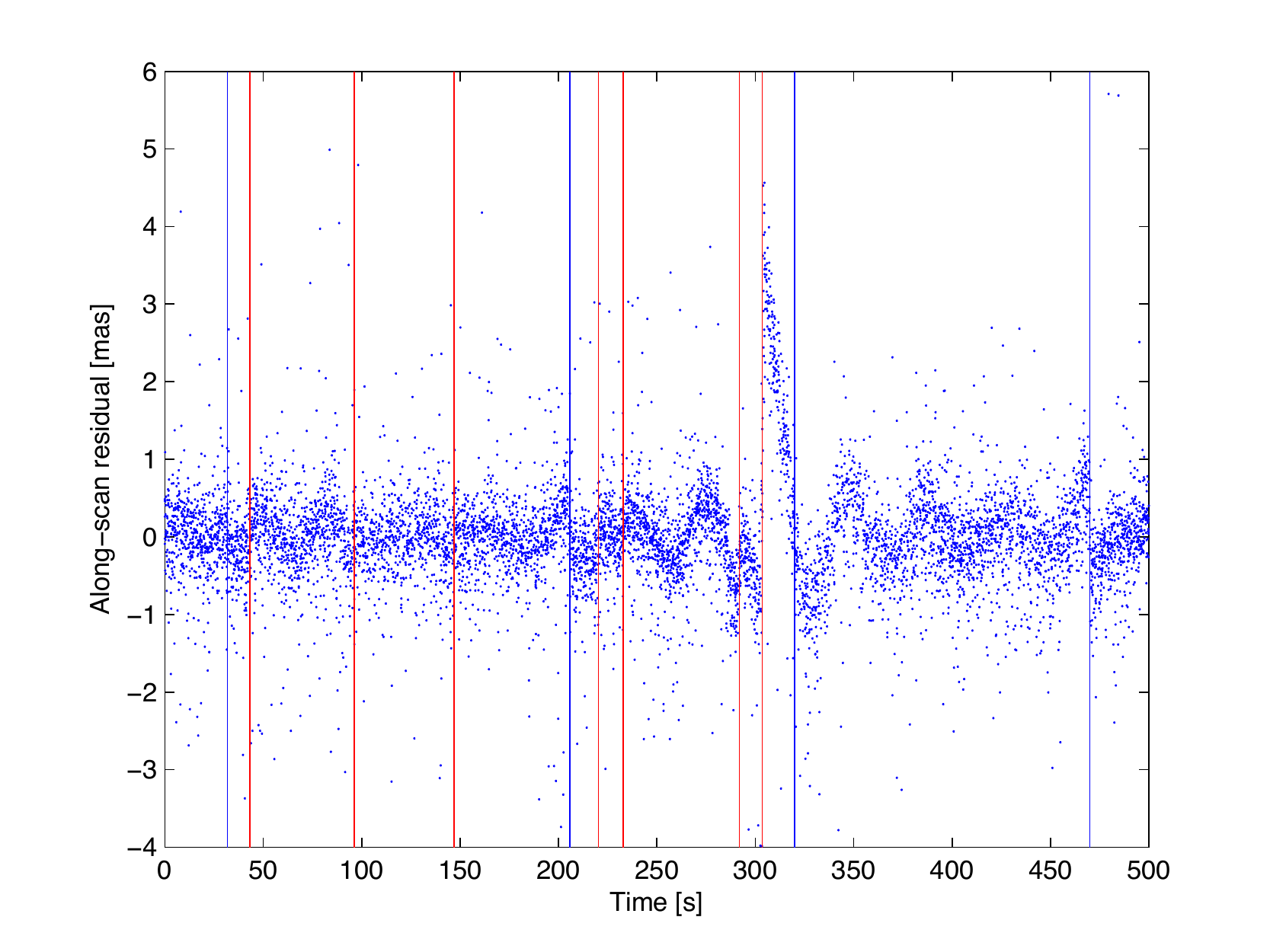} 
\hspace{12pt}\includegraphics{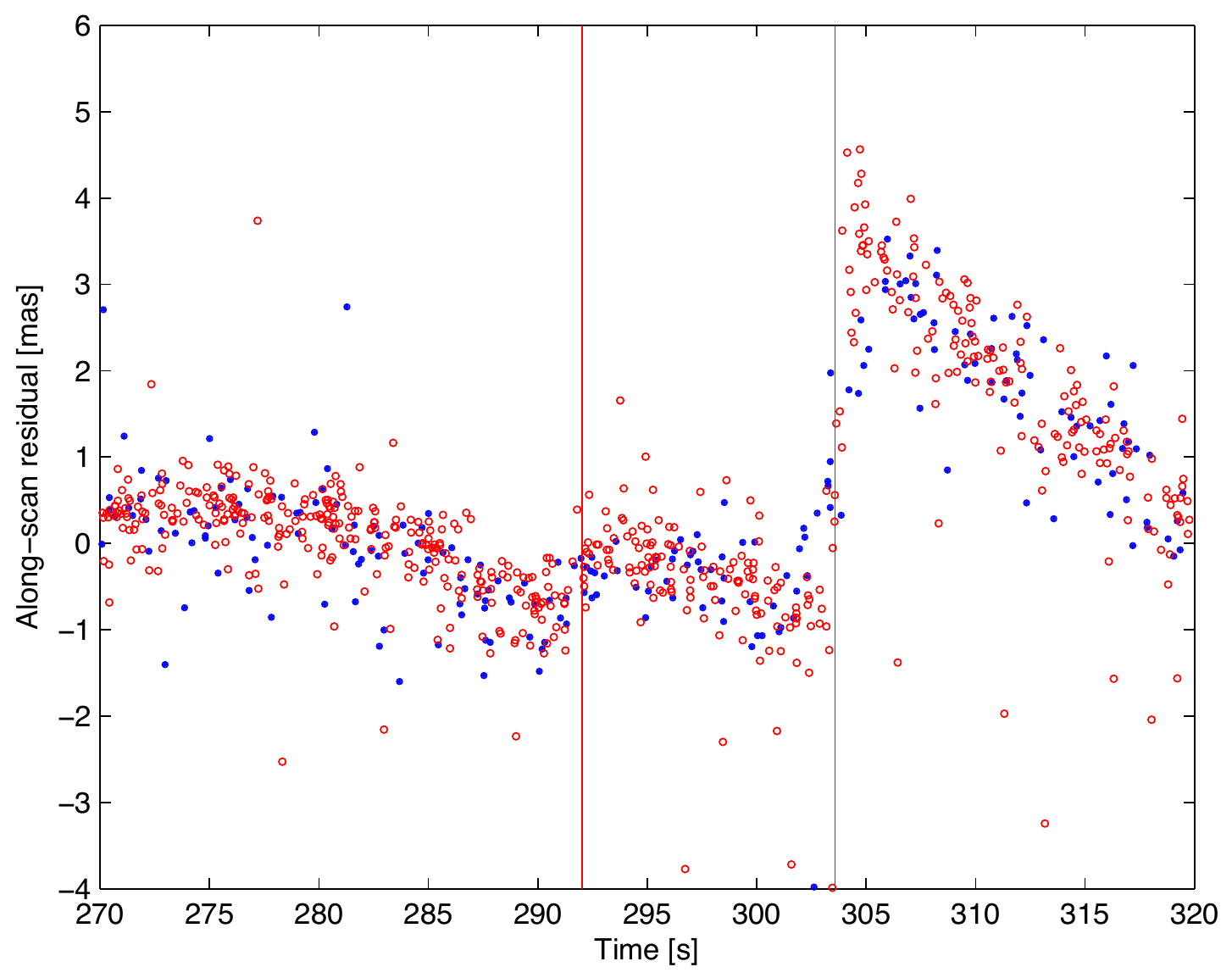}
} 
\resizebox{\hsize}{!}{%
\includegraphics{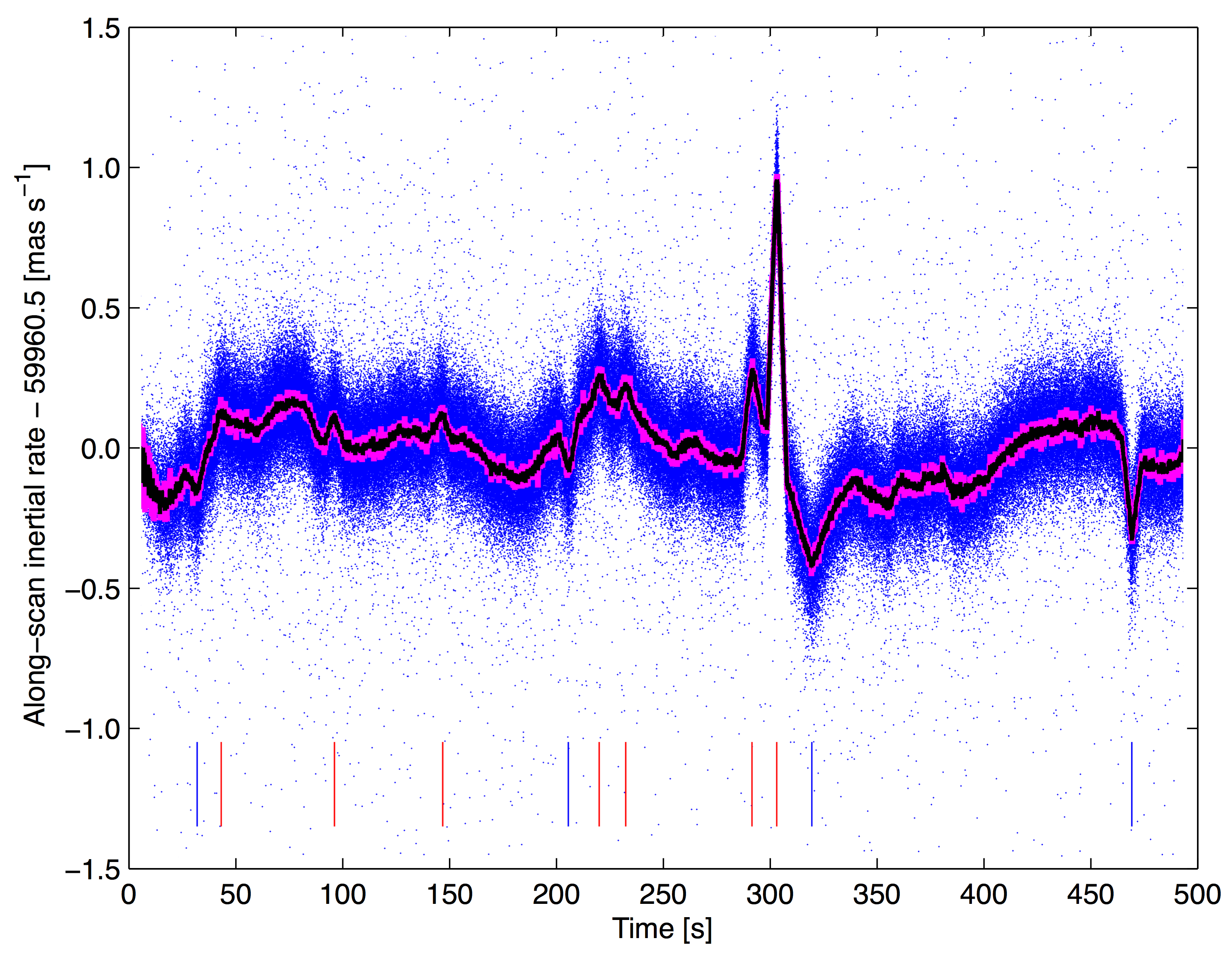} 
\includegraphics{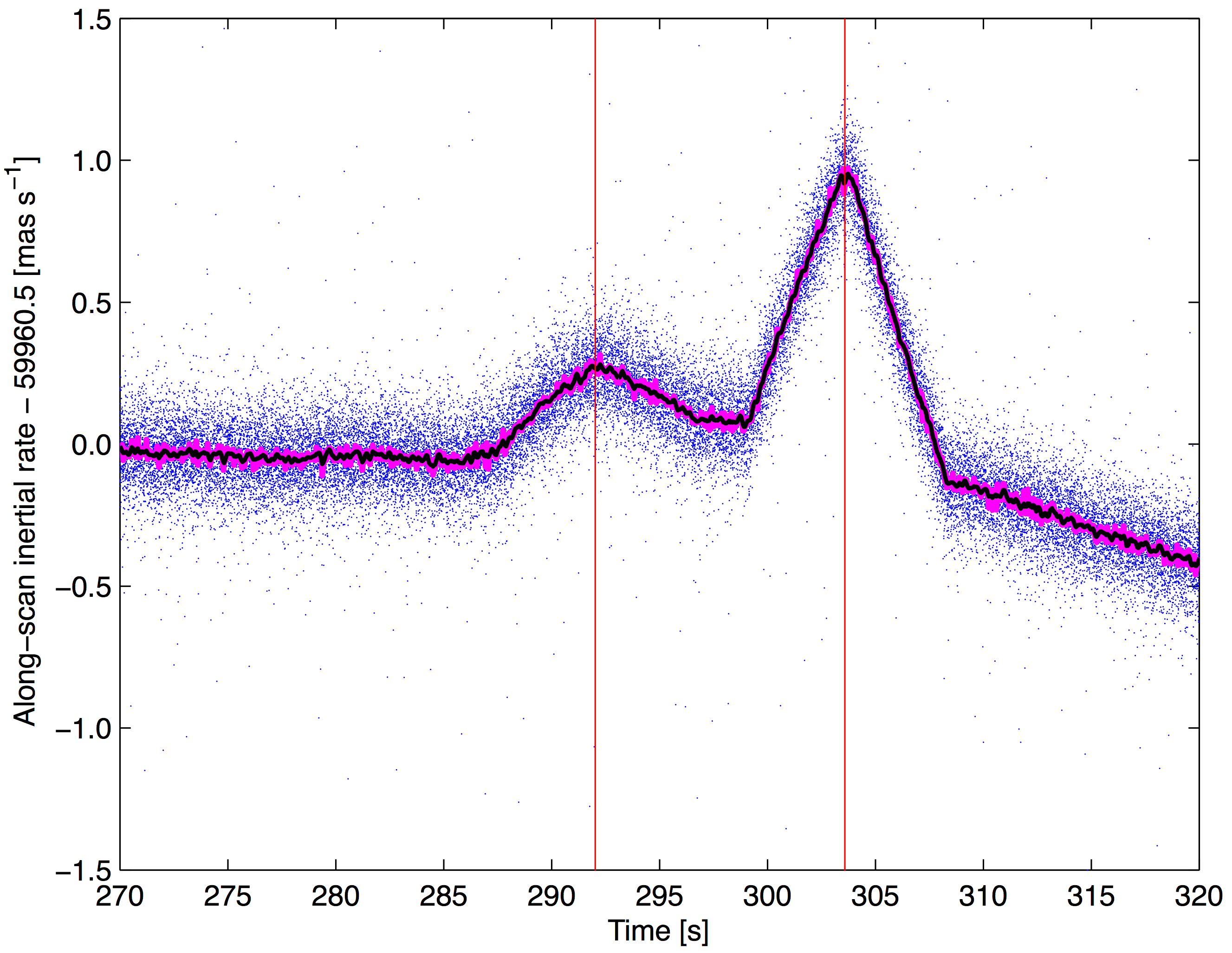}
} 
\caption{Top panels: examples of micro-clanks in the AL 
TGAS residuals during a time interval of 500~s (left) and a subinterval of 50~s (right).
The vertical lines show the times of the micro-clanks as estimated from 
rate data (red/blue for positive/negative jumps). The attitude knot interval 
is 30~s, which sets the typical period of the wiggles seen between the micro-clanks. 
In the top right panel, filled blue dots are ungated observations and 
open red circles are gated observations. The shorter integration time of the gated
observations gives a steeper slope of the residuals around the micro-clanks, as
can be seen, for example, by comparing the open red circles and filled blue
dots around the jump at 303~s in the top right panel. 
Bottom panels: AL inertial angular rates estimated from successive 
ungated CCD observations of a much larger number of faint ($\sim$15~mag) stars.
Individual rate estimates are shown as blue dots. The thick magenta curve shows 
the median rate estimate in each bin of 0.045~s duration. The black curve is a 5-point 
running triangular mean of the medians, added to better show the rate variations.
The vertical lines at the bottom of the diagram show the estimated times of the
micro-clanks detected by means of a simple matched filter.}
\label{fig:statClank}
\end{figure*}

\subsection{Micro-clanks}
\label{sec:clanks}

Soon after Gaia's launch, small rapid rotation rate changes of the spacecraft 
were discovered at a frequency of about one every few minutes, with amplitudes 
up to a few mas~s$^{-1}$. These were initially interpreted as 
micrometeoroid hits. However, based on more and better data, the observed 
rotation rate excursions were later identified to be almost all caused by sudden small 
structural changes within the spacecraft \citep{2016GaiaP}. 
They are here referred to as micro-clanks. Their physical origins are still unclear.
Micro-clanks are seen both in the AL and AC directions, and
they often repeat quasi-periodically with the spin period of the satellite. 
In the AL direction, the vast majority of them affect both fields of view 
equally and simultaneously, with no discernible effect in the BAM data, which 
suggests an origin outside of the optical instrument. For a small fraction of them, however, 
the times coincide with jumps in the BAM fringe-position data, which may be
different for the two fields of view or only seen in one of the fields. These 
micro-clanks apparently originate within the mechanical structure of the optics.

Whatever the origin may be, the effect of a micro-clank is a quasi-instantaneous discontinuity 
(on timescales $\ll 1$~s) in the physical attitude angle, while the physical attitude
rate is practically the same on either side of the discontinuity. However, since the 
physical attitude angle is not directly observable, but only a moving average over
the CCD integration time (the so-called effective attitude; \citeads{2013A&A...551A..19R}), 
the effect as seen in 
the astrometric residuals is not instantaneous but linear over the CCD integration
time, which is 4.42~s for ungated observations, and shorter for gated observations.
The left panels of Fig.~\ref{fig:statClank} show the AL TGAS residuals
in a 500~s time interval, with clear evidence of several micro-clanks. The vertical
lines show the times of all the micro-clanks detected in rate data (see below)
for this interval. In the zoomed-in plot, the effect of the CCD integration time
is clearly seen.

The micro-clanks in Gaia data were first seen in AL rate estimates, computed 
from the precise time difference of successive CCD observations, separated by 
approximately 4.85~s, of the same star. Rate estimates (both along and across scan) are very much 
easier to compute than the astrometric residuals, as they are purely differential and 
hence independent of source parameters and less sensitive to calibration errors;
they can also easily be computed for many more stars. The bottom panels in 
Fig.~\ref{fig:statClank} show early AL rate estimates for the same time
intervals as in the top panels. These were computed from ungated observations of
much fainter stars than in TGAS. The CCD integration time of 4.42~s and the time 
difference of 4.85~s between successive observations result in apparent rate 
excursions around the time of each micro-clank, with a completely predictable 
and very characteristic trapezoidal (almost triangular) profile. 
The time and amplitude of the micro-clank
can be estimated very precisely from such rate data, essentially by using a matched
filter. In this 500~s interval, no less than 11 micro-clanks were thus detected, as
indicated by the vertical lines. The astrometric residuals clearly confirm at least
ten of them. In a longer time interval of 72~min, some 120 micro-clanks were
detected with amplitudes corresponding to AL attitude discontinuities
in the range from 0.3~mas to 4.3~mas. The largest micro-clank in this interval 
is the one seen in Fig.~\ref{fig:statClank} at 303~s. It appears that they 
rarely get much bigger than this. In all the time intervals that have so far been 
investigated in detail, their frequency was similar to the numbers given above. 

Based on the limited statistics reported above, the effective attitude for
ungated observations is directly
disturbed by micro-clanks for at least $\sim$10\% of the time. If left uncorrected,
the micro-clanks are therefore a major source of attitude noise. In the present
TGAS results they enter as statistical modelling noise. For future astrometric 
solutions they must be largely eliminated. The originally foreseen strategy to handle 
clanks was to insert multiple knots in the attitude spline at the relevant times
(see Sect.~5.2.5 and Appendix~D.4 of the AGIS paper). However, with the observed 
high frequency of micro-clanks, such a procedure would weaken the attitude 
estimation considerably. It is now clear that a far better 
strategy is to apply gate-dependent corrections based on the times and amplitudes 
of micro-clanks, detected and quantified in the rate data. This will be implemented 
in the AGIS pre-processor for future Gaia data releases.

\section{Special validation solutions}
\label{app:validation}

In this appendix we briefly describe some of the special TGAS runs that were
computed in order to test the sensitivity of the baseline TGAS run to 
various modelling assumptions. Since a primary concern is the existence of 
systematic errors in the parallaxes, we focus on characterising how the parallaxes
change with respect to the baseline solution. Parallax differences are always
computed in the sense baseline solution minus special validation solution.  

\subsection{Including colour terms in the calibration}
\label{sec:valColour}

As shown in Appendix~\ref{sec:chrom}, the AL residuals for a particular 
CCD/field of view combination have an approximately linear dependence on the
colour index $V-I$ owing to the chromaticity of the instrument. This effect can therefore
largely be eliminated by including colour-dependent terms in the geometric calibration 
model (Appendix~\ref{sec:calibration_parameters}).%
\footnote{Eventually the chromaticity will be fully taken 
into account by the colour-dependent LSF and PSF calibrations, 
at which point the colour-dependent terms in the geometric calibration 
model, as determined in AGIS, should be negligible. Until that time, these terms serve to 
approximately eliminate the astrometric effects of the chromaticity.}

\begin{figure*} 
\resizebox{\hsize}{!}{
\includegraphics{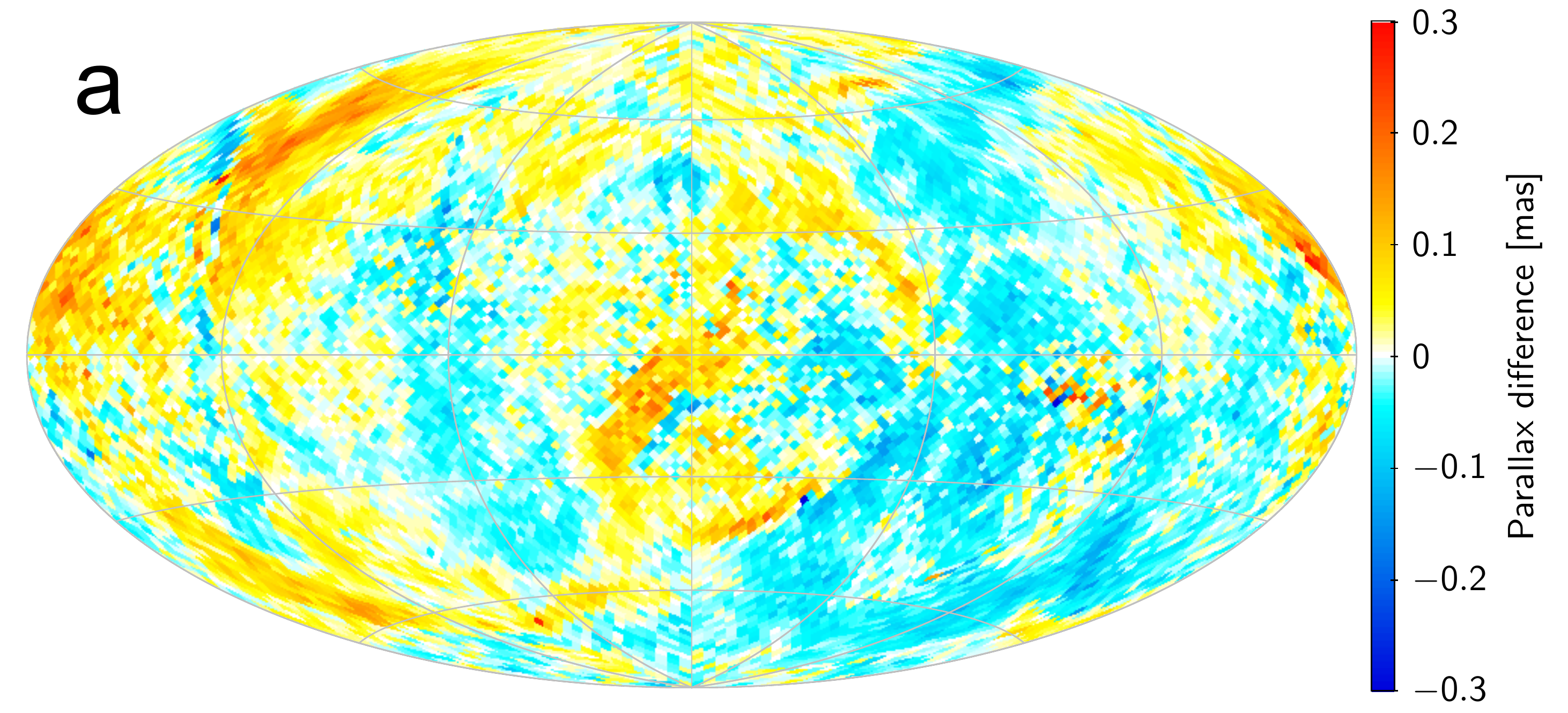}
\includegraphics{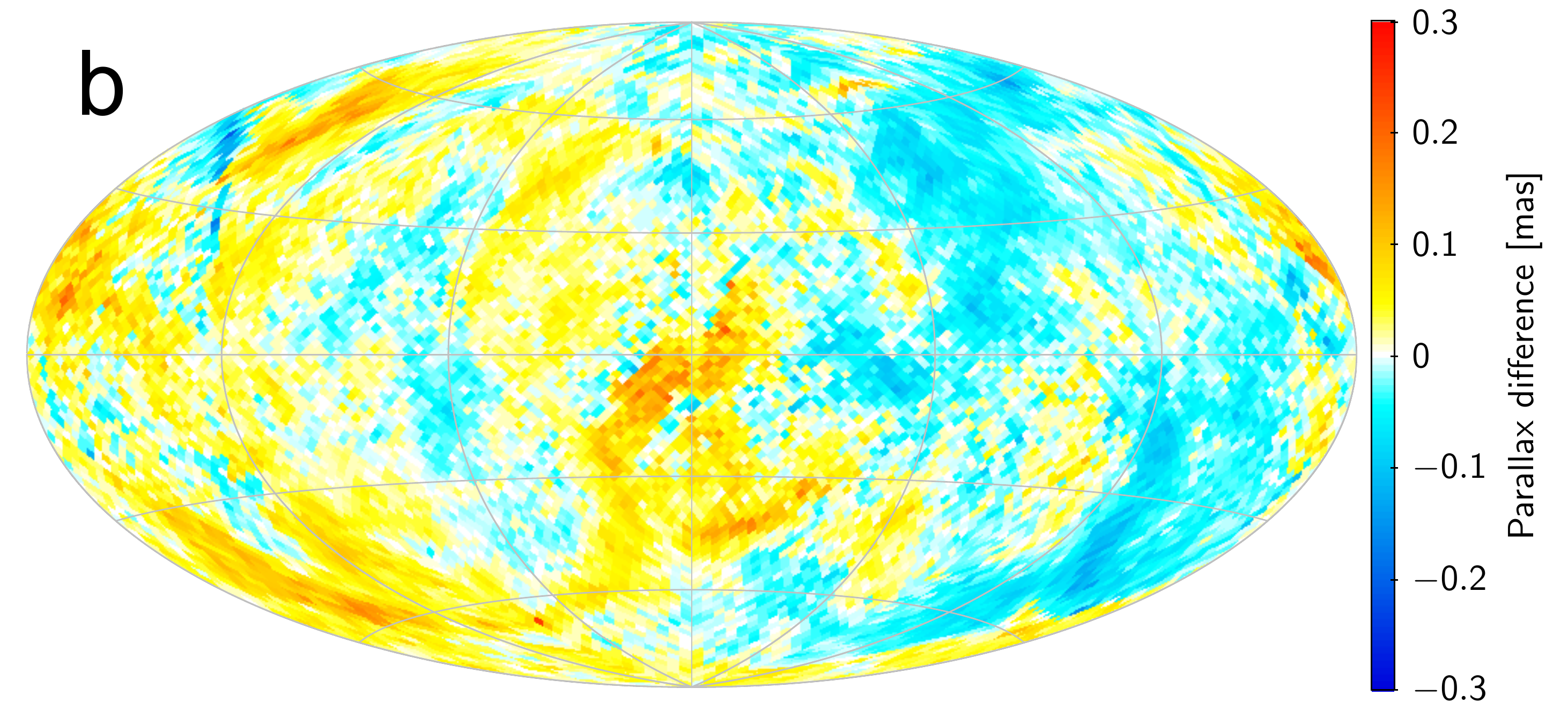}
\includegraphics{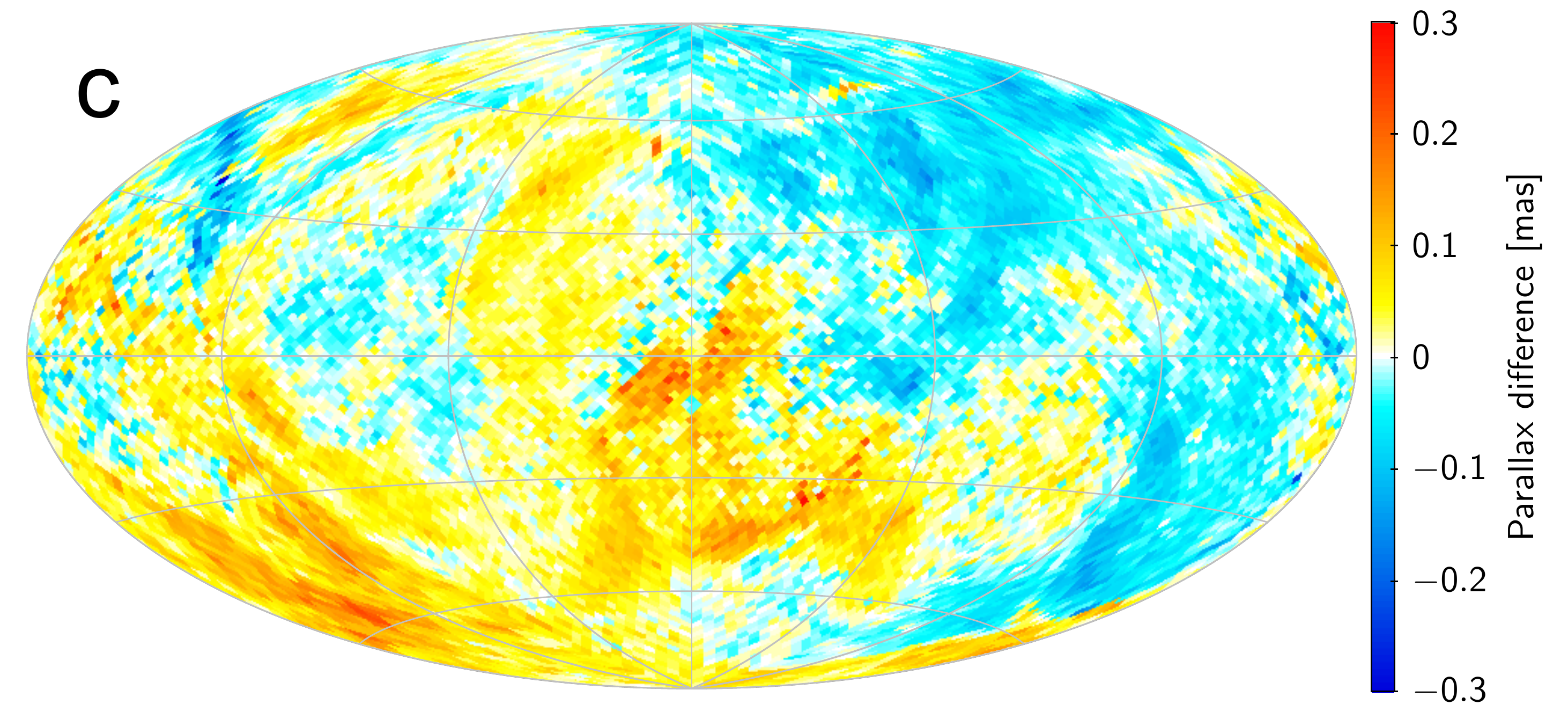}
} 
\caption{Differences in parallax between the baseline primary solution
and a special validation solution with colour terms in the calibration model. 
({\bf a}) Blue sources (colour index $C\le 0.75$). 
({\bf b}) All TGAS sources. 
({\bf c}) Red sources ($C>0.75$).
Median differences are shown in cells of about 3.36~deg$^2$. The maps use an 
Aitoff projection in equatorial (ICRS) coordinates, with origin $\alpha=\delta=0$ 
at the centre and $\alpha$ increasing from right to left.}
\label{fig:valColour}
\end{figure*}

\begin{figure*} 
\resizebox{\hsize}{!}{
\includegraphics{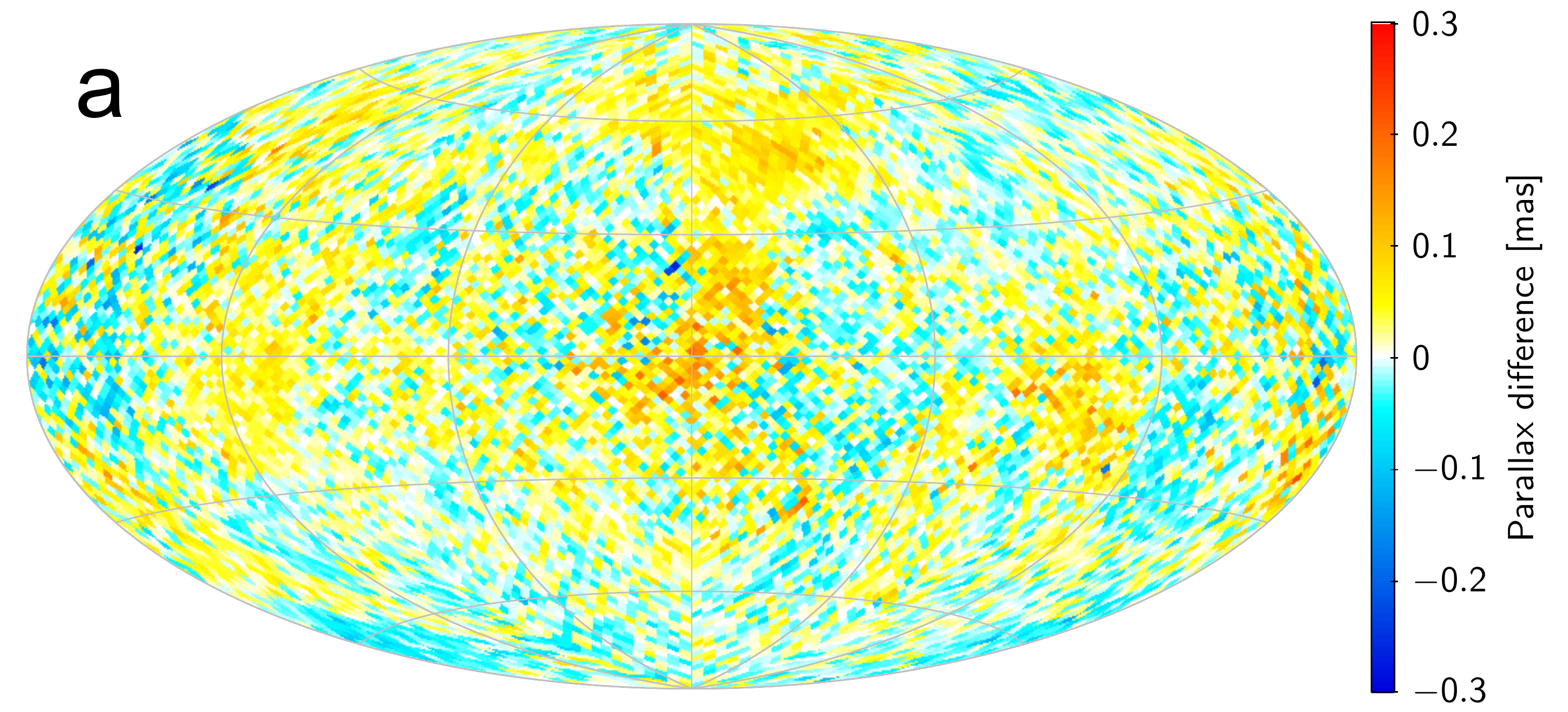} 
\includegraphics{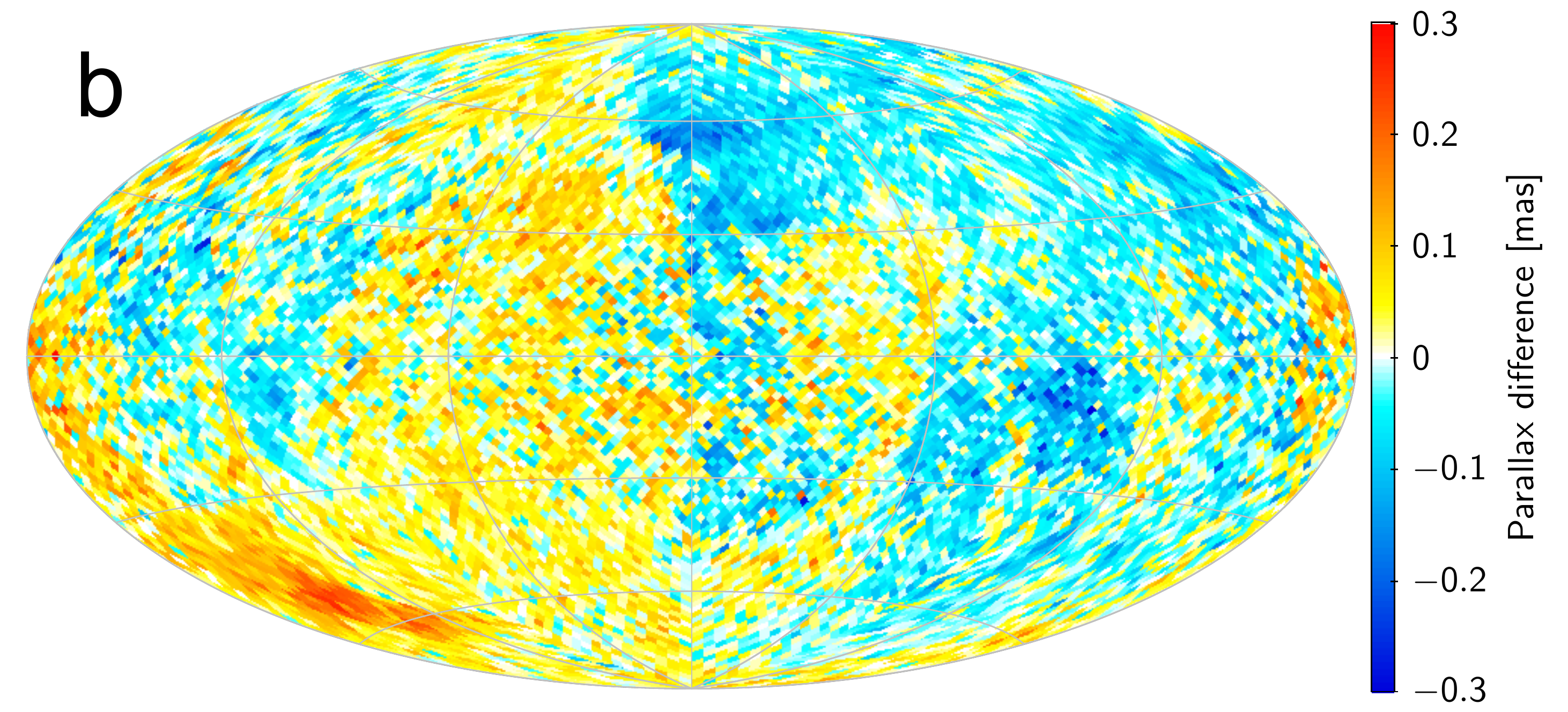} 
\includegraphics{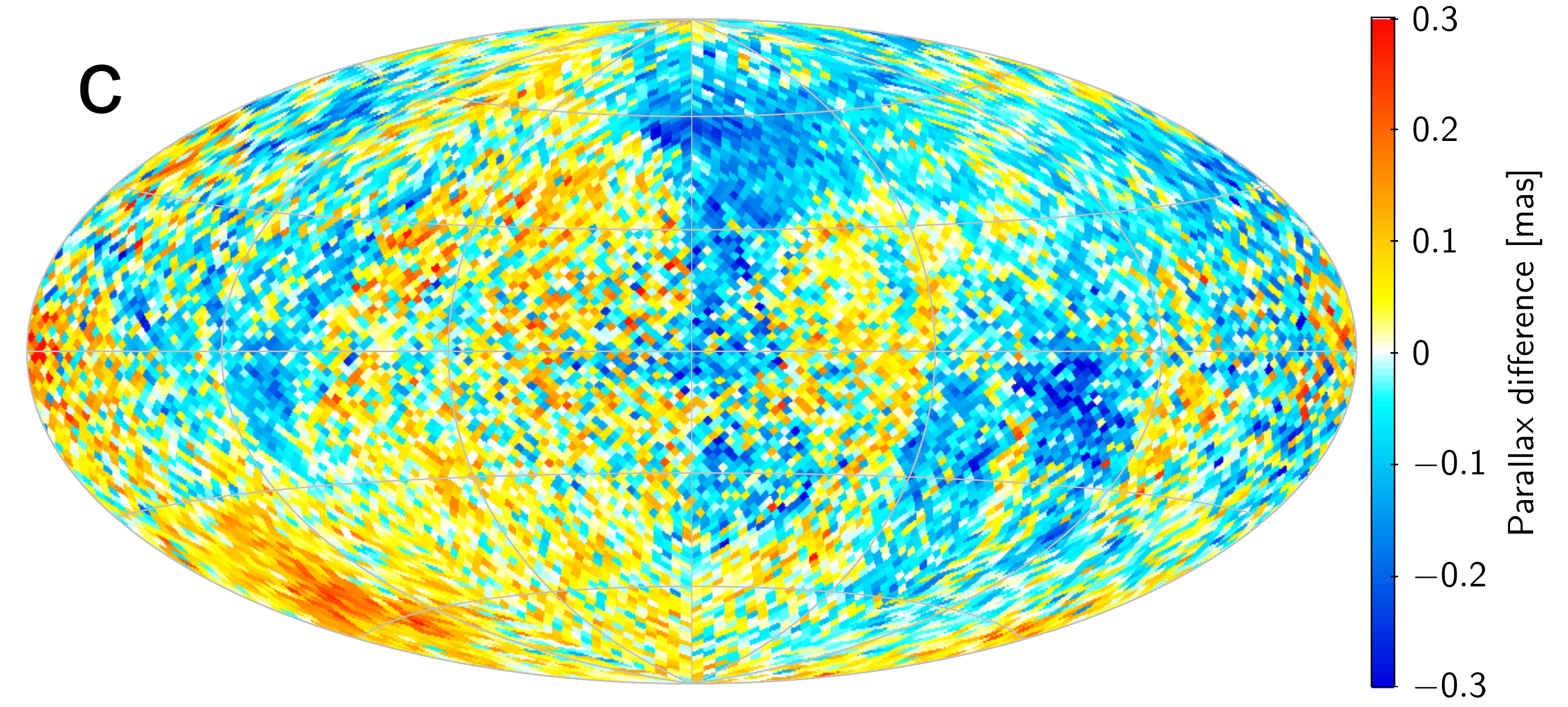} 
} 
\caption{Differences in parallax between the baseline primary solution
and a special validation solution where the data are split into early and late 
observations (within the AF). 
({\bf a}) Baseline minus late. 
({\bf b}) Baseline minus early. 
({\bf c}) Late minus early. 
Median differences are shown in cells of about 3.36~deg$^2$. The maps use an 
Aitoff projection in equatorial (ICRS) coordinates, with origin $\alpha=\delta=0$ 
at the centre and $\alpha$ increasing from right to left.}
\label{fig:valSplit}
\end{figure*}

While Gaia will eventually provide excellent colour information on all observed 
sources through its blue and red photometers 
(BP and RP in Fig.~\ref{fig:fpa}; \citealt{2016GaiaVL}), this information was not 
available at the time when the TGAS baseline and validation solutions were computed.
For the present test it was therefore necessary to compile colour information from
available external sources. For most Hipparcos stars, $V-I$ from the Hipparcos catalogue could
be used. For many \mbox{Tycho-2} stars, provisional (uncalibrated) BP and RP magnitudes 
were available, while for others 2MASS photometry ($J{-}K_s$) or Tycho colours ($B_T{-}V_T$) 
had to be used. In all cases, linear transformations were applied to put the resulting colour index 
(hereafter denoted $C$) on approximately the same scale as $V-I$. While the resulting
colours are thus of extremely inhomogeneous quality, and the applied transformations 
often quite uncertain, this $C$ is still useful for a statistical evaluation of the chromaticity. 
The median $C$ is close to 0.75~mag, roughly corresponding to a solar-type star.

For this validation solution, the geometric AL calibration model in 
Eq.~(\ref{eq:cal1}) was augmented with terms
\begin{equation}
\eta_{fngw}(\mu,\, t,\, C) = \cdots + \chi_{fnwj}(C-0.75) \, ,
\end{equation}
where $\chi_{fnwj}$ are chromaticity parameters (Appendix~\ref{sec:chrom})
depending on the field of view ($f$), CCD ($n$), window class ($w$), and time interval~($j$).
The time resolution is the same as for the large-scale AL calibration, 
i.e.\ typically 3~days. (The coefficient of the chromaticity is offset 
by 0.75~mag to reduce correlations among the calibration parameters. The choice
of offset is in principle arbitrary and does not affect the astrometric part of the solution, 
but simply means that the attitude and non-chromatic calibration parameters refer 
to sources of colour index 0.75.)

Figure~\ref{fig:valColour} shows median parallax differences (baseline \text{minus}
solution with colour terms) for three source selections: 
``blue'' (1.039~million sources with colour index $C\le 0.75$; $\text{med}(C)=0.58$),  
all (2.087~million sources; $\text{med}(C)=0.75$), and  
``red'' (1.047~million sources with $C>0.75$; $\text{med}(C)=1.13$).  
Somewhat surprisingly, the blue and red maps are not vastly different, and in particular they
are not inverted versions of each other, as could be expected if the parallax dependence
on colour, $\text{d}\varpi/\text{d}C$, was simply a function of position. Rather,
it appears that the chromaticity creates time-dependent attitude errors in the baseline solution, which then
propagate into position dependent astrometric errors that are partly independent of the colour.
The median parallax difference is small and practically the same for all three selections 
($+0.004$, $+0.004$, $+0.005$~mas, respectively, for blue, all, and red sources). 
The RSE (0.128, 0.146, and 0.168~mas, respectively) is larger for the red sources, 
which could simply be because the scatter of $C$ is larger for the red sample 
($\text{RSE}(C)=0.33$~mag) than for the blue ($\text{RSE}(C)=0.19$~mag). 

In summary, the overall effect of chromaticity on the TGAS parallaxes is of the order
of $\pm 0.15$~mas (random), with position and colour dependent systematics 
of $\pm 0.1$~mas, although the systematics may exceed $\pm 0.2$~mas in some 
parts ($\sim$1\%) of the sky.

\subsection{Partitioning the data}
\label{sec:valSplit}

During a field of view transit, a source is generally observed on all nine CCD
strips of the astrometric field (AF1 through AF9; see Fig.~\ref{fig:fpa}).
Since the sky mapper data are not used in TGAS, and all nine AF observations provide
AC measurements for bright (window class 0) sources, it is possible to
partition the observations into almost completely independent data sets based on
the CCD strip number. Here we describe the results of two validation solutions, one
using AF1--AF4 (``early'') data, the other using AF5--AF8 (``late'') data. 
Because the two solutions use completely different parts of the focal plane, they
are differently affected by any unmodelled instrumental effect that varies across the 
fields of view. For example: the parallax zero point is known to be tightly correlated 
with the harmonic coefficient $C_{1,0}$ of the basic-angle variations 
(see Appendix~\ref{sec:valVbac}).
However, the relevant basic-angle variations are an average over the actually used 
part of the AF (see Eq.~\ref{eq:DeltaGamma}), and could therefore be slightly 
different for the early and late observations. If this is the case, there might be a zero-point 
shift between the parallaxes in the two solutions.

The early/late partitioning also introduces a time shift of about 19.4~s
between the early and late data, which should make the astrometric effects
of short-range attitude modelling errors rather different in the two solutions.

The following comparisons are limited to some 2.05~million sources    
that have formal (uninflated) parallax standard uncertainties 
$\varsigma_\varpi<1$~mas in both the early and late solutions.
Figure~\ref{fig:valSplit} shows maps of the median parallax differences
between the different solutions. The global median and RSE values are
given in Table~\ref{tab:valSplit}.  
    
\begin{table}
\caption{Statistics of parallax differences between the baseline TGAS
solution ($\varpi_\text{T}$) and the two special validation solutions 
using only early ($\varpi_\text{E}$) and late ($\varpi_\text{L}$)
observations from each field-of-view transit.
\label{tab:valSplit}}
\small
\begin{tabular}{lcc}
\hline\hline
\noalign{\smallskip}
Difference & median & RSE \\
\noalign{\smallskip}
\hline
\noalign{\smallskip}
$\varpi_\text{T}-\varpi_\text{L}$ & $+0.0114\pm 0.0002$~mas & 0.235~mas \\   
$\varpi_\text{T}-\varpi_\text{E}$ & $+0.0041\pm 0.0003$~mas & 0.290~mas \\   
$\varpi_\text{L}-\varpi_\text{E}$ & $-0.0065\pm 0.0004$~mas & 0.463~mas \\   
\noalign{\smallskip}
\hline
\end{tabular}
\end{table}

Although the global parallax zero points of the three solutions differ by less than 0.02~mas,
the difference maps in Fig.~\ref{fig:valSplit} show systematic, position-dependent errors of 
$\pm 0.1$~mas or more. There are distinct similarities between these maps and the chromaticity 
maps in Fig.~\ref{fig:valColour}, which suggests that the much larger chromatic effects could
mask any possible difference in the effective basic-angle variations between the early and late data.
In conclusion these solutions mainly confirm the existence of position-dependent systematics 
at the level of $\pm 0.1$ to 0.2~mas.

\subsection{Solving the basic-angle variations}
\label{sec:valVbac}

As described in Sect.~\ref{sec:bam} and
Appendix~\ref{sec:calibration_bam}, the baseline primary (TGAS)
solution of Gaia DR1 was computed after correcting the observations
for the basic-angle variations as estimated from the BAM data. 
This was done by first fitting the harmonic
model in Eqs.~(\ref{eq:bam})--(\ref{eq:bam1}) to the BAM data, and
then using the fitted model to evaluate the correction as function of
time (or heliotropic phase angle $\Omega$). The BAM uses a dedicated
CCD located outside of the AF (see Fig.~\ref{fig:fpa}). 
Thus, although the BAM measurements are
intrinsically very precise ($<10~\mu$as per measurement), it is
possible that they do not correctly describe the basic-angle
variations relevant to the observations in the AF. This
would be the case, e.g.\ if the AL scale of the astrometric
field (angle between the successive CCD strips) also has a periodic
variation with $\Omega$. The correction relevant for a particular
observation is a combination of the basic-angle correction
$\Delta\Gamma(t)$ and a possible differential variation between the
two fields of view, $\Delta\eta(t,\,\eta,\,\zeta)$, the latter being a function
of both time and the field angles. 
From simplistic optomechanical
considerations it is reasonable to expect that the differential
correction $\Delta\eta(t,\,\eta,\,\zeta)$ is of the order
$O(\eta-\eta_\text{BAM})\sim 0.01$ times smaller than the actual basic-angle variation. 
This would not have been a problem if the basic-angle
variation itself was of the order of 10~$\mu$as as expected from
pre-launch calculations \citep{2016GaiaP}. 
However, since the variations are
now known to be of the order of 1~mas, a possible differential
variation over the field of view becomes a point of concern. A related
issue concerns the representativeness of the BAM data, given that the
laser beams of the BAM interferometer only sample a very small part
of the telescope entrance pupil.

In view of these uncertainties, it is clearly desirable to estimate as
much as possible of the short term ($\lesssim 24$~hr) basic-angle and
differential field of view variations directly from the astrometric
data.  Detailed simulations indicate that this will eventually be
possible, provided that the basic-angle variations are constrained by
suitable models, e.g.\ in the form of a generalisation of
Eqs.~(\ref{eq:bam})--(\ref{eq:bam1}). One possible exception is the
constant part of the $\cos\Omega$ coefficient, corresponding to
$C_{1,0}$ in Eq.~(\ref{eq:bam1}), which is almost completely
degenerate with respect to a global error of the parallax zero point
(\citeads{1992A&A...258...18L}, \citeads{2016A&A...586A..26M}). Further details will be 
discussed elsewhere.

\begin{table}
\caption{Fourier coefficients $C_{k,0}$, $S_{k,0}$ from the BAM data 
and as obtained in a special validation run of the TGAS primary solution. 
\label{tab:valVbacF}}
\small
\begin{tabular}{lrrrlrr}
\hline\hline
\noalign{\smallskip}
& BAM & Solution & \quad\quad &  & BAM & Solution \\
& [$\mu$as] & [$\mu$as] &&& [$\mu$as] & [$\mu$as] \\
\noalign{\smallskip}
\hline
\noalign{\smallskip}
$C_{1,0}$ & $+865.07$ & (fixed) &&
$S_{1,0}$ & $+659.83$ & $+605.66$ \\
$C_{2,0}$ & $-111.76$ & $-134.66$ &&
$S_{2,0}$ & $-85.26$ & $-77.34$ \\
$C_{3,0}$ & $-67.84$ & $-76.14$ &&
$S_{3,0}$ & $-65.91$ & $-63.34$ \\
$C_{4,0}$ & $+18.26$ & $+24.98$ &&
$S_{4,0}$ & $+17.79$ & $+19.40$ \\
$C_{5,0}$ & $+3.20$ & $+7.42$ &&
$S_{5,0}$ & $-0.20$ & $-6.44$ \\
$C_{6,0}$ & $+3.51$ & $+6.31$ && 
$S_{6,0}$ & $+0.68$ & $+1.02$ \\
$C_{7,0}$ & $+0.03$ & $+1.45$ &&
$S_{7,0}$ & $+0.34$ & $-0.31$ \\
$C_{8,0}$ & $-0.62$ & $-2.87$ &&
$S_{8,0}$ & $-0.59$ & $-6.56$ \\
\noalign{\smallskip}
\hline
\end{tabular}
\end{table}

\begin{figure} 
\resizebox{0.7\hsize}{!}{
\includegraphics{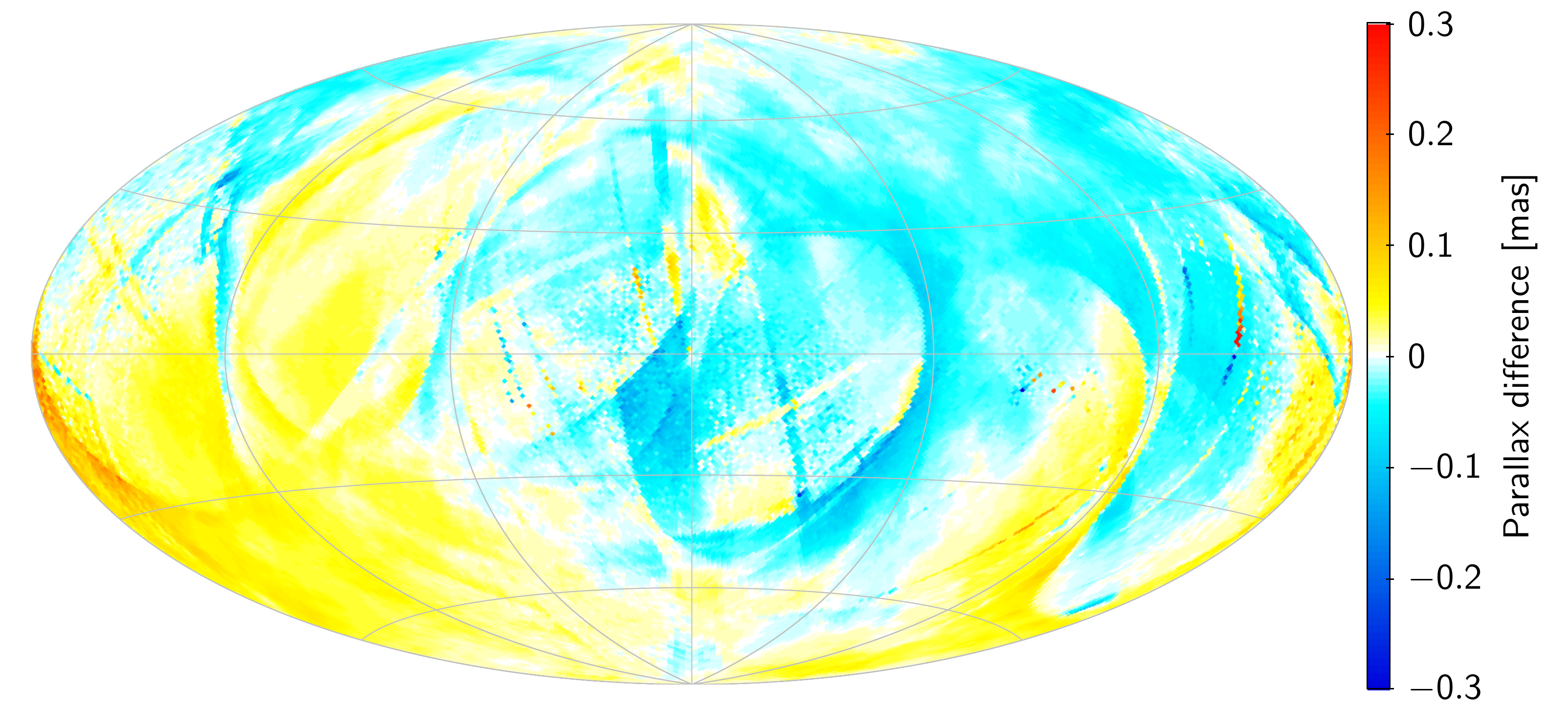}
} 
\caption{Differences in parallax between the baseline primary solution (where basic-angle 
variations are corrected based on BAM data) and a special validation solution where 
the harmonic coefficients $C_{k,0}$, $S_{k,0}$ (except $C_{1,0}$) were instead 
estimated as global parameters in the solution.
Median differences are shown in cells of about 0.84~deg$^2$. The map uses an 
Aitoff projection in equatorial (ICRS) coordinates, with origin $\alpha=\delta=0$ 
at the centre and $\alpha$ increasing from right to left.}
\label{fig:valVbacF}
\end{figure}

Special algorithms and software packages to recover both basic-angle
and differential variations have been developed and tested in
AGIS. This software will be used in the astrometric solutions of
future data releases to mitigate these effects. The software was
however not used for the current baseline solution, for which we
instead assume that the BAM provides adequate corrections.
Nevertheless, for validation purposes we have made TGAS runs
where the harmonic coefficients $C_{k,m}$, $S_{k,m}$ are estimated as
global parameters for $k=1\dots 8$ and $m=0,\,1$ (but excluding
$C_{1,0}$). The time interval covered by the current data is not long
enough to reliably estimate the linear time-dependent coefficients
($m=1$). Results for the time-independent coefficients ($m=0$) are
shown in Table~\ref{tab:valVbacF} along with the corresponding
coefficients estimated from the BAM data. In this solution $C_{1,0}$
was fixed at its value according to the BAM. In general the
coefficients obtained in the TGAS run are in good agreement with
the BAM data; the largest difference (about 0.05~mas) is obtained for
$S_{1,0}$. The corresponding parallax differences (baseline solution
minus special validation solution), shown in
Fig.~\ref{fig:valVbacF}, have a median value of $+0.006$~mas and an RSE
of 0.035~mas. The distinct asymmetry in ecliptic latitude, with an
amplitude of about 0.05~mas, is related to the particular differences in the
values of $C_{k,m}$ and $S_{k,m}$ as given in Table~\ref{tab:valVbacF}
and most importantly to the difference in $S_{1,0}$.

\subsection{Changing the attitude model}
\label{sec:valKnots}
  
The attitude model is completely defined by the knot sequence, which for the baseline
primary solution uses a regular sequence with a knot interval of 30~s, but
with additional knots inserted at certain times to allow discontinuities in the 
attitude quaternion in connection with data gaps. Changing the knot sequence
results in a different solution, which is not necessarily better, but the difference
between two such solutions (e.g.\ in parallax) gives an indication of how 
critical the attitude modelling is. For a regular knot sequence, the main 
parameters that can be changed are the time of the first knot and the 
interval between successive knots. Here we describe briefly the results of 
validation runs implementing such changes.
(The order of the spline is also a configurable parameter in AGIS,
but all solutions described in this paper use a fourth-order, or cubic, spline.)

\begin{figure}[t] 
\resizebox{\hsize}{!}{\includegraphics{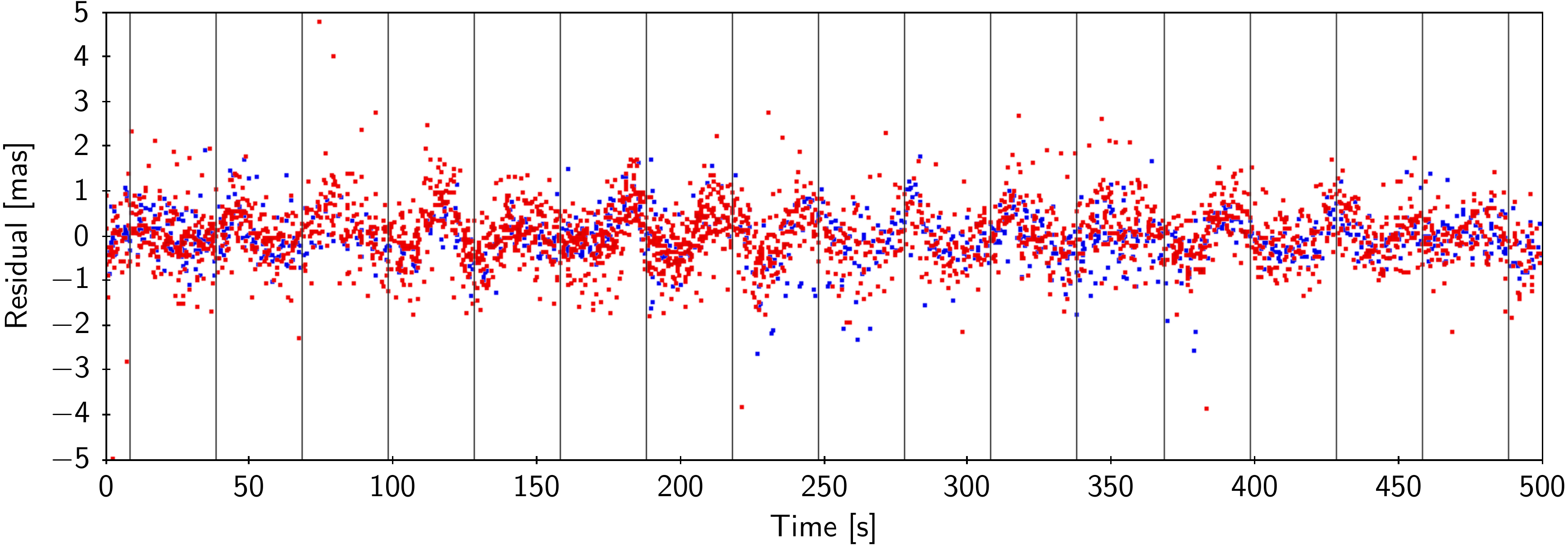}} 
\resizebox{\hsize}{!}{\includegraphics{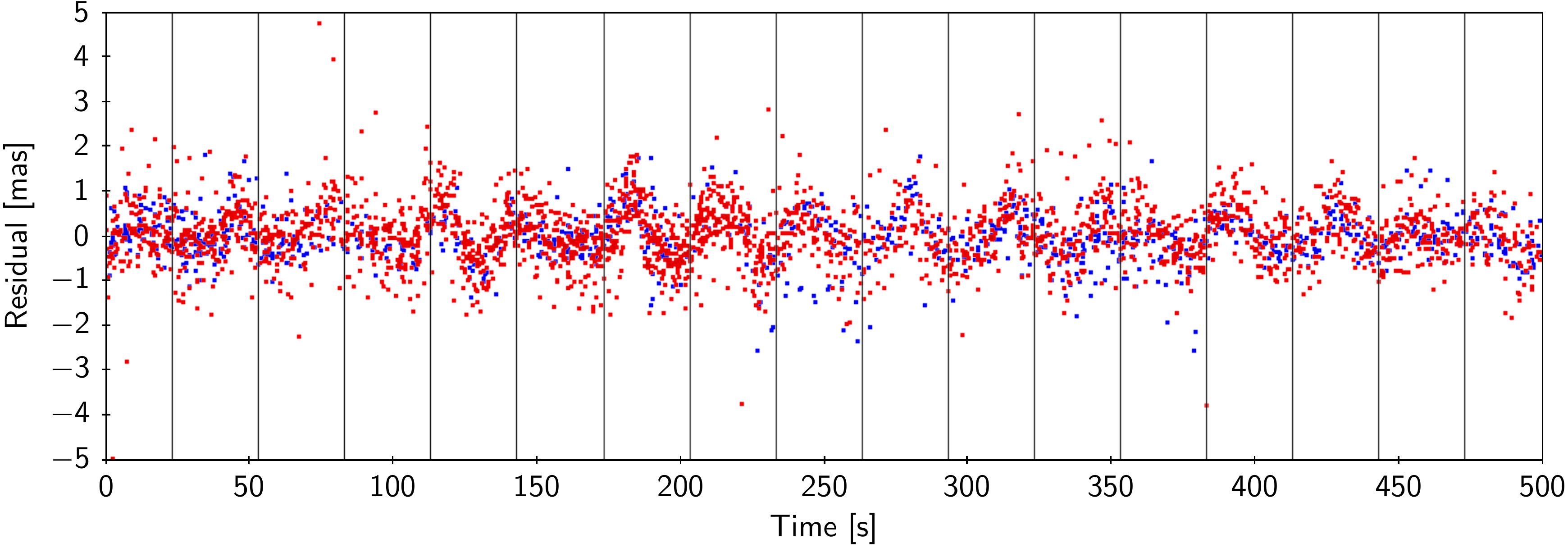}} 
\resizebox{\hsize}{!}{\includegraphics{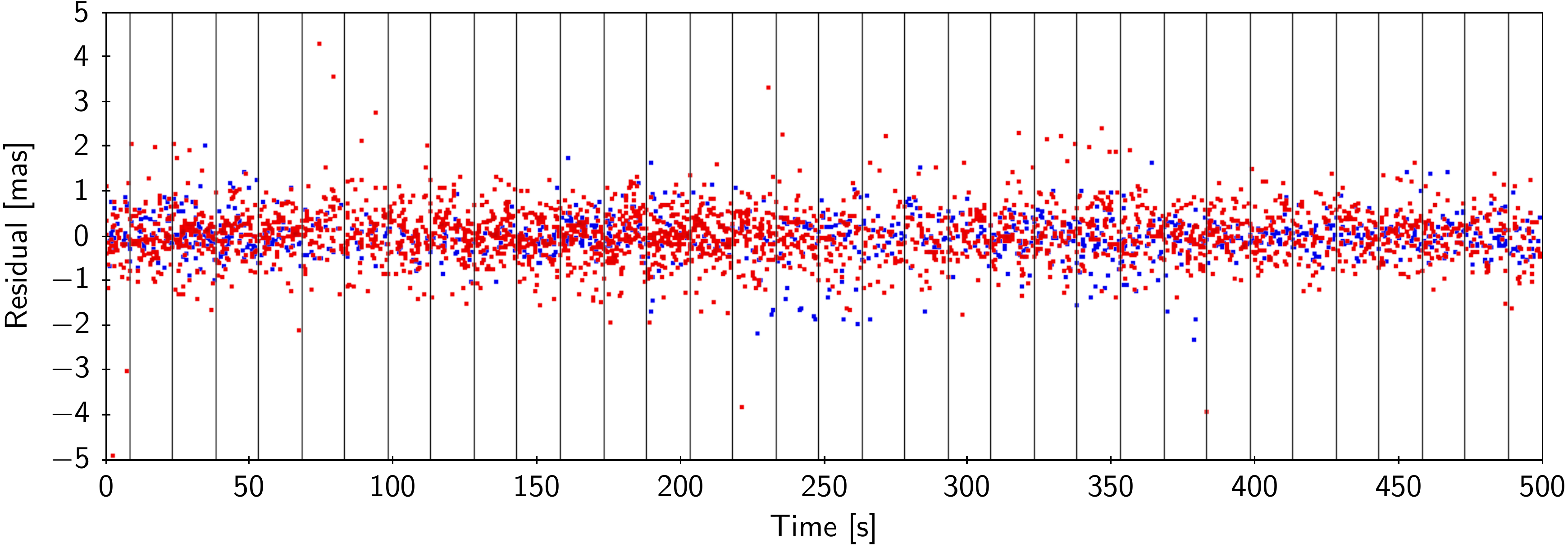}} 
\caption{Along-scan residuals of individual CCD observations vs.\ time for
an interval of 500~s. Blue and red dots represent observations in the preceding and 
following field of view, respectively. Only good observations are shown.
The vertical grey lines mark the times of the knots for the attitude spline.
Top: residuals from the baseline primary solution. Middle: residuals from the
special validation solution with shifted knot sequence.
Bottom: residuals from the special validation solution using a shorter knot interval 
of 15~s.}
\label{fig:statResTime}
\end{figure}

\paragraph{Shifting the knot sequence}

A validation solution was computed using the same 30~s knot interval as in the baseline 
solution, but shifting the knots by 15~s. For most of the time, this has very little effect
on the residuals. This can be seen by comparing the top two panels of Fig.~\ref{fig:statResTime}. 
Some details are clearly different,
but the main features, in particular the quasi-periodic wiggles, are almost identical.
This shows that these features represent real high-frequency ($\gtrsim 0.03$~Hz) 
components of the AL attitude irregularities, with the spline fitting basically
acting as a high-pass filter. Significant differences are seen in connection with the
larger micrometeoroid hits, but they affect only a small fraction of the time and sources.
The global RSE of the parallax difference between the two solutions is 0.024~mas.

\paragraph{Reducing the knot interval}

The 30~s knot interval used for the baseline primary solution was chosen because
it gives at all times a sufficient number of AL and AC observations per knot
interval. Reducing the knot interval to 15~s gives a less stable attitude solution, but
the attitude modelling errors are significantly reduced, as shown by the bottom panel
of Fig.~\ref{fig:statResTime}. Compared with the baseline solution, the RSE of the 
residuals is reduced by about 15\%.
The global RSE of the parallax difference between the 
baseline solution and the solution using a 15~s knot interval is 0.069~mas.

\end{document}